\documentclass[aps,prd,reprint,nofootinbib]{revtex4-1}
\usepackage{epsfig}
\usepackage{amsmath,amssymb,amsfonts}
\usepackage{bbm}
\usepackage{graphicx}
\usepackage[T1]{fontenc}
\usepackage[utf8]{inputenc}
\usepackage[usenames,dvipsnames]{xcolor}
\definecolor{bluecite}{HTML}{0875b7}
\usepackage[colorlinks=true,urlcolor=Emerald,linkcolor=bluecite,citecolor=bluecite]{hyperref}
\usepackage[english]{babel}
\usepackage{mathbbol}
\usepackage[normalem]{ulem}
\usepackage{acronym}

\newacro{UV}[UV]{ultraviolet}
\newacro{IR}[IR]{infrared}
\newacro{QFT}[QFT]{quantum field theory}
\newacro{EFT}[EFT]{effective field theory}
\newacro{GR}[GR]{General Relativity}
\newacro{FRG}[FRG]{Functional Renormalisation Group}
\newacro{RG}[RG]{renormalisation group}
\newacro{MES}[MES]{minimal essential scheme}

\newcommand{\GRUC}{{\textbf{\textit{[\small GR]}}}}

\newcommand{\Euler}{\ensuremath{\mathfrak E}}

\newcommand{\measure}[2]{\ensuremath{\mu\left(#1 \bigg| #2 \right)}}

\newcommand{\HKbracket}[2]{\ensuremath{\mathbf{\left\{\left\{ #1 \right\}\right\}_{#2}}}}

\newcommand{\eg}{{\textit{e.g.}}}
\newcommand{\ie}{{\textit{i.e.}}}

\allowdisplaybreaks

\begin{document}

\title{Momentum-dependent field redefinitions in Asymptotic Safety}
\author{Benjamin Knorr\,\href{https://orcid.org/0000-0001-6700-6501}{\protect \includegraphics[scale=.07]{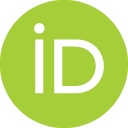}}\,}
\email[]{benjamin.knorr@su.se}
\affiliation{
Nordita, Stockholm University and KTH Royal Institute of Technology, Hannes Alfv\'ens v\"ag 12, SE-106 91 Stockholm, Sweden
}

\begin{abstract}
We discuss general momentum-dependent field redefinitions in the context of quantum-gravitational scattering amplitudes in general, and Asymptotic Safety in particular. Implementing such redefinitions at the lowest curvature order, we can bring the graviton propagator into tree-level form, avoiding issues of fiducial ghost poles and their associated violations of unitarity. We compute the beta function for Newton's constant, and find an asymptotically safe fixed point whose critical exponent changes by $0.4\%$ compared to not resolving the momentum-dependent field redefinition. This provides a strong indication that this fixed point does not feature extra degrees of freedom related to ghostly modes, and has a good chance of being related to a unitary theory.
\end{abstract}

\maketitle

\section{Introduction}\label{sec:intro}

One of the biggest open problems in theoretical physics is that of finding a single consistent description of all fundamental forces -- gravity on the one hand, and the electroweak and strong interactions on the other -- together with the observed matter like electrons. The two theories that we use to describe them, respectively -- \ac{GR} and the Standard Model of Particle Physics -- cannot be easily extended to include the interactions they are missing. In particular, the perturbative quantisation of \ac{GR} is plagued by difficulties \cite{tHooft:1974toh, Goroff:1985sz, Goroff:1985th, Stelle:1976gc}, while the Standard Model is formulated in Minkowski space.

Progress is hindered by the general expectation that quantum gravity effects are suppressed by the Planck scale, and correspondingly they are experimentally hard to probe. In such a situation with little experimental guidance, consistency conditions play a central role on the route to an all-encompassing theory. For example, one can require the theory to be unitary and (macroscopically) causal. Beyond such theoretical constraints, the theory should also satisfy a version of the correspondence principle: if the putative quantum theory of gravity and matter is not compatible with Standard Model data (\eg\ if it would predict that the Higgs boson has the same mass as the electron), it is ruled out, even though the latter is not directly related to quantum gravity.

In this work, we assume that such a successful theory of quantum gravity and matter can be a \ac{QFT}. The necessary \ac{UV} completion is assumed to come from an asymptotically safe fixed point of the \ac{RG} flow \cite{Reuter:2019byg}. In this setup, one has access to standard \ac{QFT} notions like scattering amplitudes and the effective action, allowing a straightforward connection to both theoretical consistency conditions like unitarity and causality in the form of bounds \cite{Froissart:1961ux, Cerulus:1964cjb, Alberte:2021dnj, deRham:2022hpx}, and to experimental data via \eg\ scattering cross sections.

A key long-term goal to test the Asymptotic Safety conjecture beyond the already significant evidence for the existence of the fixed point (see \cite{Knorr:2022dsx, Eichhorn:2022gku, Morris:2022btf, Martini:2022sll, Wetterich:2022ncl, Platania:2023srt, Saueressig:2023irs, Pawlowski:2023gym} for a collection of recent book chapters on the topic) is to compute suitable scattering amplitudes and simultaneously confront them with as many consistency conditions as possible. While the main motivation for this is to connect to the real world, it has been pointed out recently that causality imposes non-trivial constraints between different couplings at the fixed point \cite{Draper:2020bop}.

Such a first-principle scattering amplitude programme consists of several steps: first, for a given scattering event, one has to characterise all relevant correlation functions in the effective action that contribute. A significant amount of work has already gone into this aspect for two-to-two scattering \cite{Knorr:2019atm, Draper:2020bop, Draper:2020knh, Knorr:2022lzn, Knorr:2022dsx}. Second, the relevant momentum-dependent correlation functions have to be computed. In Asymptotic Safety, the graviton propagator \cite{Christiansen:2014raa, Christiansen:2017bsy, Bosma:2019aiu, Knorr:2021niv, Bonanno:2021squ, Fehre:2021eob}, and the three- \cite{Christiansen:2015rva} and four-graviton vertex \cite{Denz:2016qks} have been investigated, plus some matter propagators and gravity-matter vertices \cite{Meibohm:2015twa, Christiansen:2017cxa, Eichhorn:2018akn, Eichhorn:2018nda, Knorr:2019atm, Burger:2019upn, Pastor-Gutierrez:2022nki}. Recent groundbreaking work also paved the way towards Lorentzian-signature computations \cite{Bonanno:2021squ, Fehre:2021eob} which is clearly crucial for scattering amplitudes. Finally, the resulting amplitudes have to be confronted with the constraints of interest \cite{Draper:2020bop, Draper:2020knh}.

Not all parts of correlation functions are equally important for scattering amplitudes. This is due to the fact that we can perform field redefinitions without changing the physics \cite{Wegner_1974}. A trivial example of this is a rescaling of the field by a positive constant. As a consequence, only specific combinations of couplings will enter any observable. These combinations are called essential couplings. Inessential couplings in turn do not change the amplitude, and thus can be changed (almost) arbitrarily by a field redefinition. It is clear that with a suitable field redefinition, the task of computing scattering amplitudes can be drastically simplified. This observation is also central for the very definition of Asymptotic Safety: only essential couplings need to reach a fixed point \cite{Weinberg:1980gg}, and only their perturbations about a given fixed point define proper critical exponents \cite{Dietz:2013sba}.

Most previous investigations of Asymptotic Safety did not at all, or only partially, take into account field redefinitions.\footnote{In fluctuation computations, momentum-dependent but field-independent field redefinitions have been taken into account since the beginning, see \eg{} \cite{Christiansen:2014raa, Christiansen:2015rva, Denz:2016qks, Christiansen:2017bsy, Pawlowski:2020qer}, in contrast to what has been claimed recently \cite{Kawai:2023rgy}.} Only very recently, a modified non-perturbative flow equation has been derived that employs this freedom systematically \cite{Baldazzi:2021ydj}, being based on earlier related work \cite{Gies:2001nw, Pawlowski:2005xe}. Since then, it has been applied to quantum gravity by itself \cite{Baldazzi:2021orb}, and quantum gravity coupled to a shift-symmetric scalar field \cite{Knorr:2022ilz}.

From these considerations, the clear path forward towards asymptotically safe scattering amplitudes is to employ the modified flow equation to self-consistently compute the essential parts of the necessary correlation functions. As a first step in this endeavour, and to make contact to previous computations in the standard scheme, in this work we will compute the non-perturbative beta function of the Newton's constant while taking into account a momentum-dependent field redefinition of the metric that removes any non-trivial momentum dependence from the flat space graviton propagator.

This paper is structured as follows: we start by a general discussion of field redefinitions in the context of scattering amplitudes in \autoref{sec:amps}, giving us the motivation to study such redefinitions in Asymptotic Safety. In \autoref{sec:setup}, we introduce the setup and some technical choices. Then, \autoref{sec:analysis} is devoted to an in-depth discussion of the non-perturbative flow equations and the resulting phase diagram. We conclude with a brief summary and outlook in \autoref{sec:summary}. The present work illustrates how making use of field redefinitions significantly shifts the frontiers in terms of technical feasibility, and we provide two appendices where a lot of the underlying background material is collected. In appendix \ref{app:HK}, we collect the necessary non-local heat kernel coefficients and illustrate their efficient computation. In appendix \ref{app:tracecomputation}, we display the complete step-by-step derivation of the flow.

\section{Scattering amplitudes and field redefinitions}\label{sec:amps}

Before we treat the system under consideration itself, we will provide a general discussion of field redefinitions at the level of the effective action and gravity-mediated two-to-two scattering amplitudes. Similar considerations carry over to more general scattering events. For definiteness, we consider the two-to-two scattering of a massive $\mathbbm Z_2$-symmetric scalar field $\phi$ into itself, that is $\phi\phi\to\phi\phi$. In a Minkowski background, the relevant part of the effective action and the full scattering amplitude have been derived in \cite{Draper:2020knh}.\footnote{Here, we have chosen a slightly different basis that is more useful for our discussion, and that is in line with \cite{Knorr:2022ilz}.} The gravitational part of the effective action that contributes to this scattering reads
\begin{equation}\label{eq:Gammagrav}
\begin{aligned}
    \Gamma_\text{grav} \simeq \frac{1}{16\pi G_N} \int \text{d}^4x \, \sqrt{-g} \, \bigg[ -R + &R f_{RR}(\Box) R \\ &+ S^{\mu\nu} f_{SS}(\Box) S_{\mu\nu} \bigg] \, .
\end{aligned}
\end{equation}
Here, $S_{\mu\nu}$ is the trace-free part of the Ricci tensor and $R$ is the Ricci scalar of the metric $g$, $G_N$ is Newton's constant, $\Box = -D^2$ is the covariant d'Alembertian, we have set the cosmological constant to zero, and we neglected terms of cubic order in the curvature as well as the Gauss-Bonnet term since they do not contribute to the scattering event. The functions $f_{RR,SS}$ are called form factors, and they contain the information on the momentum dependence of correlation functions in a diffeomorphism-invariant way \cite{Knorr:2019atm}. The relevant scalar part of the action is
\begin{equation}\label{eq:Gammaphi}
\begin{aligned}
    \Gamma_\phi \simeq \int \text{d}^4x \, &\sqrt{-g} \, \bigg[ \frac{1}{2} \phi f_{\phi\phi}(\Box) \phi + f_{R\phi\phi}(\Box_1,\Box_2,\Box_3) \, R \, \phi \, \phi \\
    & + f_{S\phi\phi}(\Box_1,\Box_2,\Box_3) \, S^{\mu\nu} \, (D_\mu \phi) \, (D_\nu \phi) \\
    & + f_{\phi\phi\phi\phi} \left( \left\{ -D_i \cdot D_j \right\}_{1 \leq i < j \leq 4} \right) \phi \, \phi \, \phi \, \phi \bigg] \, .
\end{aligned}
\end{equation}
In this, a subscript on an operator indicates the position of the field that it acts upon, \eg
\begin{equation}
    \Box_1 \Box_2^2 \Box_3^3 S^{\mu\nu} (D_\mu \phi) (D_\nu \phi) = \left( \Box S^{\mu\nu} \right) \left( \Box^2 D_\mu \phi \right) \left( \Box^3 D_\nu \phi \right) ,
\end{equation}
and\footnote{There is an order ambiguity for the form factor $f_{\phi\phi\phi\phi}$ since its arguments do not commute. However, any difference in the ordering is proportional to a commutator of covariant derivatives, thus all choices lead to the same contribution for our two-to-two scattering process. We can thus safely ignore this subtlety for the present work.}
\begin{equation}
    D_i \cdot D_j = D_{i\alpha} D_j^\alpha \, .
\end{equation}
So far, no field redefinitions have been performed, and the $\phi\phi\to\phi\phi$ scattering amplitude in Minkowski space derived from this action is fully general.

Before we perform field redefinitions to simplify the action, and as a result the scattering amplitude, let us briefly discuss some subtleties. By definition, a field redefinition should be invertible. More concretely, no physical modes should be added to, or subtracted from, the spectrum. A clear example for an inadmissible redefinition in the case of a free massive scalar field is
\begin{equation}
    \phi \mapsto \frac{1}{\sqrt{\Box-m^2}} \phi \, .
\end{equation}
This maps the action to that of a non-dynamical field, clearly removing the propagating mode.\footnote{We hasten to add that in a more realistic scenario with interactions, the total scattering amplitude will still be the same under this redefinition. However, the standard \ac{QFT} interpretation of propagators and vertices, as well as the meaning of on-shellness, get obscured in this way. In the following, we will not consider these redefinitions to avoid such pathologies.}

From this example, it is clear that if the goal is a maximal simplification of all correlation functions, one has to make certain assumptions on the spectrum of the theory that one aims to investigate, which should be verified a posteriori. One is thus bound to a subset of theories connected to a specific universality class.\footnote{Strictly speaking, within such a subspace of the theory space with fixed spectrum, there can still be multiple fixed points defining different universality classes in the usual sense. Here we are using this generalised notion of universality class, refering to theories with a fixed spectrum.} For example, in this work (and in all previous works on the topic \cite{Baldazzi:2021orb, Knorr:2022ilz}), we will investigate what we shall call the universality class of \ac{GR} (in short \GRUC{}), which has two massless propagating graviton polarisations in Minkowski space, and no other degrees of freedom. By contrast, one could call a gravity theory with an action that is field-redefinition-equivalent to Stelle's action \cite{Stelle:1976gc, Stelle:1977ry} the Stelle universality class, described by an additional massive spin two ghost, and a regular spin zero mode.

Specifying the universality class does not completely fix the inessential couplings yet -- there are still different schemes, that is different choices for the inessential couplings within the same universality class. For example, the class of non-local higher-derivative gravity theories is in \GRUC{}, but has a non-minimal momentum dependence for the graviton propagator \cite{BasiBeneito:2022wux}. In the following, we will focus on what has been called the \ac{MES} \cite{Baldazzi:2021ydj, Baldazzi:2021orb} -- this means that all couplings that can be set to zero within the universality class will be set to zero. There are clearly other physically equally-good choices, but anticipating the complexity of amplitude computations, the minimal scheme seems to be preferred.

Let us now come back to the action \eqref{eq:Gammagrav} and \eqref{eq:Gammaphi} and perform a field redefinition. Assuming that there are no additional modes means that the form factors $f_{RR}, f_{SS}$ and $f_{\phi\phi}$ do not introduce any additional poles in the propagator, and can thus be removed. To be explicit, the scalar propagator in flat spacetime with momentum $p$ computed from \eqref{eq:Gammaphi} reads
\begin{equation}
    \mathfrak G^\phi(p^2) = \frac{1}{f_{\phi\phi}(p^2)} \, ,
\end{equation}
so that $f_{\phi\phi}(p^2)$ needs to have a unique zero at $p^2=m^2$:
\begin{equation}\label{eq:fphiphiconditions}
    f(m^2) = 0 \, , \, f(z\neq m^2) \neq 0 \, , \, z \in \mathbbm R : \frac{f(z)}{z-m^2} > 0 \, .
\end{equation}
Likewise, the flat spacetime graviton propagator computed from \eqref{eq:Gammagrav} reads
\begin{equation}
\begin{aligned}
    \mathfrak G^\text{h}(p^2) &= \frac{1}{p^2 \left( 1+p^2 f_{SS}(p^2) \right)} \Pi_2 \\
    & \quad - \frac{1}{p^2 \left( 1 - 6 p^2 f_{RR}(p^2) - \frac{p^2}{2} f_{SS}(p^2) \right)} \Pi_0 \, ,
\end{aligned}
\end{equation}
where $\Pi_{0,2}$ are the projectors onto the spin zero and spin two part \cite{Knorr:2021niv}, and we suppressed the spin one component which is pure gauge. The conditions for the form factors to not introduce extra poles in this case read
\begin{equation}\label{eq:spintwoconditions}
\begin{aligned}
    & & 1+z f_{SS}(z) &\neq 0 \, , \\
    &z \in \mathbbm R :& \qquad 1+z f_{SS}(z) &> 0 \, ,
\end{aligned}
\end{equation}
and
\begin{equation}\label{eq:spinzeroconditions}
\begin{aligned}
    & & 1 - 6 z f_{RR}(z) - \frac{z}{2} f_{SS}(z) &\neq 0 \, , \\
    &z \in \mathbbm R :& \qquad 1 - 6 z f_{RR}(z) - \frac{z}{2} f_{SS}(z) &> 0 \, .
\end{aligned}
\end{equation}
The inequalities follow from requiring that the physical modes do not turn into ghosts for real squared momenta.

Having settled these conditions, we can now remove the inessential form factors by making the field redefinitions
\begin{align}
    g_{\mu\nu} &\mapsto g_{\mu\nu} + a_R(\Box) R g_{\mu\nu} + a_S(\Box) S_{\mu\nu} \, , \label{eq:FRmetric} \\
    \phi &\mapsto a_\phi(\Box) \phi \, .
\end{align}
With the choice
\begin{equation}\label{eq:a_expression}
\begin{aligned}
    a_R(\Box) &= -\frac{1}{2\Box} \bigg( 1 - \frac{1}{3} \frac{1}{\sqrt{1 + \Box f_{SS}(\Box)}} \\
    &\qquad - \frac{2}{3} \frac{1}{\sqrt{1 - 6 \Box f_{RR}(\Box) - \frac{1}{2} \Box f_{SS}(\Box)}} \bigg)\, , \\
    a_S(\Box) &= \frac{2}{\Box} \left( \frac{1}{\sqrt{1 + \Box f_{SS}(\Box)}} - 1 \right) \, , \\
    a_\phi(\Box) &= \sqrt{\frac{\Box-m^2}{f_{\phi\phi}(\Box)}} \, ,
\end{aligned}
\end{equation}
we can remove $f_{RR}$ and $f_{SS}$ from the action and put the scalar kinetic term into a standard form. Note that by assumption, this field redefinition is well-defined due to \eqref{eq:fphiphiconditions}. Likewise, the combinations in the denominators of the metric field redefinitions are positive if no other poles are introduced by the form factors $f_{RR}$ and $f_{SS}$, see \eqref{eq:spintwoconditions} and \eqref{eq:spinzeroconditions}. We emphasise that the $a_{R,S}$ are both regular at the origin:\footnote{Here we note that terms like the one-loop logarithms deserve an individual discussion that goes beyond what we want to do here.}
\begin{equation}
\begin{aligned}
    a_R(z) &\sim f_{RR}(0) \, ,& \qquad z &\to 0 \, , \\
    a_S(z) &\sim -f_{SS}(0) \, ,& \qquad z &\to 0 \, .
\end{aligned}
\end{equation}
As a side note, we emphasise that the field redefinitions \eqref{eq:a_expression} circumvent the issue of fiducial ghosts that unavoidably appear in a derivative expansion of the effective action \cite{Platania:2020knd, Platania:2022gtt}. The viewpoint that we take here is that of performing the field redefinition at the exact level, preserving properties like unitarity while keeping computations simple, and only at the end we perform a derivative expansion if needed.

As a matter of fact, also the form factors $f_{R\phi\phi}$ and $f_{S\phi\phi}$ can be completely removed, and $f_{\phi\phi\phi\phi}$ can be simplified, but we will not show the details here. The general rule is that whenever a given operator in the effective action is proportional to the equation of motion (or can be completed to it by shifting some couplings), it is inessential. In practice, this gives rise to a bootstrap: one starts with the action that defines the underlying dynamics, \ie{}, the universality class. Then, new operators are added. If they are proportional to the equations of motion, they are inessential. If not, we add them to the original action, and use the new equations of motion that include the new operators. In a derivative or curvature expansion, this will generally not render formerly inessential operators essential. As an example, for \GRUC{}, this means that any operator involving the Ricci scalar or the trace-free Ricci tensor is inessential.

Coming back to our scattering example, via a suitable field redefinition and within \GRUC{}, we can map
\begin{equation}\label{eq:Gammagravredef}
    \Gamma_\text{grav} \mapsto \frac{1}{16\pi G_N} \int \text{d}^4x \, \sqrt{-g} \, \bigg[ -R + \mathcal O(R^3) \bigg] \, ,
\end{equation}
and
\begin{equation}
\begin{aligned}
    \Gamma_\phi \mapsto &\int \text{d}^4x \, \sqrt{-g} \, \bigg[ \frac{1}{2} \phi (\Box - m^2) \phi \\
    &+ \tilde f_{\phi\phi\phi\phi}(-D_1\cdot D_2, -D_1 \cdot D_3) \, \phi \, \phi \, \phi \, \phi + \mathcal O(\phi^6) \bigg] \, .
\end{aligned}
\end{equation}

Let us now come to the scattering amplitude for the theory after field redefinition. The full $\phi\phi\to\phi\phi$ amplitude can be split into $s$-, $t$- and $u$-channel for the gravity-mediated diagram, and the dressed vertex,
\begin{equation}
    \mathcal A = \mathcal A_s + \mathcal A_t + \mathcal A_u + \mathcal A_4 \, .
\end{equation}
Here, $s,t,u$ denote the standard Mandelstam variables, and we follow the convention of \cite{Draper:2020knh}. For the gravity-mediated contribution, after the field redefinition we get
\begin{equation}
    \mathcal A_s = 8\pi G_N \frac{t \,u}{s} \, .
\end{equation}
The partial amplitudes for the $t$- and $u$-channel follow from crossing symmetry. Clearly, this is just the expression for the amplitude stemming from \ac{GR} itself. All the non-trivial momentum dependence is contained in the vertex diagram,
\begin{equation}
    \mathcal A_4 = \mathcal A_4(s,t) \, .
\end{equation}
The bottom line of this is that by the field redefinitions, generally, (some of) the non-trivial momentum dependence of low order correlation functions is moved into higher-order correlation functions. For the two-to-two scattering, all non-trivial information is contained in the contact term. Clearly, no information is lost: after all, the full amplitude is a function of two of the three Mandelstam variables, but so is $\tilde f_{\phi\phi\phi\phi}$ which appears in $\mathcal A_4$. The key advantage is that the lower order correlation functions are trivial. This is particularly beneficial for the propagator, since in an \ac{RG} flow, it needs regularisation, which is much easier to achieve consistently for a simple momentum dependence.

The observation that most non-trivial momentum dependence in a two-to-two scattering process can be shifted into the dressed four-point vertex is rather generic. The only non-minimal contribution towards mediated diagrams can come from essential three-field operators. For such a correlator, we can always parameterise the corresponding form factor by three $\Box$ operators \cite{Knorr:2019atm}. Since equations of motion are typically related to this operator acting on a field, any non-trivial momentum dependence in a three-point correlator can be removed by a field redefinition. The only exception are purely local terms without $\Box$ operators. Let us give two concrete examples.

First, for a four-photon scattering, $\gamma\gamma\to\gamma\gamma$, there is one such local essential interaction,
\begin{equation}
    \Gamma_{\gamma}^{\{3\}\text{ess}} = \int \text{d}^4x \, \sqrt{-g} \, c_{CFF} C^{\mu\nu\rho\sigma} F_{\mu\nu} F_{\rho\sigma} \, .
\end{equation}
Here, $C$ is the Weyl tensor, $F$ is the field strength tensor of the photon, and $c_{CFF}$ is a coupling constant. This interaction term cannot be removed by a field redefinition, and clearly contributes to a photon-photon-graviton vertex. All other possible contractions of one Weyl tensor, two field strength tensors and any number of covariant derivatives can either be removed by a field redefinition, or rewritten by partial integration and Bianchi identities into terms that then can be removed by a field redefinition \cite{Knorr:2022lzn}.

The second example is that of four-graviton scattering.\footnote{There are clearly limitations on the validity of considering such a process, at least at high energies, but this is beyond the scope of this discussion.} In this case and for \GRUC{}, the only extra essential term contributing to the mediated diagrams in four dimensions\footnote{There is a second independent contraction of three Weyl tensors in dimensions larger than 5 \cite{Fulling:1992vm}, giving rise to an extra essential coupling.} is the well-known Goroff-Sagnotti term \cite{Goroff:1985sz, Goroff:1985th},
\begin{equation}
    \Gamma_\text{grav}^{\{3\}\text{ess}} = \int \text{d}^4x \, \sqrt{-g} \, c_{C^3} C_{\mu\nu}^{\phantom{\mu\nu}\rho\sigma} C_{\rho\sigma}^{\phantom{\rho\sigma}\tau\omega} C_{\tau\omega}^{\phantom{\tau\omega}\mu\nu} \, .
\end{equation}
One can once again show that all other combinations of three Weyl tensors and covariant derivatives can be reduced to inessential terms. This is also due to the identity
\begin{equation}\label{eq:BoxC}
    D^\mu D_{[\mu} C_{\nu\rho]\alpha\beta} = 0 \, ,
\end{equation}
that relates $\Box C$ to covariant derivatives of trace-free Ricci tensors and Ricci scalars via the Bianchi identity.

Let us point out that the Goroff-Sagnotti term can also be removed via a redefinition due to \eqref{eq:BoxC}, but it is \emph{non-local} \cite{Krasnov:2009ik}. Here, by non-local we mean operators that have poles at vanishing momentum when evaluated in Minkowski space. Allowing for such redefinitions would also make it possible to remove \emph{any} higher order gravitational term, so that graviton scattering would be described by just \ac{GR}. We however generally expect that such redefinitions interfere with a standard interpretation of scattering amplitudes. To see this, we construct a similar redefinition for a quartic scalar field theory described by the action
\begin{equation}
    \Gamma_\phi^\text{NLFR} = \int \text{d}^4x \, \sqrt{-g} \, \bigg[ \frac{1}{2} \phi \Box \phi - \frac{\lambda}{4!} \phi^4 \bigg] \, .
\end{equation}
If we now introduce a new field $\Phi$ via the non-local definition
\begin{equation}\label{eq:NLFR}
    \phi = \Phi + \frac{\lambda}{4!} \frac{1}{\Box} \Phi^3 + \frac{7}{2} \left(\frac{\lambda}{4!}\right)^2 \frac{1}{\Box} \left[ \Phi^2 \frac{1}{\Box} \Phi^3 \right] + \mathcal O(\lambda^3) \, ,
\end{equation}
our theory is mapped to a free theory (up to higher order terms in $\lambda$),
\begin{equation}
    \Gamma_\Phi^\text{NLFR} = \int \text{d}^4x \, \sqrt{-g} \, \bigg[ \frac{1}{2} \Phi \Box \Phi + \mathcal O(\lambda^3) \bigg] \, .
\end{equation}
We can extend \eqref{eq:NLFR} order by order in $\lambda$ to remove all higher order terms. The formally resummed redefinition reads\footnote{This expression should be understood as a power series in $\lambda$, and in this expansion, all inverse operators act on everything to their right.}
\begin{equation}\label{eq:NLFR_full}
    \Phi = \left[ 1 - \frac{\lambda}{12} \frac{1}{\Box} \phi^2 \right]^{1/2} \phi \, ,
\end{equation}
where we gave the inverse transformation of \eqref{eq:NLFR} since its closed form is simpler. The new field $\Phi$ is non-interacting, and thus scatters trivially. By contrast, we clearly have non-trivial scattering for the field $\phi$. As a consequence, all scattering information must be contained within the redefinition \eqref{eq:NLFR_full}, likely in the boundary conditions needed to appropriately define the inverse operator. For this reason, we will not consider such non-local redefinitions in the following. We stress again that the field redefinition \eqref{eq:a_expression} is not of this form, since any apparent non-locality is spurious, and all functions are regular.

The lessons of this section are as follows:
\begin{itemize}
    \item field redefinitions allow for significant simplifications in the computation of scattering amplitudes,
    \item to perform a field redefinition to bring the amplitude into the simplest form, one has to specify the spectrum of the theory -- not everything goes,
    \item most or all of the non-trivial momentum dependence of a two-to-two scattering amplitude is carried by the dressed contact term.
\end{itemize}
This sets the stage for the rest of this paper. Controlling the propagators is clearly the first important (and easiest) step in the computation of scattering amplitudes. We will thus set up a non-perturbative \ac{RG} flow in gravity for the action \eqref{eq:Gammagravredef} (with a cosmological constant and the Euler term) while taking into account running field redefinitions mimicking \eqref{eq:FRmetric}. This entails that we can track the running of the form factors $a_R$ and $a_S$, and impose that the flat graviton propagator is that of \ac{GR} at every \ac{RG} step. The relevance of our study lies in explicitly checking whether the asymptotically safe fixed point established so far indeed falls into \GRUC{}, avoiding any spurious ghosts that are unavoidable in a derivative expansion \cite{Platania:2020knd, Platania:2022gtt}. It also paves the way towards resolving the three- and four-point correlation functions that are needed to ultimately compute the full amplitudes, and confront first principle predictions from Asymptotic Safety with theoretical and experimental constraints at the level of observables.

\section{Setup}\label{sec:setup}

\subsection{Functional renormalisation group}\label{sec:FRG}

The main tool to investigate Asymptotic Safety is the \ac{FRG}, a non-perturbative formulation of the \ac{RG}. It is formulated in terms of the effective average action $\Gamma_k$, which interpolates between the microscopic action $S$ for $k\to\infty$, and the standard quantum effective action $\Gamma$ for $k\to0$. The dependence of $\Gamma_k$ on the fiducial momentum scale $k$ is governed by \cite{Wetterich:1992yh, Morris:1993qb, Ellwanger:1993mw}
\begin{equation}\label{eq:wetterich}
    \dot \Gamma_k \equiv k \partial_k \Gamma_k = \frac{1}{2} \text{Tr} \left[ \left( \Gamma_k^{(2)} + \mathfrak R_k \right)^{-1} \dot {\mathfrak R}_k \right] \, .
\end{equation}
$\Gamma_k^{(2)}$ is the second functional derivative of the effective average action with respect to the dynamical fields, $\mathfrak R_k$ is a regulator kernel, and the functional trace is a sum over discrete and an integral over continuous variables. For an up-to-date review of the \ac{FRG}, see \cite{Dupuis:2020fhh}, and for reviews of its application to Asymptotic Safety in particular, see \eg{} \cite{Reuter:2019byg, Pawlowski:2020qer} and the recent book chapters \cite{Knorr:2022dsx, Eichhorn:2022gku, Morris:2022btf, Martini:2022sll, Wetterich:2022ncl, Platania:2023srt, Saueressig:2023irs, Pawlowski:2023gym}.

The \ac{FRG} equation \eqref{eq:wetterich} can be used to derive the non-perturbative beta functions of couplings. Given a coupling $\Lambda_k$ with mass dimension $d_\Lambda$, we first make it dimensionless by multiplying with the appropriate power of $k$, so that $\Lambda_k = \lambda_k k^{d_\Lambda}$. The beta function $\beta_\lambda \equiv \dot\lambda_k$ can then be read off by a comparison of coefficients in a given operator basis that spans $\Gamma_k$. A fixed point is then any combination of couplings where all beta functions vanish, $\beta=0$. The behaviour of the flow about a fixed point is determined by the critical exponents, which are minus the eigenvalues of the stability matrix. For a single beta function of a single coupling $\lambda_k$,
\begin{equation}
    \left. \theta = -\frac{\partial \beta_\lambda}{\partial \lambda_k} \right|_{\lambda_k: \beta(\lambda_k)=0} \, .
\end{equation}
Relevant (irrelevant) operators have a positive (negative) critical exponent, and have to be fixed by experiment (are fixed by the flow).

To solve \eqref{eq:wetterich}, in practice we have to make approximations, except in special cases \cite{Knorr:2020rpm}. With our goal of computing (flat spacetime) scattering amplitudes in mind, the most appropriate approximation scheme is the curvature expansion. At order $n$, it retains all operators with up to $n$ curvature tensors, but arbitrary dependence on the covariant derivative. This is the natural scheme since it keeps the full momentum dependence of all flat correlation functions up to order $n$. In curved spacetime, this information is carried by form factors, and the techniques to work with them have been refined recently \cite{Knorr:2019atm, Knorr:2020bjm}.

Due to the gauge structure of gravity, we have to employ the background field method, splitting the metric into an arbitrary background and fluctuations about it,
\begin{equation}
    g_{\mu\nu} = \bar g_{\mu\nu} + h_{\mu\nu} \, .
\end{equation}
In this work, we will also restrict ourselves to the background field approximation, setting the fluctuation field $h$ to zero after the computing the second variation. For work going beyond this, see \eg{} \cite{Pawlowski:2020qer} for a recent review.

\subsection{Essential scheme}\label{sec:ES}

As it stands, \eqref{eq:wetterich} does not take into account the freedom to perform field redefinitions. Much more general flow equations have been known for some time, see \eg{} \cite{Wegner_1974, Pawlowski:2005xe}, and \cite{Wetterich:1997bz, Gies:2001nw} for applications. Recently, these equations have received renewed attention with a specific focus on using field redefinitions to implement the \ac{MES} \cite{Baldazzi:2021ydj}. Subsequently, Asymptotic Safety by itself \cite{Baldazzi:2021orb} and coupled to a shift-symmetric scalar field \cite{Knorr:2022ilz}, as well as the $O(N)$-model \cite{Ihssen:2023nqd} have been investigated.

At the heart of the modified flow equation implementing an essential scheme is the \ac{RG} kernel $\Psi_k$, defined as (the expectation value of) the flow of a field redefinition. With this, the new flow reads
\begin{equation}\label{eq:wetterich_ess}
    \dot \Gamma_k + \Psi_k \Gamma_k^{(1)} = \frac{1}{2} \text{Tr} \left[ \left( \Gamma_k^{(2)} + \mathfrak R_k \right)^{-1} \left( \dot {\mathfrak R}_k + 2 \Psi_k^{(1)} \mathfrak R_k \right) \right] \, .
\end{equation}
Both sides receive an extra contribution. We can then adjust the \ac{RG} kernel to impose conditions on inessential couplings. This is clear since, on the left-hand side, the \ac{RG} kernel multiplies the (full quantum) equations of motion. The \ac{MES} is then defined as that where all inessential couplings are set to zero, or a value compatible with the spectrum of the theory as discussed in \autoref{sec:amps}.

Since taking an expectation value is in general very involved, in practice, an ansatz for $\Psi_k$ is used to compute the \ac{RG} flow \eqref{eq:wetterich_ess}, assuming that it is related to a proper field redefinition. A posteriori, one would then be able to verify that this is indeed the case. So far, this has not been done in practice, and thus remains one of the open questions about this scheme.

Another open question is how to properly treat field redefinitions in the form of gauge transformations. For example, we can shift the metric by
\begin{equation}\label{eq:diffeo}
    g_{\mu\nu} \mapsto g_{\mu\nu} + D_\mu D_\nu R \, ,
\end{equation}
which is clearly a diffeomorphism along the vector $D_\nu R/2$. Such a field redefinition has no effect on the left-hand side of \eqref{eq:wetterich_ess}, but in general it can contribute to the right-hand side. This should clearly not be the case. A potential way out of this conundrum is that an appropriate field redefinition of the Faddeev-Popov ghost must be implemented so that the total contribution from this gauge transformation field redefinition drops out. In practice, this is complicated by the facts that, first, the regularisation of the graviton and ghost sectors cannot be independently chosen, but must be tuned to allow for this cancellation. Second, due to the use of the background field method, the flow equation breaks diffeomorphism symmetry by itself, making such a cancellation even more difficult. We leave this task for a future investigation, and for now we simply do not consider field redefinitions of this type. This is to prevent introducing additional parameters that, due to the reasons explained above, would artificially allow to impose more conditions on the flow than one should be able to.

\subsection{Ansatz}\label{sec:ansatz}

Let us now proceed by presenting our ansatz for the effective average action and the \ac{RG} kernel for which we will solve the flow equation. As mentioned earlier, our systematic expansion scheme is the curvature expansion, where one retains the full dependence on the covariant derivatives of individual correlation functions with up to a certain number of curvatures. It is equivalent to an expansion around a flat background that retains the momentum dependence of the correlation functions. Concretely, we will resolve all operators with up to two spacetime curvatures. From now on, we will restrict ourselves to Euclidean signature -- recent work established that a Wick rotation is possible \cite{Bonanno:2021squ}, and is consistent with a direct computation in Lorentzian signature \cite{Fehre:2021eob}.

Implementing the \ac{MES}, discarding boundary terms and restricting ourselves to \GRUC{}, we find that all second order curvature terms are inessential, except for the topological Euler term. This entails that our ansatz for $\Gamma_k$ reads
\begin{equation}
 \Gamma_k = \int \text{d}^4x \, \sqrt{g} \, \left\{ \frac{1}{16\pi G_k} \bigg[ 2\Lambda_k - R \bigg] + \Theta_k \Euler{} \right\} \, ,
\end{equation}
where $\Euler$ is the Euler density. This ansatz agrees with previous work \cite{Baldazzi:2021orb, Knorr:2022ilz}. Where we differ is in the \ac{RG} kernel, which in our case captures the non-trivial momentum dependence of the graviton propagator:
\begin{equation}
 \Psi_{k,\mu\nu} = \gamma_g g_{\mu\nu} + \gamma_R(\Delta) \, R \, g_{\mu\nu} + \gamma_S(\Delta_2)_{\mu\nu}^{\phantom{\mu\nu}\rho\sigma} S_{\rho\sigma} \, .
\end{equation}
The operator $\Delta_2$ includes a convenient endomorphism that is spelled out in appendix \ref{app:tracecomputation}. In the language of an expansion about a flat spacetime, this \ac{RG} kernel is equivalent to a rescaling of the metric fluctuation by a tensor-valued, momentum-dependent wave function renormalisation akin to \cite{Christiansen:2014raa, Knorr:2021niv}. In this language, $\gamma_S$ captures the non-trivial momentum dependence of the spin two propagator, whereas a combination of $\gamma_R$ and $\gamma_S$ encodes that of the (physical, off-shell) spin zero part. By investigating the extra term on the left-hand side of the flow equation stemming from the field redefinition, see \eqref{eq:LHSextraterm} in the appendix, it is clear that we can adjust $\gamma_{R,S}$ to match terms on the right-hand side of the flow so that no form factors $f_{RR,SS}$ are generated in the effective average action during the flow, thus implementing the \ac{MES}. Note that both gamma functions $\gamma_{R,S}$ depend on $k$, but we omit indicating this for better readability.

At this point we note that to linear order in curvature, there is a third independent term that we could add to the \ac{RG} kernel,
\begin{equation}
 \Delta\Psi_{k,\mu\nu}^\text{diff} = D_\mu D_\nu \gamma_{DDR}(\Delta) \, R \, .
\end{equation}
However, this term is clearly related to the diffeomorphism \eqref{eq:diffeo}. Following our earlier discussion, we will discard this term, as a proper treatment likely needs a careful study of how to relate field redefinitions of the metric with those of the corresponding Faddeev-Popov ghosts.

We will employ a harmonic (or Feynman-de-Donder) gauge fixing, which brings the kinetic term into minimal form, and a type II regulator with the natural endomorphism in all sectors \cite{Codello:2008vh}. In doing so, we follow the same conventions as \cite{Knorr:2022ilz}. More details, including on our notation, are collected in appendix \ref{app:tracecomputation}.

Finally, we introduce dimensionless couplings via
\begin{equation}
    g = G_k k^2 \, , \qquad \lambda = \Lambda_k k^{-2} \, ,
\end{equation}
and drop the subscript $k$ everywhere in the following to improve readability. Likewise, we introduce dimensionless counterparts for $\gamma_{R,S}$, but since in the following we only talk about the dimensionless versions, we do not introduce new symbols for them.

Following \cite{Baldazzi:2021orb}, we fix the cosmological constant by
\begin{equation}
    \lambda = 8 \pi g \lim_{g\to0} \mathcal F \bigg|_{R=0} \, ,
\end{equation}
where $\mathcal F$ is the full right-hand side of the flow equation, and we set all curvatures to zero. This makes it so that we flow to $\Lambda=0$ for $k\to0$, thus implementing that the physical cosmological constant vanishes.

\section{Results}\label{sec:analysis}

Putting together all the ingredients specified in the last section, one can compute the beta functions for $g$ and $\Theta$ as well as the expressions for all gamma functions. Due to the specific choices made in the setup, the whole computation can be carried out by hand (up to contractions of large tensorial expressions for which it is convenient to use computer tensor algebra; specifically we have used \emph{xAct} \cite{xActwebpage, Brizuela:2008ra, Nutma:2013zea}). We present all major steps of the computation of the necessary heat kernel coefficients in appendix \ref{app:HK}, and the computation of the \ac{RG} flow in appendix \ref{app:tracecomputation}. The result of this endeavour is collected in equations \eqref{eq:flow1}, \eqref{eq:flowR}, \eqref{eq:flowFFR}, \eqref{eq:flowFFS} and \eqref{eq:flowE}, which represent one of the main results of this work. We also want to highlight the computation of one specific non-local heat kernel coefficient, given in \eqref{eq:the_big_one}, that to our knowledge has not been computed before.

Let us briefly collect the main results for the reader who is not interested in the technical details of the analysis:
\begin{itemize}
    \item The fixed point found previously in the \ac{MES} persists \cite{Baldazzi:2021orb}, and is extremely stable upon including the gamma functions $\gamma_{R,S}$. The single critical exponent in the system changes by $0.4\%$. This is because both $\gamma_{R,S}$ remain small over the whole momentum range.
    \item Momentum locality \cite{Christiansen:2015rva}, \ie{} the requirement that the high-energy flow of correlation functions in units of itself goes to zero which is related to a well-defined Wilsonian block spinning, is not fulfilled in this setup. This emphasises the need to take into account the flow of fluctuation correlation functions.
    \item The topological coupling $\Theta$ has a positive beta function at the fixed point, in agreement with previous findings \cite{Knorr:2021slg}. This suggests that at this fixed point, spacetimes with ``complicated'' topologies, \ie{} with strongly negative Euler characteristic, contribute most to the Euclidean path integral.
\end{itemize}

\subsection{Consistency checks}

We first make sure that we can reproduce earlier results. This includes first the well-known one-loop running induced by \ac{GR}, coming (in a standard scheme) in the form of a logarithmic running of the couplings multiplying terms quadratic in curvature \cite{tHooft:1974toh}. In the \ac{MES}, the corresponding divergences are accounted for in the leading term of the gamma functions in an expansion about vanishing Newton's constant \cite{Baldazzi:2021orb},
\begin{equation}\label{eq:gamma_small_g}
\begin{aligned}
    \gamma_R(0) &\sim -\frac{23}{120\pi} g \, , \\
    \gamma_S(0) &\sim \frac{7}{10\pi} g \, .
\end{aligned}
\end{equation}
We find this result independent of the chosen regulator function, as it must be. In turn, for the beta function of the coupling of the Euler term, we have
\begin{equation}
    \dot\Theta \sim \frac{1}{16\pi^2} \frac{53}{45} \, .
\end{equation}
All these expressions are in agreement with previous results. Second, we checked that upon truncating the form factors to constant terms, we reproduce the results reported in \cite{Knorr:2022ilz}, where the same setup was used. 

\subsection{Large momentum behaviour}

Let us next investigate the large momentum behaviour of the gamma functions $\gamma_{R,S}$. A term-by-term analysis reveals that both of them fall off asymptotically,
\begin{equation}
    \gamma_R(z) \sim \frac{c^\infty_R}{z} \, , \qquad \gamma_S(z) \sim \frac{c^\infty_S}{z} \, , \qquad z \to \infty \, .
\end{equation}
The precise form of the coefficients involves threshold integrals including the gamma functions themselves over all momenta. We thus have to assume that these are finite. Our numerical analysis below however shows that this is indeed the case, at least for the window of interesting values for $g$.

We can however find closed-form expressions for $\gamma_{R,S}$, and as a consequence for $c_{R,S}^\infty$, in a series expansion in powers of $g$. Identifying the regulator shape functions of all modes, the leading term reads
\begin{equation}
\begin{aligned}
    \gamma_R(z) &\sim \frac{g}{\pi} \int_0^{\frac{1}{4}} \! \text{d}u \, \measure{\frac{7}{16}, \frac{15}{4}, -\frac{1}{4}}{u} \frac{\mathcal R(u z) - u z \mathcal R'(u z)}{u z + \mathcal R(u z)} \, , \\
    \gamma_S(z) &\sim \frac{g}{\pi} \int_0^{\frac{1}{4}} \! \text{d}u \, \measure{2,8,2}{u} \frac{\mathcal R(u z) - u z \mathcal R'(u z)}{u z + \mathcal R(u z)} \, .
\end{aligned}
\end{equation}
Letting $z\to0$, we reconfirm \eqref{eq:gamma_small_g}. On the other hand, taking $z$ large, we conclude
\begin{equation}
\begin{aligned}
    c_R^\infty &\sim -\frac{35g}{12\pi} \int_0^\infty \text{d}z \frac{\mathcal R(z) - z \mathcal R'(z)}{z+\mathcal R(z)} \, , \\
    c_S^\infty &\sim -\frac{11g}{3\pi} \int_0^\infty \text{d}z \frac{\mathcal R(z) - z \mathcal R'(z)}{z+\mathcal R(z)} \, .
\end{aligned}
\end{equation}
Note that the threshold integral appearing in these expressions is strictly positive since the numerator of the integrand originates from the \ac{RG}-derivative of the regulator, which has to be positive for a well-defined coarse-graining. Consequently, $c_{R,S}^\infty<0$, and thus both gamma functions approach zero from below for small $g$. Taken together with the behaviour for small arguments \eqref{eq:gamma_small_g}, we conclude that $\gamma_S$ needs to cross zero at least once for small enough $g$. Incidentally, the ratio of the gamma functions at large arguments and small $g$ goes to a regulator-independent constant,
\begin{equation}\label{eq:ratio_large_momenta}
    \lim_{z\to\infty} \lim_{g\to0} \frac{\gamma_R(z)}{\gamma_S(z)} = \frac{35}{44} \, .
\end{equation}
It is not clear whether there is any physical meaning in this observation, so for now we treat it as a mere curiosity.

An interesting notion that is intertwined with the flow of correlation functions at large momenta is that of momentum locality, introduced in \cite{Christiansen:2015rva}. In brief, momentum locality is the condition that the flow of any $n$-point correlation function, measured in units of itself, tends to zero as all of the momenta go to infinity. For example, for a two-point function with a suitable norm $|\cdot|$,
\begin{equation}\label{eq:momlocality}
    \lim_{p^2/k^2\to\infty} \frac{|\dot\Gamma^{(2)}(p^2)|}{|\Gamma^{(2)}(p^2)|} = 0 \, .
\end{equation}
This ensures that a coarse-graining step does not influence the physics at momenta that are larger than the scale that is integrated out \cite{Pawlowski:2020qer}. For a perturbatively non-renormalisable theory like gravity, this is a non-trivial condition -- counting powers of momenta shows that, generically, momentum locality is expected to be violated. Surprisingly, the property has been found to hold for some non-trivial graviton correlation functions in fluctuation computations \cite{Christiansen:2014raa, Christiansen:2015rva}.

Since the present investigation is the first complete momentum-dependent analysis in the background field approximation, it is an important question to ask whether momentum locality is fulfilled. For this, we will investigate the running of the \emph{background} two-point function in a flat background. Due to our general setup, we can disentangle the momentum dependence of the spin two and zero parts of the propagator. For simplicity, we focus on an expansion in powers of $g$, and only keep the leading term. Moreover, in interpreting \eqref{eq:momlocality} in the essential scheme, we include the extra term proportional to the \ac{RG} kernel in the numerator. With this in mind, in the spin two sector, we find
\begin{equation}
\begin{aligned}
    \lim_{p^2/k^2\to\infty} \frac{|\dot\Gamma^{(2)}(p^2)|_2}{|\Gamma^{(2)}(p^2)|_2} &\sim \frac{4g}{\pi} \int_0^\infty \text{d}z \frac{\mathcal R(z) - z \mathcal R'(z)}{z+\mathcal R(z)} \neq 0 \, .
\end{aligned}
\end{equation}
In the spin zero sector, we have
\begin{equation}
\begin{aligned}
    \lim_{p^2/k^2\to\infty} \frac{|\dot\Gamma^{(2)}(p^2)|_0}{|\Gamma^{(2)}(p^2)|_0} &\sim \frac{95g}{6\pi} \int_0^\infty \text{d}z \frac{\mathcal R(z) - z \mathcal R'(z)}{z+\mathcal R(z)} \neq 0 \, .
\end{aligned}
\end{equation}
We conclude that, even in this simple limit, momentum locality is \emph{not} fulfilled in our setup. This highlights the importance of going beyond the background field approximation, and gives a concrete motivation to extend the essential scheme to fluctuation computations.

\begin{figure}
	\includegraphics[width=\columnwidth]{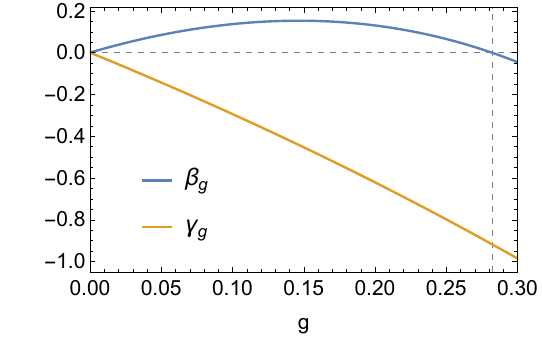}
	\caption{\label{fig:beta+gamma}Beta function $\beta_g$ and gamma function $\gamma_g$ as functions of $g$, the effect of $\gamma_{R,S}$ being included. The horizontal dashed line indicates zero, and the vertical dashed line indicates the fixed point value.}
\end{figure}

\subsection{Numerical solutions}

To investigate the phase diagram, the beta and gamma functions have to be evaluated numerically. Note that the general structure of the set of equations is
\begin{align}
    \beta_g &= a_\beta(g) + b_\beta(g) \beta_g + c_\beta(g) \gamma_g \notag \\
    &\qquad\qquad + d_\beta(g)[\gamma_R] + e_\beta(g)[\gamma_S] \, , \\
    \gamma_g &= a_g(g) + b_g(g) \beta_g + c_g(g) \gamma_g \notag \\
    &\qquad\qquad + d_g(g)[\gamma_R] + e_g(g)[\gamma_S] \, , \\
    \gamma_R(z) &= a_R(g,z) + b_R(g,z) \beta_g + c_R(g,z) \gamma_g \notag \\
    &\qquad \qquad + d_R(g,z)[\gamma_R] + e_R(g,z)[\gamma_S] \, , \\
    \gamma_S(z) &= a_S(g,z) + b_S(g,z) \beta_g + c_S(g,z) \gamma_g \notag \\
    &\qquad \qquad + d_S(g,z)[\gamma_R] + e_S(g,z)[\gamma_S] \, .
\end{align}
Here, $a_x,b_x,c_x$ are threshold integrals, and $d_x,e_x$ are linear integral operators (also in the form of threshold integrals). This makes it clear that we are dealing with a linear system of mixed algebraic and integral equations. By formally solving the first two equations for $\beta_g$ and $\gamma_g$, we can set up two integral equations for $\gamma_{R,S}$, whose solution can then be fed back to compute $\beta_g$ and $\gamma_g$.

There are different strategies to solve this set of equations. The most straightforward way is to expand the functions $\gamma_R, \gamma_S$ in a suitable set of orthogonal functions, and reducing the system to a purely algebraic one using either the inner product related to the basis, or a collocation method \cite{Boyd:ChebyFourier}. We will do so in the following. While this comes with many benefits and has been well-tested \cite{Borchardt:2015rxa, Borchardt:2016pif}, it is a purely numerical approach. Alternatives that allow to keep some formal generality, \eg\ regulator dependence, are a systematic expansion in powers of $g$,\footnote{In previous work \cite{Baldazzi:2021orb, Knorr:2022ilz}, the fixed point was found to be at rather small values of $g$ so that one likely has to retain only a small number of terms to obtain a satisfactory precision.} and the use of the Liouville-Neumann series. Both these methods are extremely tedious in the present case, so we refrain from using them here.

To make our numerical setup concrete, we identify all regulator shape functions, and specifically pick
\begin{equation}
    \mathcal R(z) = e^{-z} \, ,
\end{equation}
since it is numerically well-behaved. We will focus on the fixed point with the smallest positive value for $g$. For all numerical results, we will use pseudo-spectral methods, using rational Chebyshev functions \cite{Boyd:ratCheb} as a basis in the momentum argument to expand $\gamma_{R,S}$, together with a collocation method. As a compactification parameter, we chose $L=10$. Within a truncation of order 15, we 
obtain the gamma functions numerically by using another collocation grid in $g$ of order 10 in the range $g\in\left[0,3/10\right]$. This gives us a high precision interpolation for both $\gamma_{R,S}$ for all $z$ in the specified range for $g$.\footnote{We have used 32 digits in the computation, and used Mathematica's \emph{NIntegrate} routine to evaluate integrals numerically. The results are quantitatively stable upon increasing the pseudo-spectral truncation order to 18 in $z$-direction, to 24 in $g$-direction, and the numerical precision to 48 digits. Differences between these two choices do not exceed $4\times 10^{-7}$ for $\gamma_{R,S}$ and $3\times 10^{-10}$ for $\beta_g$ and $\gamma_g$, and the digits given for all fixed point quantities remain unchanged. Evaluating the defining integral equations for $\gamma_{R,S}$ using our pseudo-spectral solution at the lower resolution, we find a maximal global error of about $2\times10^{-8}$ in the whole specified $g$-range and for all momenta.} From that, we can evaluate both $\gamma_g$ and the actual beta function of $g$.

\begin{figure}
	\includegraphics[width=\columnwidth]{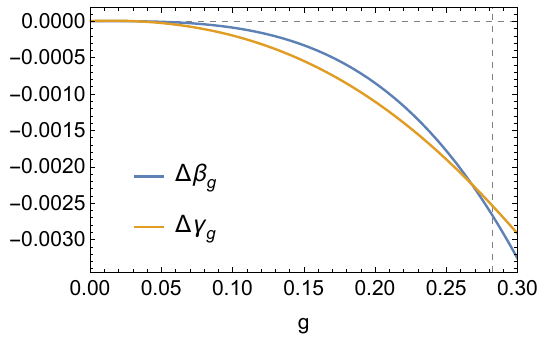}
    \caption{\label{fig:Delta_beta+gamma}Differences of the beta function $\Delta\beta_g$ and gamma function $\Delta\gamma_g$ as functions of $g$ as defined in \eqref{eq:difference}. The horizontal dashed line indicates zero, and the vertical dashed line indicates the fixed point value. The difference is at the per mille level in the shown range for $g$.}
\end{figure}

\begin{figure*}
	\includegraphics[width=\columnwidth]{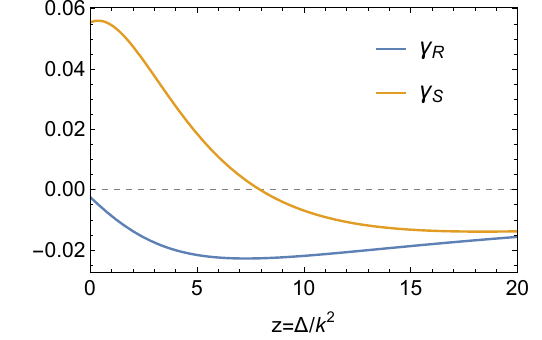} \hfill \includegraphics[width=\columnwidth]{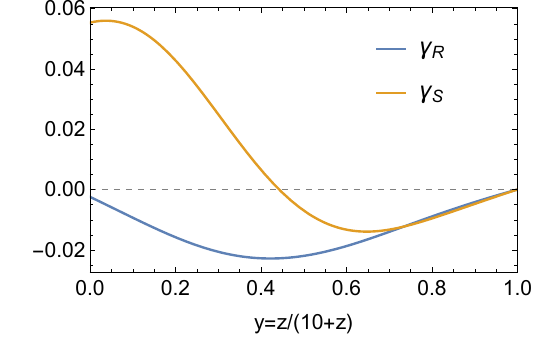}
	\caption{\label{fig:FFgammaFP}Gamma functions $\gamma_{R,S}$ evaluated at the fixed point, for small momentum arguments (left panel) and globally (right panel). Both functions are numerically small over the whole momentum range, and go to zero at infinity, as expected.}
\end{figure*}

Finally, we can discuss the results. To set a benchmark, we start with a truncation where we completely neglect both $\gamma_{R,S}$. While this level of truncation has been done before, our regulator choice differs, so we will get slightly different numbers for the fixed point value and the critical exponent. The fixed point values for $g$ and $\gamma_g$ read
\begin{equation}
    g = 0.2830 \, , \qquad \gamma_g = -0.9182 \, ,
\end{equation}
the critical exponent being
\begin{equation}
    \theta = 2.338 \, .
\end{equation}
Values of couplings and gamma functions at fixed points are not universal, so we should not expect results that are necessarily close to the ones in \cite{Baldazzi:2021orb}. Indeed, our value for $g$ at the fixed point is about half of theirs. The gamma function $\gamma_g$ is about $-1$, but also there the difference is roughly $20\%$. By contrast, the critical exponent is a universal quantity, and indeed, our estimate is extremely close to the one in \cite{Baldazzi:2021orb} at the same level of truncation, with a difference of $1\%$. This is despite significant differences in the choice of the regulator.

Let us now include the momentum-dependent gamma functions $\gamma_{R,S}$. In \autoref{fig:beta+gamma}, we show the beta function $\beta_g$ as well as the gamma function $\gamma_g$ in the range $g\in\left[0,3/10\right]$, which includes the relevant fixed point. The fixed point is now located at
\begin{equation}
    g = 0.2819 \, , \qquad \gamma_g = -0.9164 \, ,
\end{equation}
with the critical exponent
\begin{equation}
    \theta = 2.347 \, .
\end{equation}
This result is extremely remarkable: the inclusion of $\gamma_{R,S}$ is practically without effect, which means that these functions are completely unimportant to obtain precision results. The critical exponent changes by a mere $0.4\%$. To illustrate this point even more, in \autoref{fig:Delta_beta+gamma}, we show the difference of both $\beta_g$ and $\gamma_g$ upon including $\gamma_{R,S}$, that is
\begin{equation}\label{eq:difference}
    \Delta\beta_g = \beta_g - \left( \beta_g \bigg|_{\gamma_{R,S}=0} \right) \, , \quad \Delta\gamma_g = \gamma_g - \left( \gamma_g \bigg|_{\gamma_{R,S}=0} \right) \, .
\end{equation}
The difference is at the low per mille level over the whole range $g\in\left[0,3/10\right]$.

Moving on to the gamma functions themselves, their fixed point form is indicated in \autoref{fig:FFgammaFP}, for both small values of the argument, and globally with the help of a compactification. We can see some of the features that we discussed analytically earlier: for small momenta, $\gamma_S$ is positive while $\gamma_R$ is negative, and both tend to zero from below at large arguments. As another quality check of the numerics, we checked that \eqref{eq:ratio_large_momenta} is fulfilled. After subtracting a constant of the order of $10^{-7}$ that is numerical noise due to the gamma functions not going exactly to zero,\footnote{Here we have used a Gauss grid. By using a Lobatto grid instead, one could impose that the gamma functions vanish exactly asymptotically, but this makes practically no difference for all other results.} we can reproduce \eqref{eq:ratio_large_momenta} at a level of $0.3\%$. This discrepancy is relatively large (from the perspective of pseudo-spectral methods) since we are probing sub-leading effects.

\begin{figure*}
	\includegraphics[width=\columnwidth]{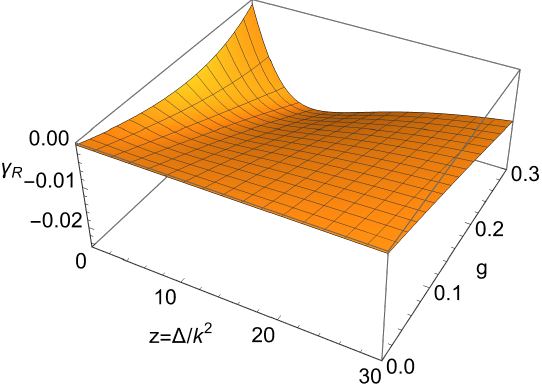} \hfill \includegraphics[width=\columnwidth]{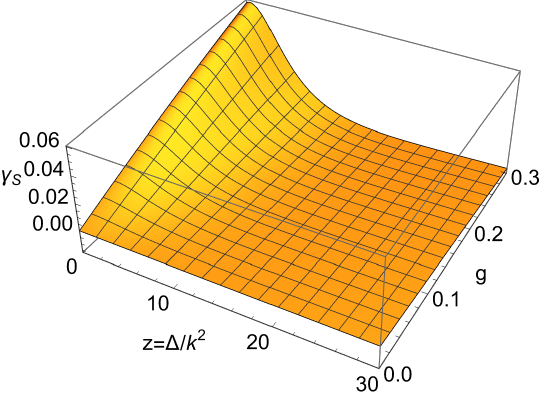}
	\caption{\label{fig:FFgamma3d}Gamma functions $\gamma_{R,S}$ in the range $g\in\left[0,3/10\right]$ and $z\in\left[0,30\right]$. In the limit $g\to0$, both gamma functions vanish as required. Increasing $g$ they both increase approximately linearly in the whole displayed region, but remain small.}
\end{figure*}

In \autoref{fig:FFgamma3d} we also show the two gamma functions in the full $g$-range that we investigated. As required, for $g\to0$, both gamma functions vanish, and they slowly build up upon increasing $g$. The largest absolute value is reached around vanishing argument for $\gamma_S$, and at $z\approx 6$ for $\gamma_R$, but both of them stay rather small, not exceeding $0.06$ in absolute value. This is of course also consistent with our earlier findings that the system is virtually unchanged under the inclusion of these gamma functions.

\begin{figure}
	\includegraphics[width=\columnwidth]{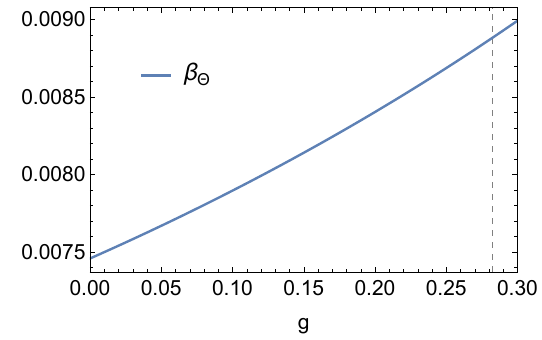}
    \caption{\label{fig:beta_Theta}Beta function of the coupling of the topological Euler term as functions of $g$. The vertical dashed line indicates the fixed point value. Over the whole $g$-range, it is positive.}
\end{figure}

Finally, we briefly discuss the beta function for the coupling of the Euler density. Its beta function is extremely compact,
\begin{equation}
    \dot \Theta = \frac{1}{32\pi^2} \left[ \frac{19}{18} \frac{2 - \frac{\beta_g-2g}{g} + 2 \gamma_g}{1-2\lambda} + \frac{11}{90} \left( 2 - \frac{\beta_g-2g}{2g} \right) \right] \, .
\end{equation}
As noted in the appendix, there is no direct contribution coming from the gamma functions $\gamma_{R,S}$, although they contribute indirectly through $\beta_g$ and $\gamma_g$. We depict it in \autoref{fig:beta_Theta}. In the whole $g$-range, it is positive. This agrees with the findings of \cite{Knorr:2021slg} in the standard scheme, indicating that ``complicated'' topologies (\ie{}, those with strongly negative Euler characteristic) contribute most to the Euclidean path integral at this fixed point.

\section{Summary and outlook}\label{sec:summary}

In this work, we investigated general momentum-dependent field redefinition in quantum gravity. Such field redefinitions are extremely useful to efficiently compute quantum-gravitational scattering amplitudes from first principles, but also play an important role in the definition of critical exponents in the gravitational Asymptotic Safety scenario.

In \autoref{sec:amps}, we first discussed some general aspects of field redefinitions in the context of scattering amplitudes. We have shown that for a two-to-two scattering event, we can perform a field redefinition so that almost all non-trivial information is contained in the contact diagram. We also spelled out some requirements for the discussion of the topic: the spectrum of the theory has to be fixed a priori, and field redefinitions should not introduce or remove any degrees of freedom, which has to be checked a posteriori. Finally, we discussed a concrete example of a non-local field redefinition, in analogy to a similar proposal arguing that the Goroff-Sagnotti term can be removed \cite{Krasnov:2009ik}. In this example, the interaction term in a scalar field theory could be removed, leading to a trivial scattering amplitude for the newly defined field. From this example we conclude that such non-local field redefinitions should be used with extreme care as they can significantly change the standard physical interpretation of any computation.

In \autoref{sec:setup}, we briefly reviewed the \ac{FRG} and spelled out our setup. Many details were put into the appendices, including the full derivation of all flow equations, and the computation of a specific heat kernel coefficient. We point out the remarkable fact that thanks to the \ac{MES}, practically all computations could be carried out by hand.

Following this, in \autoref{sec:analysis}, we analysed the flow equations. We showed that the fixed point found previously in the \ac{MES} \cite{Baldazzi:2021orb} is extremely stable upon taking the momentum-dependent field redefinitions into account. The single critical exponent of our system changed at the sub-percent level. This is a strong indication for the excellent convergence of the fixed point in the \ac{MES}, and suggests that indeed the fixed point resides in \GRUC{}.

To progress further on the path to asymptotically safe scattering amplitudes, the next step is to either include operators with more than two curvatures (to resolve graviton scattering), or to include non-minimal momentum-dependent gravity-matter vertices (to resolve gravity-mediated scattering). Some open questions about the essential scheme also have to be addressed, including how to properly treat field redefinitions that are (partially) a gauge transformation, and how to perform the a posteriori check that the field redefinition itself, and not only the \ac{RG} kernel, is well-defined along the flow. Along a different direction, it would be worthwhile, and extremely challenging, to repeat the current computation in the standard scheme to understand the differences between the standard and the \ac{MES}, and the phase diagram of quantum gravity more globally, not restricted to \GRUC{}. We hope to come back to these interesting questions at a future point.

\acknowledgments{I would like to thank Chris Ripken for collaboration during an early stage of this project, Alessio Baldazzi, Luca Buoninfante, Kevin Falls, Yannick Kluth and Frank Saueressig for interesting discussions, and Alessia Platania and Manuel Reichert for useful comments on the manuscript. Nordita is supported in part by NordForsk.}

\clearpage

\appendix

\begin{widetext}
\section{Heat kernel}\label{app:HK}

In this appendix we collect the necessary heat kernel formulas. To our knowledge, one of them, \eqref{eq:the_big_one}, did not appear in the literature before, so we use momentum space techniques to compute it. To perform the contractions of the tensors, we used the Mathematica package suite \emph{xAct} \cite{xActwebpage, Brizuela:2008ra, Nutma:2013zea}.

\subsection{Definitions and useful formulas}

We start by recalling some definitions. The general heat kernel integral kernel is defined as
\begin{equation}\label{eq:HKf}
 f(z) = \int_0^1 \text{d}\xi \, e^{-z \, \xi (1-\xi)} = \sum_{j\geq0} \frac{1}{\left(\frac{3}{2}\right)_j} \left(-\frac{z}{4} \right)^j \, ,
\end{equation}
where
\begin{equation}
    \left( a \right)_b = \frac{\Gamma\left(a+1\right)}{\Gamma\left(a-b+1\right)} \, ,
\end{equation}
is the Pochhammer symbol. Next, we are often working with the inverse Laplace transform, denoted by $\mathcal L^{-1}$:
\begin{equation}
 g(z) = \int_0^\infty \text{d}s \, \mathcal L^{-1}[g(z)](s) \, e^{-sz} \, .
\end{equation}
Generally, we only use $\mathcal L^{-1}$ in a formal way, that is in intermediate steps of computations. We will assume that all such transformations are, or can be made, well-defined. Some well-known formulas \cite{Benedetti:2010nr} are
\begin{align}
 \int_0^\infty \text{d}s \, \mathcal L^{-1}[g(z)](s) \, s^{-n} &= \frac{1}{\Gamma(n)} \int_0^\infty \text{d}z \, z^{n-1} \, g(z) \, , \label{eq:snegtoint} \\
 \int_0^\infty \text{d}s \, \mathcal L^{-1}[g(z)](s) &= g(0) \, , \label{eq:szerotoval} \\
 \int_0^\infty \text{d}s \, \mathcal L^{-1}[g(z)](s) \, s^n &= (-1)^n \, g^{(n)}(0) \, , \label{eq:spostoval}
\end{align}
where $n\geq1$. The first relation can be proven by inserting the inverse Laplace transform on the right-hand side and integrating over $z$, whereas the other two formulas follow directly from the definition of the inverse Laplace transform.

Next, we have the following formula:
\begin{equation}\label{eq:stou}
 \int_0^\infty \text{d}s \, \mathcal L^{-1}[g(z)](s) \frac{f(s\Delta) - \sum_{k=0}^n \frac{1}{\left(\frac{3}{2}\right)_k} \left( -\frac{s\Delta}{4} \right)^k}{(s\Delta)^{n+1}} = - \frac{1}{\left(\frac{3}{2}\right)_{n}} \left(-\frac{1}{4}\right)^n \int_0^\frac{1}{4} \left( 1 - 4u \right)^{n+\frac{1}{2}} \, g(u\Delta) \, , \quad n\geq-1 \, .
\end{equation}
Note how in the numerator on the left-hand side, we subtract the first $n$ terms of the Taylor series of the integral kernel $f$, so that there are no poles for small $\Delta$. This gives a natural way of grouping specific terms of the non-local heat kernel. The formula can be proven by inserting the inverse Laplace transform on the right-hand side and performing the integration over $u$. Concretely, the first few cases read
\begin{equation}
\begin{aligned}
 \int_0^\infty \text{d}s \, \mathcal L^{-1}[g(z)](s) \, f(s\Delta) &= 2 \int_0^\frac{1}{4} \frac{1}{\sqrt{1-4u}} \, g(u\Delta) \, , \\
 \int_0^\infty \text{d}s \, \mathcal L^{-1}[g(z)](s) \, \frac{f(s\Delta)-1}{s\Delta} &= - \int_0^\frac{1}{4} \sqrt{1 - 4u} \, g(u\Delta) \, , \\
 \int_0^\infty \text{d}s \, \mathcal L^{-1}[g(z)](s) \, \frac{f(s\Delta)-1+\frac{s\Delta}{6}}{(s\Delta)^2} &= \frac{1}{6} \int_0^\frac{1}{4} (1 - 4u)^{3/2} \, g(u\Delta) \, , \\
 \int_0^\infty \text{d}s \, \mathcal L^{-1}[g(z)](s) \, \frac{f(s\Delta)-1+\frac{s\Delta}{6}-\frac{(s\Delta)^2}{60}}{(s\Delta)^3} &= -\frac{1}{60} \int_0^\frac{1}{4} (1 - 4u)^{5/2} \, g(u\Delta) \, .
\end{aligned}
\end{equation}
As a useful generalisation to \eqref{eq:stou}, we can derive
\begin{equation}\label{eq:sjinvstou}
\begin{aligned}
 &\int_0^\infty \text{d}s \, \mathcal L^{-1}[g(z)](s) \, \frac{1}{s^j} \frac{f(s\Delta) - \sum_{k=0}^n \frac{1}{\left(\frac{3}{2}\right)_k} \left( -\frac{s\Delta}{4} \right)^k}{(s\Delta)^{n+1}} \\
 &\hspace{1cm} = - \frac{1}{\left(\frac{3}{2}\right)_{n+j}} \left(-\frac{1}{4}\right)^{n+j} \Delta^j \int_0^\frac{1}{4} \left( 1 - 4u \right)^{(n+j)+\frac{1}{2}} \, g(u\Delta) \\
 &\hspace{2cm} + \frac{1}{\left(\frac{3}{2}\right)_{n+1}} \left(- \frac{1}{4} \right)^{n+1} \frac{1}{\Gamma(j)} \int_0^\infty \text{d}z \, z^{j-1} \, {}_2F_1\left( 1, 1-j ; n+\frac{5}{2} \bigg| \frac{\Delta}{4z} \right) \, g(z) \, , \quad n\geq-1 \, , j \geq 0 \, .
\end{aligned}
\end{equation}
This can be proven by adding zero to the numerator on the left-hand side in the form of adding and subtracting additional terms of the Taylor series, then using \eqref{eq:snegtoint} and \eqref{eq:stou}, and finally performing a sum. We emphasise the difference in the factors in front of the integrals in the two terms. As a matter of fact, the formula \eqref{eq:sjinvstou} includes \eqref{eq:snegtoint}, \eqref{eq:szerotoval} and \eqref{eq:stou} as special cases.

Finally, \eqref{eq:stou} motivates the introduction of the short-hand
\begin{equation}
    \HKbracket{c_0, \dots, c_n}{z} = \sum_{j=0}^n c_j \frac{f(z) - \sum_{k=0}^{n-1} \frac{1}{\left(\frac{3}{2}\right)_k} \left( -\frac{z}{4} \right)^k}{z^n} = c_0 f(z) + c_1 \frac{f(z)-1}{z} + \dots \, ,
\end{equation}
and the integral measure
\begin{equation}
 \mu(c_0, \dots, c_n | u) = \sum_{j=0}^n \frac{4 c_j}{\left(\frac{3}{2}\right)_{j-1}} \left( -\frac{1}{4} \right)^j (1-4u)^{j-\frac{1}{2}} = \frac{2c_0}{\sqrt{1-4u}} - c_1 \sqrt{1-4u} + \frac{c_2}{6} (1-4u)^{3/2} + \dots \, .
\end{equation}
This allows us to easily express the non-local heat kernel coefficients, and to perform the inverse Laplace transform,
\begin{equation}
    \int_0^\infty \text{d}s \, \mathcal L^{-1}[g(z)](s) \HKbracket{c_0, \dots, c_n}{s\Delta} = \int_0^\frac{1}{4} \text{d}u \, \mu(c_0, \dots, c_n | u) g(u \Delta) \, ,
\end{equation}
in a compact fashion. Note that $\mu$ is homogeneous of degree one in the $c_j$, so that
\begin{equation}
    \mu(\alpha c_0, \dots, \alpha c_n | u) = \alpha \, \mu(c_0, \dots, c_n | u) \, .
\end{equation}
We will sometimes use this property to normalise the arguments in such a fashion that the measure is $1$ at $u=0$.

\subsection{Momentum space techniques}

Let us briefly describe how to compute non-local heat kernel coefficients with the help of momentum space techniques. To set the stage, we follow the notation of \cite{Groh:2011dw} and start out with a Laplace-type operator $\Delta = - D^2$ that shall include a general connection within the covariant derivative $D$. The bundle curvature corresponding to this general connection is defined via
\begin{equation}
    \mathcal F_{\mu\nu} = -\left[ D_\mu, D_\nu \right] \, .
\end{equation}
To keep some generality, we will also allow for an endomorphism $\mathbbm E$ to be added to this operator.

The heat kernel $H$ of the operator $\Delta + \mathbbm E$ solves the heat equation for some fiducial heat kernel ``time'' $s$,
\begin{equation}
    \left( \partial_s + \Delta + \mathbbm E \right) H(x,y;s) = 0 \, , \qquad H(x,y;0) = \delta(x-y) \mathbbm 1 \, .
\end{equation}
Inspired by the solution for a flat manifold, we make the general ansatz
\begin{equation}
    H(x,y;s) = \left( \frac{1}{4\pi s} \right)^{\frac{d}{2}} e^{-\frac{\sigma(x,y)}{2s}} \Omega(x,y;s) \, .
\end{equation}
Here, $\sigma(x,y)$ is one half of the geodesic distance, and $\Omega$ has to be computed. The connection to the trace of the exponential of our differential operator then comes in the form of taking the coincidence limit of the heat kernel,
\begin{equation}
    \text{Tr} \left[ e^{-s\left( \Delta + \mathbbm E \right)} \right] = \text{tr} \int \text{d}^dx \, \sqrt{g} \, H(x,x;s) = \left( \frac{1}{4\pi s} \right)^\frac{d}{2} \text{tr} \, \int \text{d}^dx \, \sqrt{g} \, \overline{\Omega}(s) \, .
\end{equation}
Here, Tr indicates a functional trace and tr a trace over bundle indices, and we introduced the short-hand for the coincidence limit $\overline{\Omega}(s) = \Omega(x,x;s)$.

From previous computations \cite{Barvinsky:1990up, Barvinsky:1993en, Vassilevich:2003xt, Codello:2012kq} as well as symmetry considerations, we know that $\overline{\Omega}(s)$ has an asymptotic expansion in powers of the curvature of the form
\begin{equation}\label{eq:HKRHS}
 \overline{\Omega}(s) \sim \mathbbm 1 + s \, a_1(s\Delta) \, R \, \mathbbm 1 + s \, a_2(s\Delta) \, \mathbbm E + \dots \, .
\end{equation}

An efficient way to compute the coefficients $a_i$ is to expand both the trace and the general form in an expansion about a flat background metric as well as a vanishing endomorphism and gauge connection, and then to compute the trace in momentum space. Concretely, we use an expansion of the Laplacian of the form
\begin{equation}
 \Delta + \mathbbm E = \square + \left( \Delta + \mathbbm E - \square \right) \equiv \square + \sum_{i\geq1} \mathbbm d_i \, ,
\end{equation}
where $\square=-\partial^2$ is the flat background Laplacian and $\mathbbm d_i$ is an operator with exactly $i$ fluctuation fields. To expand the exponential of this operator, we then use \cite{Knorr:2019atm}
\begin{equation}\label{eq:expexp}
\begin{aligned}
 e^{-s(\Delta+\mathbbm E)} &= \Bigg[ \mathbbm 1 + \sum_{j\geq0} \frac{(-s)^{j+1}}{(j+1)!} \, \left[ \square, \mathbbm d_1 \right]_j + \sum_{j\geq0} \frac{(-s)^{j+1}}{(j+1)!} \, \left[ \square, \mathbbm d_2 \right]_j \\
 & + \frac{1}{2} \left( \sum_{j,k\geq0} \frac{(-s)^{j+k+2}}{(j+1)!(k+1)!} \, \left[ \square, \mathbbm d_1 \right]_j \, \left[ \square, \mathbbm d_1 \right]_k + \sum_{j\geq0} \sum_{k\geq1} \frac{(-s)^{j+k+2}}{(j+k+2)!} \left[ \square, \left[ \mathbbm d_1, \left[ \square, \mathbbm d_1 \right]_k \right] \right]_j \right) + \dots \Bigg] \, e^{-s\square} \, .
\end{aligned}
\end{equation}
Here, we have introduced the multi-commutator defined by
\begin{equation}
    \left[ X, Y \right]_n = \left[ X, \left[ X, Y \right]_{n-1} \right] \, , \qquad \left[ X, Y \right]_1 = \left[ X, Y \right] = XY - YX \, , \qquad \left[ X, Y \right]_0 = Y \, .
\end{equation}
In the final step, we assign momenta $p_j$ to the fluctuation fields contained in the $\mathbbm d_i$, and we call the traced-over momentum $q$. The functional trace then boils down to a momentum integral,
\begin{equation}
 \text{Tr} = \int \text{d}^dx \, \sqrt{\eta} \, \, \, \text{tr} \, \int \, \frac{\text{d}^dq}{(2\pi)^d} \, .
\end{equation}
Here, $\eta$ is the flat metric. Clearly, the first term in the expansion \eqref{eq:expexp} then gives the flat term in \eqref{eq:HKRHS} via
\begin{equation}
 \int \frac{\text{d}^dq}{(2\pi)^d} \, e^{-s q^2} = \left( \frac{1}{4\pi s} \right)^\frac{d}{2} \, .
\end{equation}

Let us now briefly illustrate how the curvature terms can be computed in practice. For this, we compute the non-local diagonal heat kernel to linear order in curvature, that is the coefficients $a_1$ and $a_2$. To this effect, we consider a general field $\Phi^A$ where $A$ is any collection of bundle or spacetime indices. To first order in fluctuations, we then have
\begin{equation}
 \left[ \left( \Delta + \mathbbm E \right) \Phi \right]^A \simeq \square \, \Phi^A + \mathbbm d_{1\phantom{A}B}^{\phantom{1}A} \, \Phi^B \, ,
\end{equation}
with
\begin{equation}
 \mathbbm d_{1\phantom{A}B}^{\phantom{1}A} = h^{\mu\nu} \partial_\mu \partial_\nu \delta_B^{\phantom{B}A} + \left( \partial_\alpha h^{\alpha\mu} \delta_B^{\phantom{B}A} - \frac{1}{2} \partial^\mu h \, \delta_B^{\phantom{B}A} - 2 \mathcal A^{A\mu}_{\phantom{A\mu}B} \right) \partial_\mu - \partial_\mu \mathcal A^{A\mu}_{\phantom{A\mu}B} + \mathbbm E^A_{\phantom{A}B} \, .
\end{equation}
Here $\mathcal A$ is the connection in the bundle that $\Phi^A$ lives in. Calling the momentum of each of the fluctuation fields $p$, we have
\begin{equation}
\begin{aligned}
 &\text{Tr} \sum_{j\geq0} \frac{(-s)^{j+1}}{(j+1)!} \, \left[ \square, \mathbbm d_1 \right]_j \, e^{-s\square} = \int \text{d}^dx \, \sqrt{\eta} \, \text{tr} \, \int \frac{\text{d}^dq}{(2\pi)^d} \sum_{j\geq0} \frac{(-s)^{j+1}}{(j+1)!} \, \left( p^2+2pqx \right)^j  \times \\
 &\hspace{3cm} \left[ -h^{\mu\nu} q_\mu q_\nu \delta_B^{\phantom{B}A} - \left( p_\alpha h^{\alpha\mu} \delta_B^{\phantom{B}A} - \frac{1}{2} p^\mu h \, \delta_B^{\phantom{B}A} + 2 \mathbf{i} \mathcal A^{A\mu}_{\phantom{A\mu}B} \right) q_\mu - \mathbf{i} p_\mu \mathcal A^{A\mu}_{\phantom{A\mu}B} + \mathbbm E^A_{\phantom{A}B} \right] \, e^{-s q^2} \, .
\end{aligned}
\end{equation}
In this, $x$ is the cosine of the angle between $p$ and $q$, so that $p^\mu q_\mu = p q x$. We now need the integral relations
\begin{align}
 \int \text{d}^dq \, f(p,q,x) q_\mu &= p_\mu \, \int \text{d}^dq \, f(p,q,x) \frac{p q x}{p^2} \, , \\
 \int \text{d}^dq \, f(p,q,x) q_\mu q_\nu &= \left( \eta_{\mu\nu} - \frac{p_\mu p_\nu}{p^2} \right) \, \int \text{d}^dq \, f(p,q,x) \frac{1}{d-1} (1-x^2) q^2 + \frac{p_\mu p_\nu}{p^2} \int \text{d}^dq \, f(p,q,x) q^2 x^2 \, .
\end{align}
These can be derived by contracting both sides with either momenta or metrics. We also note that the sum together with the exponential can be written in the following way:
\begin{equation}\label{eq:HKhelperequation}
 \sum_{j\geq0} \frac{(-s)^{j+1}}{(j+1)!} \, \left( p^2+2pqx \right)^j e^{-s q^2}  = -s \int_0^1 \text{d}a \, e^{-s\left( a(p^2+2pqx)+q^2 \right)} \, .
\end{equation}
This rewriting as an integral over an exponential is generally helpful to perform the momentum integral over $q$. Inserting all expressions and rewritings, exchanging integrals in a convenient way (assuming without proof that this is valid) and performing them, we get
\begin{equation}
\begin{aligned}
 &\text{Tr} \sum_{j\geq0} \frac{(-s)^{j+1}}{(j+1)!} \, \left[ \square, \mathbbm d_1 \right]_j \, e^{-s\square} \\
 &\hspace{1cm} = \left( \frac{1}{4\pi s} \right)^\frac{d}{2} \int \text{d}^dx \, \sqrt{\eta} \, \text{tr} \, \left[ \left( \frac{1}{2} h + \left( \frac{1}{4} f(sp^2) + \frac{1}{2} \frac{f(sp^2)-1}{sp^2} \right) \left( h p^2 - h^{\mu\nu} p_\mu p_\nu \right) \right) \, \mathbbm 1 - s \, f(sp^2) \, \mathbbm E \right] \, .
\end{aligned}
\end{equation}
Here $f$ is the general heat kernel function introduced in \eqref{eq:HKf}. This expression can now be compared to the expansion of the covariant expression \eqref{eq:HKRHS} to fix $a_1$ and $a_2$. The first term comes from the expansion of the determinant of the metric and the identity operator, the second term is the linearised Ricci scalar, and the last term clearly corresponds to the endomorphism. We thus read off
\begin{align}
    a_1(z) &= \HKbracket{\frac{1}{4}, \frac{1}{2}}{z} \, , \\
    a_2(z) &= \HKbracket{-1}{z} \, ,
\end{align}
reconfirming earlier computations, see \eg\ \cite{Codello:2012kq}.

To compute higher orders of the heat kernel, it is also useful to note the relation
\begin{equation}
\begin{aligned}
 &\frac{1}{2} \left( \sum_{j,k\geq0} \frac{(-s)^{j+k+2} (-1)^j}{(j+1)!(k+1)!} (p^2+2pqx)^{j+k} + \sum_{k\geq1} \frac{(-s)^{k+2}}{(k+2)!} \left( 1 - (-1)^k \right) (p^2+2pqx)^k \right) \, e^{-sq^2} \\
 &\hspace{10cm} = \, s^2 \int_0^1 \text{d}a \int_0^a \text{d}b \, e^{-s\left( b(p^2+2pqx)+q^2 \right)} \, ,
\end{aligned}
\end{equation}
which is similar to \eqref{eq:HKhelperequation}.

Before we collect the heat kernel formulas that we need, let us finally note that in this way, one cannot compute topological terms at the lowest possible order. Concretely, the four-dimensional Euler density is quadratic in curvature, but the first non-vanishing flat-space correlation function in any dimension is cubic in $h$. Since we already know its heat kernel coefficient, we will simply take it from the literature and not compute the heat kernel to order $h^3$.

\subsection{Non-local heat kernel coefficients}

In this section we collect all heat kernel formulas that we will need. Wherever we spell out the traces that we need specifically for our system, we will restrict to the case $d=4$. We will eventually use the completely trace-free basis, with the trace-free Ricci tensor $S$ defined by
\begin{equation}
    S_{\mu\nu} = R_{\mu\nu} - \frac{1}{d} g_{\mu\nu} R \, ,
\end{equation}
and the Weyl tensor $C$ defined in the usual way. Additionally, wherever we have a choice, we will use the option with fewer indices. For example, due to
\begin{equation}
    D^\alpha D_{[\alpha} C_{\mu\nu]\rho\sigma} = 0 \, ,
\end{equation}
we can use the Bianchi identity to remove any occurrence of $\Delta C$ and replace it by derivatives acting on $S$ or the Ricci scalar, and higher order curvature terms. As a consequence, we have that
\begin{equation}
    \int \text{d}^dx \sqrt{g} \, \left[ C^{\mu\nu\rho\sigma} \Delta^n C_{\mu\nu\rho\sigma} - 4 \frac{d-3}{d-2} S^{\mu\nu} \Delta^n S_{\mu\nu} + \frac{(d-2)(d-3)}{d(d-1)} R \Delta^n R \right] = \mathcal O\left(\mathcal R^3\right) \, , \qquad n\geq1 \, .
\end{equation}
For $n=0$, we just have the integral over the four-dimensional Euler density $\Euler$ on the left-hand side, which cannot be eliminated from our basis in this way. For the form factors, we will keep $R$ and $S$ in our basis, and use the above equation to remove any occurrence of $C$.

\subsubsection{Zero derivatives}

We will start with the standard heat kernel for an operator of Laplace type with endomorphism $\mathbbm E$. It reads
\begin{equation}\label{eq:HKdiagonal}
\begin{aligned}
 \text{Tr} \left[ e^{-s(\Delta+\mathbbm E)} \right] &\simeq \left( \frac{1}{4\pi s} \right)^\frac{d}{2} \int \text{d}^dx \, \sqrt{g} \, \text{tr} \, \Bigg[ \mathbbm 1 + \HKbracket{\frac{1}{4}, \frac{1}{2}}{s\Delta} s \, R \, \mathbbm 1 + \HKbracket{-1}{s\Delta} s \, \mathbbm E + \frac{s^2}{180} \Euler{} \, \mathbbm 1 \\
 &\hspace{3.75cm} + s^2 \, R \HKbracket{ \frac{1}{32}, \frac{1}{8} , - \frac{1}{8}}{s\Delta} R \, \mathbbm 1 + s^2 \, R^{\mu\nu} \HKbracket{ 0,0,1}{s\Delta} R_{\mu\nu} \, \mathbbm 1 \\
 &\hspace{3.75cm} + s^2 \, R \HKbracket{ - \frac{1}{4} , - \frac{1}{2} }{s\Delta} \mathbbm E + s^2 \, \mathbbm E \HKbracket{ \frac{1}{2} }{s\Delta} \mathbbm E + s^2 \, \mathcal F^{\mu\nu} \HKbracket{ 0, -\frac{1}{2} }{s\Delta} \mathcal F_{\mu\nu} \Bigg] \, .
\end{aligned}
\end{equation}
Here, tr indicates the remaining trace over bundle indices. This result agrees with previous work, see \eg{} \cite{Codello:2012kq}.

We will now present the traced heat kernel coefficients for the individual modes in $d=4$ that appear in our computation, transformed into our basis.

\paragraph{Trace mode}

For the trace mode of the graviton, we have $\mathbbm E=0$ and $\mathcal F=0$, so that
\begin{equation}
 \text{Tr}_0 \left[ e^{-s\Delta} \right] \simeq \left( \frac{1}{4\pi s} \right)^2 \int \text{d}^4x \, \sqrt{g} \, \Bigg[ 1 + \frac{1}{6} \, s \, R + \frac{s^2}{180} \Euler{} + s^2 \, R \HKbracket{ \frac{1}{32}, \frac{1}{8}, \frac{1}{8}}{s\Delta} R + s^2 \, S^{\mu\nu} \HKbracket{ 0, 0, 1}{s\Delta} S_{\mu\nu} \Bigg] \, .
\end{equation}

\paragraph{Ghost}

For the ghost (see \eqref{eq:gammaghost} below), we have $\mathbbm E = -\text{Ric}$ and $\text{tr}_c \mathcal F^2 = -\text{Riem}^2$, so that
\begin{equation}\label{eq:HKc}
 \text{Tr}_c \left[ e^{-s(\Delta-\text{Ric})} \right] \simeq \left( \frac{1}{4\pi s} \right)^2 \int \text{d}^4x \, \sqrt{g} \, \Bigg[ 4 + \frac{5}{3} s \, R - \frac{11}{180} s^2 \, \Euler{} + s^2 \, R \HKbracket{ \frac{1}{2}, 1, \frac{1}{2}}{s\Delta} R + s^2 \, S^{\mu\nu} \HKbracket{ \frac{1}{2}, 2, 4}{s\Delta} S_{\mu\nu} \Bigg] \, .
\end{equation}

\paragraph{Traceless mode}

For the traceless mode of the graviton (see \eqref{eq:Delta2} below), $\mathbbm E = \frac{2}{3}R \Pi^\text{TL} - 2 \mathbbm C$, $\text{tr}_\text{TL} \mathbbm E^2 = 3\text{Riem}^2 - 6\text{Ric}^2+5R^2$ and $\text{tr}_\text{TL} \mathcal F^2 = -6\text{Riem}^2$. This gives
\begin{equation}
\begin{aligned}
 \text{Tr}_\text{TL} \left[ e^{-s(\Delta+\mathbbm E)} \right] &\simeq \left( \frac{1}{4\pi s} \right)^2 \int \text{d}^4x \, \sqrt{g} \, \Bigg[ 9 - \frac{9}{2} \, s \, R + \frac{21}{20} s^2 \, \Euler{} \\
 &\hspace{3.75cm} + s^2 \, R \HKbracket{ \frac{17}{32}, - \frac{15}{8} , \frac{9}{8}}{s\Delta} R + s^2 \, S^{\mu\nu} \HKbracket{ 3, 12, 9}{s\Delta} S_{\mu\nu} \Bigg] \, .
\end{aligned}
\end{equation}

\subsubsection{One derivative}

For our setup, the heat kernel with one derivative has to be computed only up to linear order in curvature. This is because all curvature tensors and the metric have an even number of indices, so that there is at least one other derivative in the insertion in the functional trace, which can only act on a curvature tensor. To this order, and up to total derivatives, one can easily derive the relation
\begin{equation}
 \text{Tr} \left[ X^\mu \, D_\mu \, e^{-s(\Delta+\mathbbm E)} \right] \simeq -\frac{1}{2} \text{Tr} \left[ \left( D_\mu X^\mu \right) \, e^{-s(\Delta+\mathbbm E)} \right] + \left( \frac{1}{4\pi s} \right)^\frac{d}{2} \int \text{d}^dx \, \sqrt{g} \, \text{tr} \left[ X^\mu D^\nu \HKbracket{0,1}{s\Delta} \, s \, \mathcal F_{\mu\nu} \right] \, .
\end{equation}

\subsubsection{Two derivatives I}

We will furthermore need the heat kernel with two derivatives up to linear order, with a curvature insertion. Concretely, we need
\begin{equation}
 \text{Tr} \left[ X^{\alpha\beta} D_{(\alpha} D_{\beta)} \, e^{-s(\Delta+\mathbbm E)} \right] \, ,
\end{equation}
where $X$ is a tensor which is symmetric in the index pair $(\alpha\beta)$. Note that $X$ in general also has bundle indices that we suppress, and we assume that it is linear in curvature. For this trace, we find
\begin{equation}
\begin{aligned}
 \text{Tr} \left[ X^{\alpha\beta} D_{(\alpha} D_{\beta)} \, e^{-s(\Delta+\mathbbm E)} \right] &\simeq \left( \frac{1}{4\pi s} \right)^\frac{d}{2} \int \text{d}^dx \, \sqrt{g} \, \text{tr} \, X^{\alpha\beta} \, \Bigg[ -\frac{1}{2s} g_{\alpha\beta} \, \mathbbm 1 -\frac{1}{2} g_{\alpha\beta} \, \HKbracket{\frac{1}{4}, \frac{1}{2}}{s\Delta} R \, \mathbbm 1 \\
 & + \frac{1}{2} g_{\alpha\beta} \, \HKbracket{1}{s\Delta} \, \mathbbm E + \HKbracket{0,-1}{s\Delta} R_{\alpha\beta} \, \mathbbm 1 + s \HKbracket{\frac{1}{16}, 0, -\frac{3}{4}}{s\Delta} D_{(\alpha} D_{\beta)} R \, \mathbbm 1 \\
 & + s \HKbracket{-\frac{1}{4}, \frac{1}{2}}{s\Delta} D_{(\alpha} D_{\beta)} \mathbbm E + s \HKbracket{0,1}{s\Delta} D_{(\alpha} D^\gamma \mathcal F_{\beta)\gamma} \Bigg] \, .
\end{aligned}
\end{equation}

\subsubsection{Two derivatives II}

Finally, we need the heat kernel with two derivatives up to quadratic order in curvature. For this, we consider
\begin{equation}\label{eq:DDH}
 \text{Tr} \left[ X^{\alpha\beta} D_{(\alpha} D_{\beta)} \, e^{-s(\Delta+\mathbbm E)} \right] \, ,
\end{equation}
where $X$ is a \emph{covariantly constant} tensor (that is, constructed from metrics and deltas only) that is symmetric in the index pair $(\alpha\beta)$. Note that $X$ in general also has bundle indices. The condition of covariant constancy simplifies the computation of the necessary quadratic heat kernel coefficients tremendously.

This trace splits into two contributions, since the two derivatives can either act both on the exponential of the world function, or on $\Omega$ -- the cross-term vanishes in the coincidence limit due to the properties of the world function. This readily gives
\begin{equation}
 \text{Tr} \left[ X^{\alpha\beta} D_{(\alpha} D_{\beta)} \, e^{-s(\Delta+\mathbbm E)} \right] = -\frac{1}{2s} \text{Tr} \left[ X^{\alpha\beta} g_{\alpha\beta} \, e^{-s(\Delta+\mathbbm E)} \right]  + \left( \frac{1}{4\pi s} \right)^\frac{d}{2} \text{tr} \left[ X^{\alpha\beta} \overline{D_{(\alpha} D_{\beta)} \Omega}(s) \, \right] \, .
\end{equation}
The first term only needs the standard diagonal heat kernel coefficients. The second is the most complicated heat kernel coefficient needed, and to our knowledge has not been reported previously in the literature. Neglecting total derivatives, it reads
\begin{equation}\label{eq:the_big_one}
\begin{aligned}
 &\text{tr} \left[ X^{\alpha\beta} \overline{D_{(\alpha} D_{\beta)} \Omega}(s) \, \right] = \int \text{d}^dx \, \sqrt{g} \, \text{tr} \, X^{\alpha\beta} \, \Bigg[ \frac{1}{6} R_{\alpha\beta} \, \mathbbm 1 + s \, R \, \left( -\frac{1}{15} + \HKbracket{\frac{1}{8}, \frac{1}{4}}{s\Delta} \right) \, R_{\alpha\beta} \, \mathbbm 1 \\
 &\hspace{2.75cm} + s \, R_{(\alpha}^{\phantom{(\alpha}\gamma} \, \left( 0 \right) \, R_{\beta)\gamma}^{\phantom{)}} \, \mathbbm 1 + s \, R_{\gamma\delta} \, \left( \frac{1}{30} \right) \, R_{(\alpha\phantom{\gamma}\beta)}^{\phantom{(\alpha}\gamma\phantom{\beta)}\delta} \, \mathbbm 1 \\
 &\hspace{2.75cm} + s \, c_{\Euler{}} \, \left[ R_{(\alpha}^{\phantom{(\alpha}\gamma\delta\kappa} \, R_{\beta)\gamma\delta\kappa} - 2 R_{\gamma\delta} \, R_{(\alpha\phantom{\gamma}\beta)}^{\phantom{(\alpha}\gamma\phantom{\beta)}\delta} - 2 R_{(\alpha}^{\phantom{(\alpha}\gamma} \, R_{\beta)\gamma}^{\phantom{)}} + R \, R_{\alpha\beta} \right] \, \mathbbm 1 \\
 &\hspace{2.75cm} + s \, R_{\alpha\beta} \, \left( \frac{1}{3} - \HKbracket{\frac{1}{2}}{s\Delta} \right) \, \mathbbm E + s \, R_{(\alpha}^{\phantom{(\alpha}\gamma} \, \left( \frac{1}{6} + \HKbracket{0,1}{s\Delta} \right) \, \mathcal F_{\beta)\gamma}^{\phantom{)}} + s \, \mathcal F_{(\alpha}^{\phantom{(\alpha}\gamma} \, \HKbracket{0,-1}{s\Delta} \, \mathcal F_{\beta)\gamma}^{\phantom{)}} \\
 &\hspace{2.75cm} + s^2 \, R \, \HKbracket{\frac{1}{128}, \frac{7}{64}, \frac{5}{32}, -\frac{1}{16}}{s\Delta} \, D_{(\alpha} D_{\beta)} R \, \mathbbm 1 + s^2 \, R^{\gamma\delta} \, \HKbracket{0,0,\frac{1}{4}, \frac{1}{2}}{s\Delta} \, D_{(\alpha} D_{\beta)} R_{\gamma\delta} \, \mathbbm 1 \\
 &\hspace{2.75cm} + s^2 \, R \, \HKbracket{-\frac{1}{16}, -\frac{1}{2}, -\frac{1}{4}}{s\Delta} \, D_{(\alpha} D_{\beta)} \mathbbm E + s^2 \, \mathbbm E \, \HKbracket{\frac{1}{8}, \frac{1}{4}}{s\Delta} \, D_{(\alpha} D_{\beta)} \mathbbm E \\
 &\hspace{2.75cm} + s^2 \, R \, \HKbracket{0,0,1}{s\Delta} \, D_{(\alpha} D^\gamma \mathcal F_{\beta)\gamma} + s^2 \, \mathbbm E \, \HKbracket{0,\frac{1}{2},1}{s\Delta} \, D_{(\alpha} D^\gamma \mathcal F_{\beta)\gamma} \\
 &\hspace{2.75cm} + s^2 \, D_{(\alpha} D^\gamma \mathcal F_{\beta)\gamma} \, \HKbracket{0,-\frac{1}{2},-1}{s\Delta} \, \mathbbm E + s^2 \, \mathcal F^{\gamma\delta} \, \HKbracket{0,-\frac{1}{8}, -\frac{3}{4}}{s\Delta} \, D_{(\alpha} D_{\beta)} \mathcal F_{\gamma\delta} \Bigg] \, .
\end{aligned}
\end{equation}
To construct the basis, we used that $\Delta R_{\alpha\beta\gamma\delta}$ can be written in terms of Ricci tensors and scalars so that no form factor can appear with any term involving a Riemann tensor. Moreover, we used the Bianchi identity for the gauge field strength, $D_{[\alpha} \mathcal F_{\beta\gamma]}=0$.

The coefficient $c_{\Euler{}}$ cannot be computed via a direct comparison with the flat expansion, since the expansion of the tensor structure to second order in $h$ around a flat background vanishes. This is related to the fact that the contraction of this tensor structure with the metric gives the four-dimensional Euler density, which even in general dimension only contributes at order $h^3$ around a flat background. We can however fix it by requiring consistency of the formula for a specific choice of $X^{\alpha\beta}$. Concretely, if we choose
\begin{equation}
 X^{\alpha\beta} = g^{\alpha\beta} \, \mathbbm 1 \, ,
\end{equation}
the total heat kernel \eqref{eq:DDH} reduces to a diagonal heat kernel,
\begin{equation}\label{eq:HKconsistency}
 \text{Tr} \left[ g^{\alpha\beta} \, \mathbbm 1 \, D_{(\alpha} D_{\beta)} \, e^{-s(\Delta+\mathbbm E)} \right]  = \text{Tr} \left[ -\Delta \, e^{-s(\Delta+\mathbbm E)} \right]  = \partial_s \, \text{Tr} \left[ e^{-s(\Delta+\mathbbm E)} \right] + \text{Tr} \left[ \mathbbm E \, e^{-s(\Delta+\mathbbm E)} \right] \, .
\end{equation}
Since we only need to compare a pure curvature term, we can set $\mathbbm E=0$. We can clearly also restrict to local terms with exactly four derivatives, and to $d=4$. Let us call this trace $\mathcal T$. This entails on the one hand
\begin{equation}
\begin{aligned}
 \mathcal T &= - \frac{4}{2s} \text{Tr} \left[ e^{-s\Delta} \right] + \frac{1}{(4\pi s)^2} \text{tr} \left[ \, \overline{-\Delta \Omega}(s) \, \right] \bigg|_{\mathcal R^2} \\
 &= -\frac{4}{2s} \frac{1}{(4\pi s)^2} \int \sqrt{g} \, \text{tr} \left[ s^2 \, \left( \frac{1}{80} R^2 + \frac{1}{60} S^{\mu\nu} S_{\mu\nu} + \frac{1}{180} \Euler{} \right) \mathbbm 1 \right] + \frac{1}{(4\pi s)^2} \int \sqrt{g} \, \text{tr} \left[ s \left( \frac{1}{60} R^2 + \frac{1}{30} R^{\mu\nu} R_{\mu\nu} + c_{\Euler{}} \Euler{} \right) \, \mathbbm 1 \right] \\
 &= \frac{s}{(4\pi s)^2} \int \sqrt{g} \, \text{tr} \left[ \left( c_{\Euler{}} - \frac{1}{90} \right) \Euler{} \, \mathbbm 1 \right] \, .
\end{aligned}
\end{equation}
On the other hand, since the local four-derivative terms come with a power of $s^0$ in $d=4$, we directly have
\begin{equation}
 \mathcal T = \partial_s \, \text{Tr} \left[ e^{-s\Delta} \right] \bigg|_{\mathcal R^2} = 0 \, .
\end{equation}
This implies that the final coefficient is
\begin{equation}
 c_{\Euler{}} = \frac{1}{90} \, .
\end{equation}
One can easily check that the coefficient $c_{\Euler{}}$ is independent of the dimension by repeating the comparison for arbitrary $d$. As an additional check, a direct comparison with the local heat kernel with two derivatives in arbitrary dimension gives the same result.

The complete result \eqref{eq:the_big_one} was cross-checked with the local heat kernel up to six derivatives, and we also checked \eqref{eq:HKconsistency} by comparing with the diagonal non-local heat kernel \eqref{eq:HKdiagonal}.

\clearpage

\section{Computation of the functional trace}\label{app:tracecomputation}

In this appendix we display the computation of the full trace. Let us collect the main ingredients of our setup first. Our starting action is
\begin{equation}
 \Gamma_k = \int \text{d}^4x \, \sqrt{g} \, \left\{ \frac{1}{16\pi G_k} \bigg[ 2\Lambda_k - R \bigg] + \Theta_k \Euler{} \right\} \, ,
\end{equation}
together with a harmonic gauge fixing condition,
\begin{equation}
 \Gamma^\text{gf} = \frac{1}{32\pi G_k} \int \text{d}^4x \, \sqrt{\bar g} \, \bar g^{\mu\nu} \mathcal F_\mu \mathcal F_\nu \, , \qquad
 \mathcal F_\mu = \bar D^\alpha h_{\mu\alpha} - \frac{1}{2} \bar D_\mu h \, .
\end{equation}
An overbar indicates background quantities built from the background metric $\bar g$. The resulting ghost action reads
\begin{equation}\label{eq:gammaghost}
    \Gamma^c = \frac{1}{\sqrt{G_k}} \int \text{d}^4x \, \sqrt{\bar g} \, \bar c^\mu \bar{\Delta}_{c\mu}^{\phantom{c\mu}\nu} c_\nu \, , \qquad \bar{\Delta}_{c\mu}^{\phantom{c\mu}\nu} = \bar \Delta \, \delta_\mu^{\phantom{\mu}\nu} - \bar R_\mu^{\phantom{\mu}\nu} \, .
\end{equation}
Following \cite{Knorr:2022ilz}, we pick a regulator that is fully adapted to this action and of type II \cite{Codello:2008vh}. Suppressing indices, we have
\begin{equation}\label{eq:reg}
    \mathfrak R^\text{h} = \frac{1}{32\pi G_k} \bigg[ \mathcal R^\text{TL}(\bar{\Delta}_2) \Pi^\text{TL} - \mathcal R^\text{Tr}(\bar \Delta) \Pi^\text{Tr} \bigg] \, , \qquad \mathfrak R^\text{c} = \frac{1}{\sqrt{G_k}} \mathcal R^\text{c}(\bar{\Delta}_c) \mathbbm 1 \, ,
\end{equation}
with
\begin{equation}\label{eq:Delta2}
    \bar{\Delta}_2 = \left[ \bar \Delta + \frac{2}{3} \bar R \right] \Pi^\text{TL} - 2\bar{\mathbbm C} \, .
\end{equation}
In this, we have used the projectors
\begin{equation}
    \Pi^{\text{Tr}\mu\nu}_{\phantom{\text{Tr}\mu\nu}\rho\sigma} = \frac{1}{4} \bar g^{\mu\nu} \bar g_{\rho\sigma} \, , \qquad \Pi^{\text{TL}\mu\nu}_{\phantom{\text{TL}\mu\nu}\rho\sigma} = \mathbbm 1^{\mu\nu}_{\phantom{\mu\nu}\rho\sigma} - \Pi^{\text{Tr}\mu\nu}_{\phantom{\text{Tr}\mu\nu}\rho\sigma} = \frac{1}{2} \left( \delta^\mu_{\phantom{\mu}\rho} \delta^\nu_{\phantom{\nu}\sigma} + \delta^\mu_{\phantom{\mu}\sigma} \delta^\nu_{\phantom{\nu}\rho} \right) - \frac{1}{4} \bar g^{\mu\nu} \bar g_{\rho\sigma} \, ,
\end{equation}
and the symmetrised Weyl tensor
\begin{equation}
 \bar{\mathbbm C}^{\mu\phantom{\rho}\nu}_{\phantom{\mu}\rho\phantom{\nu}\sigma} = \frac{1}{2} \left( \bar{C}^{\mu\phantom{\rho}\nu}_{\phantom{\mu}\rho\phantom{\nu}\sigma} + \bar{C}^{\nu\phantom{\rho}\mu}_{\phantom{\nu}\rho\phantom{\mu}\sigma} \right) \, .
\end{equation}
From this, we obtain the propagator
\begin{equation}\label{eq:prop}
 \mathfrak G^\text{h} = 32\pi G_k \bigg[ \mathcal G^\text{TL}(\bar{\Delta}_2) \Pi^\text{TL} - \mathcal G^\text{Tr}(\bar\Delta) \Pi^\text{Tr} \bigg] \, , \qquad \mathfrak G^\text{c} = \sqrt{G_k} \, \mathcal G^\text{c}(\bar{\Delta}_c) \mathbbm 1 \, ,
\end{equation}
with
\begin{equation}\label{eq:propfuncs}
 \mathcal G^\text{TL,Tr}(x) = \frac{1}{x+\mathcal R^\text{TL,Tr}(x)-2\Lambda_k} \, , \qquad \mathcal G^\text{c}(x) = \frac{1}{x+\mathcal R^\text{c}(x)} \, .
\end{equation}
Since the prefactor of the regulator \eqref{eq:reg} includes Newton's constant, it is convenient to introduce
\begin{equation}
 \mathring{\mathcal R}^\text{TL,Tr}(z) = \left(2 - \frac{\dot g - 2g}{g} \right) \mathcal R^\text{TL,Tr}(z) - 2 z \, \mathcal R^{\text{TL,Tr}\prime}(z) \, , \qquad \mathring{\mathcal R}^\text{c}(z) = \left( 2 - \frac{\dot g - 2g}{2g} \right) \mathcal R^\text{c}(z) - 2 z \, \mathcal R^{\text{c}\prime}(z) \, ,
\end{equation}
where $\dot g$ is the beta function of the dimensionless Newton's constant. This takes into account the cancellation of the same inverse prefactors in the propagators \eqref{eq:prop}.

To take into account the propagator form factors in the \ac{MES}, we define the \ac{RG} kernel
\begin{equation}
 \Psi_{\mu\nu} = \gamma_g g_{\mu\nu} + g_{\mu\nu} \gamma_R(\Delta) R + \gamma_S(\Delta_2)_{\mu\nu}^{\phantom{\mu\nu}\alpha\beta} S_{\alpha\beta} \, .
\end{equation}
The modified flow equation reads
\begin{equation}\label{eq:modflowapp}
 \dot \Gamma_k + \Psi_{\mu\nu} \frac{\delta \Gamma_k}{\delta g_{\mu\nu}} = \frac{1}{2} \text{Tr} \left[ \mathfrak G_{\mu\nu\rho\sigma}^\text{h} \left\{ \mathbbm 1^{\rho\sigma}_{\phantom{\rho\sigma}\tau\omega} k \partial_k + 2 \frac{\delta \Psi_{\tau\omega}}{\delta g_{\rho\sigma}} \right\} \mathfrak R^{\text{h}\,\tau\omega\kappa\lambda} \right] - \text{Tr} \left[ \mathfrak G_{\mu\nu}^\text{c} \,  \dot{\mathfrak R}^{\text{c}\,\nu\rho} \right] \, .
\end{equation}
The extra term on the left-hand side reads explicitly
\begin{equation}\label{eq:LHSextraterm}
\begin{aligned}
 \Psi_{\mu\nu} \frac{\delta \Gamma_k}{\delta g_{\mu\nu}} &= \frac{1}{16\pi G_N} \int \text{d}^4x \, \sqrt{g} \, \left[ \left( 4 \Lambda - R \right) \left( \gamma_g + \gamma_R(\Delta) R \right) + S^{\mu\nu}  \gamma_S(\Delta_2)_{\mu\nu}^{\phantom{\mu\nu}\alpha\beta} S_{\alpha\beta} \right] \\
 &= \frac{1}{16\pi G_N} \int \text{d}^4x \, \sqrt{g} \, \left[ \Big\{ 2 \gamma_g \Big\} \, 2 \Lambda - \Big\{ \gamma_g - 4\Lambda \gamma_R(0) \Big\} \, R - R \gamma_R(\Delta) R + S^{\mu\nu}  \gamma_S(\Delta_2)_{\mu\nu}^{\phantom{\mu\nu}\alpha\beta} S_{\alpha\beta} \right] \, .
\end{aligned}
\end{equation}
To project onto $\gamma_S$, we note that in a curvature expansion, it is enough to compute the full momentum dependence of terms quadratic in curvature. We thus can neglect the endomorphism for the projection, $S \gamma_S(\Delta_2) S \simeq S \gamma_S(\Delta) S$.

In the following, we will need an expression for the variation of a form factor. Such a general variation reads
\begin{equation}\label{eq:FFvariation}
 \left( \delta f(\mathcal O) \right) X = - \int_0^\infty \text{d}s \, \mathcal L^{-1}[f(\mathcal O)](s) \, s \, \int_0^1 \text{d}\alpha \, e^{-s\alpha\mathcal O} \left( \delta \mathcal O \right) e^{-s(1-\alpha)\mathcal O} X \, .
\end{equation}
To perform some of the tensor contractions, we used the Mathematica package suite \emph{xAct} \cite{xActwebpage, Brizuela:2008ra, Nutma:2013zea}. From now on, we will suppress overbars, since all quantities below are background quantities.

\subsection{Flow overview}

We will start with an overview of the different contributions to the trace, \ie\ the right-hand side of \eqref{eq:modflowapp}. The first part consists of the ``standard'' term and a specific part of the \ac{RG} kernel, namely that where the variation hits the explicit metric factors, which gives an identity. Concretely, this reads
\begin{equation}
 \mathcal F_1 = \frac{1}{2} \text{Tr} \left[ \mathfrak G_{\mu\nu\rho\sigma}^\text{h} \left\{ k \partial_k + 2 \gamma_g + 2 \left( \gamma_R(\Delta) R \right) \right\} \mathfrak R^{\text{h}\,\rho\sigma\kappa\lambda} \right] \, .
\end{equation}
Inserting the explicit expressions for the propagator and regulator, \eqref{eq:prop} and \eqref{eq:reg}, we find
\begin{equation}
\begin{aligned}
 \mathcal F_1 &= \frac{1}{2} \text{Tr} \bigg[ \left( \mathring{\mathcal R}^\text{TL}(\Delta_2) + 2 \left( \gamma_g + \left( \gamma_R(\Delta) R \right) \right) \mathcal R^\text{TL}(\Delta_2) \right) \mathcal G^\text{TL}(\Delta_2) \Pi^\text{TL} \\
 &\hspace{2cm} + \left( \mathring{\mathcal R}^\text{Tr}(\Delta) + 2 \left( \gamma_g + \left( \gamma_R(\Delta) R \right) \right) \mathcal R^\text{Tr}(\Delta) \right) \mathcal G^\text{Tr}(\Delta) \Pi^\text{Tr} \bigg] \, .
\end{aligned}
\end{equation}
Using the inverse Laplace transform $\mathcal L^{-1}$ w.r.t. $\Delta_2$ or $\Delta$, we split this further into individual contributions where we simply can insert the heat kernel coefficients. Concretely,
\begin{align}
 \mathcal F_1^{\text{TL},1} &= \frac{1}{2} \int_0^\infty \text{d}s \, \mathcal L^{-1}\left[ \left(\mathring{\mathcal R}^\text{TL}(\Delta_2) + 2\gamma_g \mathcal R^\text{TL}(\Delta_2)\right) \mathcal G^\text{TL}(\Delta_2) \right](s) \, \text{Tr}_\text{TL} \left[ e^{-s\Delta_2} \right] \, , \\
 \mathcal F_1^{\text{TL},2} &= \int_0^\infty \text{d}s \, \mathcal L^{-1}\left[ \mathcal R^\text{TL} (\Delta_2) \mathcal G^\text{TL}(\Delta_2) \right](s) \, \text{Tr}_\text{TL} \left[ \left( \gamma_R(\Delta) R \right) e^{-s\Delta_2} \right] \, , \\
 \mathcal F_1^{\text{Tr},1} &= \frac{1}{2} \int_0^\infty \text{d}s \, \mathcal L^{-1}\left[ \left(\mathring{\mathcal R}^\text{Tr}(\Delta) + 2\gamma_g \mathcal R^\text{Tr}(\Delta) \right) \mathcal G^\text{Tr}(\Delta) \right](s) \, \text{Tr}_\text{0} \left[ e^{-s\Delta} \right] \, , \\
 \mathcal F_1^{\text{Tr},2} &= \int_0^\infty \text{d}s \, \mathcal L^{-1}\left[ \mathcal R^\text{Tr}(\Delta) \mathcal G^\text{Tr}(\Delta) \right](s) \, \text{Tr}_\text{0} \left[ \left( \gamma_R(\Delta) R \right) e^{-s\Delta} \right] \, .
\end{align}
Here, the label on the trace indicates the type of trace that has to be performed: the label $0$ indicates a scalar trace, whereas the label TL indicates a traceless trace, that is
\begin{equation}
 \text{Tr}_\text{TL} \, X = \text{Tr} \, \Pi^\text{TL} X \, .
\end{equation}

Moving on to the other contributions, we next discuss the variation of the Ricci scalar (without the variation of the form factor which will be dealt with next) in the \ac{RG} kernel. This reads
\begin{equation}
 \mathcal F_2 = \text{Tr} \left[ \mathfrak G_{\mu\nu\rho\sigma}^\text{h} g_{\tau\omega} \frac{\delta R}{\delta g_{\rho\sigma}} \gamma_R(\Delta) \mathfrak R^{\text{h}\,\tau\omega\kappa\lambda} \right] \, .
\end{equation}
Due to the appearance of the explicit metric factor, this reduces to
\begin{equation}
 \mathcal F_2 = \text{Tr}_0 \left[ g_{\rho\sigma} \frac{\delta R}{\delta g_{\rho\sigma}} \gamma_R(\Delta) \mathcal R^\text{Tr}(\Delta) \mathcal G^\text{Tr}(\Delta) \right] \, .
\end{equation}
The necessary variation can be computed easily:
\begin{equation}
 g_{\rho\sigma} \frac{\delta R}{\delta g_{\rho\sigma}} = 3\Delta - R \, .
\end{equation}
Splitting the resulting trace into two parts again, we have
\begin{align}
 \mathcal F_2^1 &= 3 \int_0^\infty \text{d}s \, \mathcal L^{-1}\left[ \Delta \, \gamma_R(\Delta) \mathcal R^\text{Tr}(\Delta) \mathcal G^\text{Tr}(\Delta) \right](s) \, \text{Tr}_0 \left[ e^{-s\Delta} \right] \, , \\
 \mathcal F_2^2 &= - \int_0^\infty \text{d}s \, \mathcal L^{-1}\left[ \gamma_R(\Delta) \mathcal R^\text{Tr}(\Delta) \mathcal G^\text{Tr}(\Delta) \right](s) \, \text{Tr}_0 \left[ R \, e^{-s\Delta} \right] \, .
\end{align}

To finish off the contribution stemming from $\gamma_R$, we now discuss the variation of the gamma function itself. Following the general formula \eqref{eq:FFvariation}, we have
\begin{equation}
 \mathcal F_3 = - \text{Tr} \left[ \mathfrak G_{\mu\nu\rho\sigma}^\text{h} g_{\tau\omega}  \int_0^\infty \text{d}t \, \mathcal L^{-1}\left[ \gamma_R(\Delta) \right](t) \, t \, \int_0^1 \text{d}\alpha \, \left\{ e^{-t\alpha\Delta} \frac{\delta\Delta}{\delta g_{\rho\sigma}} e^{-t(1-\alpha)\Delta} R \right\} \mathfrak R^{\text{h}\,\tau\omega\kappa\lambda} \right] \, .
\end{equation}
Once again the appearance of the explicit metric factor restricts this contribution to the trace component. With the variation
\begin{equation}
\begin{aligned}
 g_{\rho\sigma} \left( \frac{\delta \Delta}{\delta g_{\rho\sigma}} R \right) &= g_{\rho\sigma} \left[ \left( D^\rho D^\sigma R \right) - D^{(\rho} \left( D^{\sigma)} R \right) + \frac{1}{2} g^{\rho\sigma} D^\kappa \left( D_\kappa R \right) \right] \\
 &= \left( D^\kappa R \right) D_\kappa - 2 \left( \Delta R \right) \, ,
\end{aligned}
\end{equation}
we find
\begin{equation}
\begin{aligned}
 \mathcal F_3 &= - \int_0^\infty \text{d}t \, \mathcal L^{-1}\left[ \gamma_R(\Delta) \right](t) \, t \, \int_0^1 \text{d}\alpha \int_0^\infty \text{d}s \, \mathcal L^{-1}\left[ e^{-t\alpha\Delta}\mathcal R^\text{Tr}(\Delta) \mathcal G^\text{Tr}(\Delta) \right](s) \, \times \\
 &\hspace{7cm} \text{Tr}_0 \left[ \left\{ D^\kappa e^{-t(1-\alpha)\Delta} R \right\} D_\kappa e^{-s\Delta} - 2 \left\{ \Delta e^{-t(1-\alpha)\Delta} R \right\} e^{-s\Delta} \right] \, .
\end{aligned}
\end{equation}
For this contribution in particular, there is a further simplification. Here we are only interested in the scalar heat kernel to linear order in the curvature. To this order, we have the identity
\begin{equation}
 \text{Tr}_0 \left[ X^\mu D_\mu e^{-s\Delta} \right] \simeq -\frac{1}{2} \text{Tr}_0 \left[ \left( D_\mu X^\mu \right) e^{-s\Delta} \right] \, .
\end{equation}
This entails that
\begin{equation}
 \mathcal F_3 = \frac{3}{2} \int_0^\infty \text{d}t \, \mathcal L^{-1}\left[ \gamma_R(\Delta) \right](t) \, t \, \int_0^1 \text{d}\alpha \int_0^\infty \text{d}s \, \mathcal L^{-1}\left[ e^{-t\alpha\Delta}\mathcal R^\text{Tr}(\Delta) \mathcal G^\text{Tr}(\Delta) \right](s) \text{Tr}_0 \left[ \left\{ \Delta e^{-t(1-\alpha)\Delta} R \right\} e^{-s\Delta} \right] \, .
\end{equation}

Now we are moving on to the contribution stemming from $\gamma_S$. We will first discuss the contribution stemming from the variation of $S$ only. For this, we find
\begin{equation}
 \mathcal F_4 = \text{Tr} \left[ \mathfrak G_{\mu\nu\rho\sigma}^\text{h} \frac{\delta S_{\alpha\beta}}{\delta g_{\rho\sigma}} \gamma_S(\Delta_2)_{\tau\omega}^{\phantom{\tau\omega}\alpha\beta} \mathfrak R^{\text{h}\,\tau\omega\kappa\lambda} \right] \, .
\end{equation}
Due to the projector property of $\gamma_S$, this receives contributions only from the traceless sector. Moreover, the explicit variation reads
\begin{equation}
\begin{aligned}
 \Pi^\text{TL}_{\phantom{\text{TL}}\mu\nu\rho\sigma} \frac{\delta S_{\alpha\beta}}{\delta g_{\rho\sigma}} \Pi^{\text{TL}\alpha\beta\gamma\delta} &= \frac{1}{4} \left( 2\Delta - R \right) \Pi^{\text{TL}\phantom{\mu\nu}\gamma\delta}_{\phantom{\text{TL}}\mu\nu} + \Pi^{\text{TL}\phantom{\mu\nu}\kappa\lambda}_{\phantom{\text{TL}}\mu\nu} \left[ D_\kappa \delta_\lambda^{\phantom{\lambda}\tau} D^\omega \right] \Pi^{\text{TL}\phantom{\tau\omega}\gamma\delta}_{\phantom{\text{TL}}\tau\omega} \\
 &\equiv \left[ \frac{1}{2} \Delta_2 - \frac{7}{12} R + \mathbbm C \right] \Pi^{\text{TL}\phantom{\mu\nu}\gamma\delta}_{\phantom{\text{TL}}\mu\nu} + \left[ \tilde{\mathbbm X}^{\alpha\beta}\right]^{\phantom{\mu\nu}\gamma\delta}_{\mu\nu} D_\alpha D_\beta \\
 &\equiv \left[ \frac{1}{2} \Delta_2 - \frac{5}{12} R + \frac{1}{2} \mathbbm C + \mathbbm S \right] \Pi^{\text{TL}\phantom{\mu\nu}\gamma\delta}_{\phantom{\text{TL}}\mu\nu} + \left[ \mathbbm X^{\alpha\beta}\right]^{\phantom{\mu\nu}\gamma\delta}_{\mu\nu} D_{(\alpha} D_{\beta)} \, .
\end{aligned}
\end{equation}
Here we collected terms into an explicit $\Delta_2$, and introduced the covariantly constant tensor
\begin{equation}
  \left[ \mathbbm X^{\alpha\beta}\right]^{\phantom{\mu\nu}\gamma\delta}_{\mu\nu} = \Pi^{\text{TL}\phantom{\mu\nu}\kappa(\alpha}_{\phantom{\text{TL}}\mu\nu} \Pi^{\text{TL}\beta)\phantom{\kappa}\gamma\delta}_{\phantom{\text{TL}\beta}\kappa} \, .
\end{equation}
Furthermore, we introduced the traceless tensor
\begin{equation}
 \mathbbm S^{\mu\phantom{\rho}\nu}_{\phantom{\mu}\rho\phantom{\nu}\sigma} = \frac{1}{2} \left( S^\mu_{\phantom{\mu}\rho} \delta^\nu_{\phantom{\nu}\sigma} + S^\nu_{\phantom{\nu}\rho} \delta^\mu_{\phantom{\mu}\sigma} - \frac{1}{2} S^{\mu\nu} g_{\rho\sigma} - \frac{1}{2} g^{\mu\nu} S_{\rho\sigma} \right) \, .
\end{equation}
With this, we can split the trace into
\begin{align}
 \mathcal F_4^1 &= \frac{1}{2} \int_0^\infty \text{d}s \, \mathcal L^{-1}\left[ \Delta_2 \, \gamma_S(\Delta_2) \mathcal R^\text{TL}(\Delta_2) \mathcal G^\text{TL}(\Delta_2) \right](s) \, \text{Tr}_\text{TL} \left[ e^{-s\Delta_2} \right] \, , \\
 \mathcal F_4^2 &= -\frac{5}{12} \int_0^\infty \text{d}s \, \mathcal L^{-1}\left[ \gamma_S(\Delta_2) \mathcal R^\text{TL}(\Delta_2) \mathcal G^\text{TL}(\Delta_2) \right](s) \, \text{Tr}_\text{TL} \left[ R \, e^{-s\Delta_2} \right] \, , \\
 \mathcal F_4^3 &= \frac{1}{2} \int_0^\infty \text{d}s \, \mathcal L^{-1}\left[ \gamma_S(\Delta_2) \mathcal R^\text{TL}(\Delta_2) \mathcal G^\text{TL}(\Delta_2) \right](s) \, \text{Tr}_\text{TL} \left[ \mathbbm C \, e^{-s\Delta_2} \right] \, , \\
 \mathcal F_4^4 &= \int_0^\infty \text{d}s \, \mathcal L^{-1}\left[ \gamma_S(\Delta_2) \mathcal R^\text{TL}(\Delta_2) \mathcal G^\text{TL}(\Delta_2) \right](s) \, \text{Tr}_\text{TL} \left[ \mathbbm S \, e^{-s\Delta_2} \right] \, , \\
 \mathcal F_4^5 &= \int_0^\infty \text{d}s \, \mathcal L^{-1}\left[ \gamma_S(\Delta_2) \mathcal R^\text{TL}(\Delta_2) \mathcal G^\text{TL}(\Delta_2) \right](s) \, \text{Tr}_\text{TL} \left[ \mathbbm X^{\alpha\beta} D_{(\alpha} D_{\beta)} e^{-s\Delta_2} \right] \, .
\end{align}
The contribution $\mathcal F_4^5$ is the most difficult to compute since it needs the complete non-local heat kernel with two derivatives.

The final piece is the variation of $\gamma_S$. Once again following the general variation \eqref{eq:FFvariation}, we first have
\begin{equation}
 \mathcal F_5 = - \text{Tr} \left[ \mathfrak G_{\mu\nu\rho\sigma}^\text{h} \int_0^\infty \text{d}t \, \mathcal L^{-1}\left[ \gamma_S(\Delta_2) \right](t) \, t \, \int_0^1 \text{d}\alpha \, \left\{ e^{-t\alpha\Delta_2} \frac{\delta\Delta_{2\tau\omega}^{\phantom{2\tau\omega}\alpha\beta}}{\delta g_{\rho\sigma}} e^{-t(1-\alpha)\Delta_2} S_{\alpha\beta} \right\} \mathfrak R^{\text{h}\,\tau\omega\kappa\lambda} \right] \, .
\end{equation}
Once again due to the projector properties of all objects involved, this expression only involves the traceless sector. We can thus write it as
\begin{equation}
 \mathcal F_5 = - \text{Tr}_\text{TL} \left[ \int_0^\infty \text{d}t \, \mathcal L^{-1}\left[ \gamma_S(\Delta_2) \right](t) \, t \, \int_0^1 \text{d}\alpha \, \left\{ \frac{\delta\Delta_{2\tau\omega}^{\phantom{2\tau\omega}\alpha\beta}}{\delta g_{\rho\sigma}} e^{-t(1-\alpha)\Delta_2} S_{\alpha\beta} \right\} e^{-t\alpha\Delta_2} \mathcal R^\text{TL}(\Delta_2) \mathcal G^\text{TL}(\Delta_2) \right] \, .
\end{equation}
The necessary variation reads
\begin{equation}
\begin{aligned}
 \Pi^{\text{TL}\phantom{\rho\sigma}\kappa\lambda}_{\phantom{\text{TL}}\rho\sigma} \left[ \frac{\delta\Delta_{2\tau\omega}^{\phantom{2\tau\omega}\alpha\beta}}{\delta g_{\rho\sigma}} S_{\alpha\beta} \right] &= \left[A_1^{\phantom{0}\alpha\beta}\right]_{\tau\omega}^{\phantom{\tau\omega}\kappa\lambda} S_{\alpha\beta} + \left[A_2^{\phantom{0}\alpha\beta\gamma\delta}\right]_{\tau\omega}^{\phantom{\tau\omega}\kappa\lambda} \left( D_{(\gamma} D_{\delta)} S_{\alpha\beta} \right) \\
 &\qquad + \left[A_3^{\phantom{0}\alpha\beta\gamma\delta}\right]_{\tau\omega}^{\phantom{\tau\omega}\kappa\lambda} \left( D_\gamma S_{\alpha\beta} \right) D_\delta + \left[A_4^{\phantom{0}\alpha\beta\gamma\delta}\right]_{\tau\omega}^{\phantom{\tau\omega}\kappa\lambda} S_{\alpha\beta} D_{(\gamma} D_{\delta)} \, .
\end{aligned}
\end{equation}
Here $A_1$ is linear in curvature and $A_{2,3,4}$ are constructed from metrics alone. The explicit expressions for the $A_i$ is rather lengthy, but straightforward to compute. By construction, they are all traceless in the index pairs $(\alpha\beta)$, $(\tau\omega)$ and $(\kappa\lambda)$. With this, we can split the final trace contribution into
\begin{align}
 &\begin{aligned}
  \mathcal F_5^1 &= -\int_0^\infty \text{d}t \, \mathcal L^{-1}\left[ \gamma_S(\Delta_2) \right](t) \, t \int_0^1 \text{d}\alpha \int_0^\infty \text{d}s \, \mathcal L^{-1}\left[ e^{-t\alpha\Delta_2} \mathcal R^\text{TL}(\Delta_2) \mathcal G^\text{TL}(\Delta_2) \right](s) \times \\
  &\hspace{8cm} \text{Tr}_\text{TL} \left[ \left\{ e^{-t(1-\alpha)\Delta_2} S_{\alpha\beta} \right\} A_1^{\alpha\beta} e^{-s\Delta_2} \right] \, ,
 \end{aligned} \\
 &\begin{aligned}
  \mathcal F_5^2 &= -\int_0^\infty \text{d}t \, \mathcal L^{-1}\left[ \gamma_S(\Delta_2) \right](t) \, t \int_0^1 \text{d}\alpha \int_0^\infty \text{d}s \, \mathcal L^{-1}\left[ e^{-t\alpha\Delta_2} \mathcal R^\text{TL}(\Delta_2) \mathcal G^\text{TL}(\Delta_2) \right](s) \times \\
  &\hspace{8cm} \text{Tr}_\text{TL} \left[ \left\{ D_{(\gamma} D_{\delta)} e^{-t(1-\alpha)\Delta_2} S_{\alpha\beta} \right\} A_2^{\alpha\beta\gamma\delta} e^{-s\Delta_2} \right] \, ,
 \end{aligned} \\
 &\begin{aligned}
  \mathcal F_5^3 &= -\int_0^\infty \text{d}t \, \mathcal L^{-1}\left[ \gamma_S(\Delta_2) \right](t) \, t \int_0^1 \text{d}\alpha \int_0^\infty \text{d}s \, \mathcal L^{-1}\left[ e^{-t\alpha\Delta_2} \mathcal R^\text{TL}(\Delta_2) \mathcal G^\text{TL}(\Delta_2) \right](s) \times \\
  &\hspace{8cm} \text{Tr}_\text{TL} \left[ \left\{ D_\gamma e^{-t(1-\alpha)\Delta_2} S_{\alpha\beta} \right\} A_3^{\alpha\beta\gamma\delta} D_\delta e^{-s\Delta_2} \right] \, ,
 \end{aligned} \\
 &\begin{aligned}
  \mathcal F_5^4 &= -\int_0^\infty \text{d}t \, \mathcal L^{-1}\left[ \gamma_S(\Delta_2) \right](t) \, t \int_0^1 \text{d}\alpha \int_0^\infty \text{d}s \, \mathcal L^{-1}\left[ e^{-t\alpha\Delta_2} \mathcal R^\text{TL}(\Delta_2) \mathcal G^\text{TL}(\Delta_2) \right](s) \times \\
  &\hspace{8cm} \text{Tr}_\text{TL} \left[ \left\{ e^{-t(1-\alpha)\Delta_2} S_{\alpha\beta} \right\} A_4^{\alpha\beta\gamma\delta} D_{(\gamma} D_{\delta)} e^{-s\Delta_2} \right] \, .
 \end{aligned}
\end{align}
Note that since $A_1$ is linear in curvature, $\mathcal F_5^1$ only needs the flat heat kernel. As a matter of fact, one finds
\begin{equation}
 \mathcal F_5^1 \propto \left[A_1^{\phantom{0}\alpha\beta}\right]_{\tau\omega}^{\phantom{\tau\omega}\kappa\lambda} \Pi^{\text{TL}\phantom{\kappa\lambda}\tau\omega}_{\phantom{\text{TL}}\kappa\lambda} = 0 \, .
\end{equation}
We also note that for the other contributions, we only need the heat kernel linear in curvature. The flat part drops out, because $S$ is traceless, and we are not interested in total derivatives which could result from contractions of the form $D^\mu S_{\mu\nu} = \frac{1}{2} D_\nu R$. For $\mathcal F_5^4$ one might expect that there could be a contribution from the flat heat kernel since $\Delta_2$ also has curvature terms. It turns however out that the corresponding contraction vanishes.

Besides the gravitational contribution, there is also the ghost contribution $\mathcal F_c$, which we display in \autoref{sec:ghosttrace}.

\subsection{Graviton contribution}

We now compute the gravitational contribution to the \ac{RG} flow term by term, showing all intermediate steps.

\subsubsection{\texorpdfstring{$\mathcal F_1$}{F1}}
\paragraph{\texorpdfstring{$\mathcal F_1^{\text{TL},1}$}{F1TL1}}

The first contribution is simply given by the trace of the operator $\Delta_2$:
\begin{equation}
\begin{aligned}
 \mathcal F_1^{\text{TL},1} = \frac{1}{2} &\int_0^\infty \text{d}s \, \mathcal L^{-1}\left[ \left(\mathring{\mathcal R}^\text{TL}(\Delta_2) + 2\gamma_g \mathcal R^\text{TL}(\Delta_2)\right) \mathcal G^\text{TL}(\Delta_2) \right](s) \times \\
 & \left( \frac{1}{4\pi s} \right)^2 \int \text{d}^4x \, \sqrt{g} \Bigg[ 9 - \frac{9}{2} s \, R + \frac{21}{20} s^2 \, \Euler{} + s^2 \, R \, \HKbracket{\frac{17}{32}, -\frac{15}{8}, \frac{9}{8}}{s\Delta} R + s^2 \, S^{\mu\nu} \, \HKbracket{3,12,9}{s\Delta} S_{\mu\nu} \Bigg] \, .
\end{aligned}
\end{equation}
With the formulas for the inverse Laplace transform collected in the previous appendix, we can rewrite this as
\begin{equation}
\begin{aligned}
 \mathcal F_1^{\text{TL},1} &= \frac{1}{32\pi^2} \int \text{d}^4x \, \sqrt{g} \Bigg[ 9 \int_0^\infty \text{d}z \, z \, \left(\mathring{\mathcal R}^\text{TL}(z) + 2\gamma_g \mathcal R^\text{TL}(z)\right) \mathcal G^\text{TL}(z) \\
 &\hspace{3cm} - \frac{9}{2} R \int_0^\infty \text{d}z \, \left(\mathring{\mathcal R}^\text{TL}(z) + 2\gamma_g \mathcal R^\text{TL}(z)\right) \mathcal G^\text{TL}(z) + \frac{21}{20} \left(\mathring{\mathcal R}^\text{TL}(0) + 2\gamma_g \mathcal R^\text{TL}(0)\right) \mathcal G^\text{TL}(0) \, \Euler{} \\
 &\hspace{3cm} + \frac{1}{32} R \int_0^\frac{1}{4} \text{d}u \, \measure{17, -60, 36}{u} \, \left(\mathring{\mathcal R}^\text{TL}(u\Delta) + 2\gamma_g \mathcal R^\text{TL}(u\Delta)\right) \mathcal G^\text{TL}(u\Delta) \, R \\
 &\hspace{3cm} + 3 \, S^{\mu\nu} \int_0^\frac{1}{4} \text{d}u \, \measure{1,4,3}{u} \, \left(\mathring{\mathcal R}^\text{TL}(u\Delta) + 2\gamma_g \mathcal R^\text{TL}(u\Delta)\right) \mathcal G^\text{TL}(u\Delta) \, S_{\mu\nu} \Bigg] \, .
\end{aligned}
\end{equation}
Here we pulled out factors so that the measure has integer arguments which are relatively prime and as small as possible.

\paragraph{\texorpdfstring{$\mathcal F_1^{\text{TL},2}$}{F1TL2}}

The second contribution reads
\begin{equation}
 \mathcal F_1^{\text{TL},2} = \int_0^\infty \text{d}s \, \mathcal L^{-1}\left[ \mathcal R^\text{TL} (\Delta_2) \mathcal G^\text{TL}(\Delta_2)\right](s) \, \left( \frac{1}{4\pi s} \right)^2 \int \text{d}^4x \, \sqrt{g} \, \left( \gamma_R(\Delta) R \right) \Bigg[ 9 + 9\, s \, \HKbracket{-\frac{5}{12}, \frac{1}{2}}{s\Delta} R \Bigg] \, .
\end{equation}
This reduces to
\begin{equation}
\begin{aligned}
 \mathcal F_1^{\text{TL},2} &= \frac{1}{16\pi^2} \int \text{d}^4x \, \sqrt{g} \Bigg[ 9 \, \gamma_R(0) \, R \int_0^\infty \text{d}z \, z \, \mathcal R^\text{TL}(z) \mathcal G^\text{TL}(z) \\
 &\hspace{1.5cm} + \frac{9}{2} \left( \gamma_R(\Delta) R \right) \left[ \int_0^\frac{1}{4} \text{d}u \, \measure{0, -\frac{5}{6}, 1}{u} \, \Delta \, \mathcal R^\text{TL}(u\Delta) \, \mathcal G^\text{TL}(u\Delta) - \int_0^\infty \text{d}z \, \mathcal R^\text{TL}(z) \mathcal G^\text{TL}(z) \right] \, R \Bigg] \, .
\end{aligned}
\end{equation}
Here we pulled out a factor so that the integrals in the last line have a unit prefactor within the brackets.

\paragraph{\texorpdfstring{$\mathcal F_1^{\text{Tr},1}$}{F1Tr1}}

The third contribution is:
\begin{equation}
\begin{aligned}
 \mathcal F_1^{\text{Tr},1} = \frac{1}{2} \int_0^\infty &\text{d}s \, \mathcal L^{-1}\left[ \left(\mathring{\mathcal R}^\text{Tr}(\Delta) + 2\gamma_g \mathcal R^\text{Tr}(\Delta)\right) \mathcal G^\text{Tr}(\Delta) \right](s) \times \\
 & \left( \frac{1}{4\pi s} \right)^2 \int \text{d}^4x \, \sqrt{g} \Bigg[ 1 + \frac{1}{6} s \, R + \frac{s^2}{180} \Euler{} + s^2 \, R \, \HKbracket{\frac{1}{32},\frac{1}{8},\frac{1}{8}}{s\Delta} R + s^2 \, S^{\mu\nu} \, \HKbracket{0,0,1}{s\Delta} S_{\mu\nu} \Bigg] \, .
\end{aligned}
\end{equation}
This simplifies into
\begin{equation}
\begin{aligned}
 \mathcal F_1^{\text{Tr},1} &= \frac{1}{32\pi^2} \int \text{d}^4x \, \sqrt{g} \Bigg[ \int_0^\infty \text{d}z \, z \, \left(\mathring{\mathcal R}^\text{Tr}(z) + 2\gamma_g \mathcal R^\text{Tr}(z)\right) \mathcal G^\text{Tr}(z) \\
 &\hspace{3cm} + \frac{1}{6} R \int_0^\infty \text{d}z \, \left(\mathring{\mathcal R}^\text{Tr}(z) + 2\gamma_g \mathcal R^\text{Tr}(z)\right) \mathcal G^\text{Tr}(z) + \frac{1}{180} \left(\mathring{\mathcal R}^\text{Tr}(0) + 2\gamma_g \mathcal R^\text{Tr}(0)\right) \mathcal G^\text{Tr}(0) \, \Euler{} \\
 &\hspace{3cm} + \frac{1}{32} R \int_0^\frac{1}{4} \text{d}u \, \measure{1, 4, 4}{u} \, \left(\mathring{\mathcal R}^\text{Tr}(u\Delta) + 2\gamma_g \mathcal R^\text{Tr}(u\Delta)\right) \mathcal G^\text{Tr}(u\Delta) \, R \\
 &\hspace{3cm} + S^{\mu\nu} \int_0^\frac{1}{4} \text{d}u \, \measure{0,0,1}{u} \, \left(\mathring{\mathcal R}^\text{Tr}(u\Delta) + 2\gamma_g \mathcal R^\text{Tr}(u\Delta)\right) \mathcal G^\text{Tr}(u\Delta) \, S_{\mu\nu} \Bigg] \, .
\end{aligned}
\end{equation}

\paragraph{\texorpdfstring{$\mathcal F_1^{\text{Tr},2}$}{F1Tr2}}

Finally, the last contribution to $\mathcal F_1$ is
\begin{equation}
 \mathcal F_1^{\text{Tr},2} = \int_0^\infty \text{d}s \, \mathcal L^{-1}\left[ \mathcal R^\text{Tr} (\Delta_2) \mathcal G^\text{Tr}(\Delta_2)\right](s) \, \left( \frac{1}{4\pi s} \right)^2 \int \text{d}^4x \, \sqrt{g} \, \left( \gamma_R(\Delta) R \right) \Bigg[ 1 + s \, \HKbracket{\frac{1}{4},\frac{1}{2}}{s\Delta} R \Bigg] \, .
\end{equation}
This gives
\begin{equation}
\begin{aligned}
 \mathcal F_1^{\text{Tr},2} &= \frac{1}{16\pi^2} \int \text{d}^4x \, \sqrt{g} \Bigg[ \gamma_R(0) \, R \int_0^\infty \text{d}z \, z \, \mathcal R^\text{Tr}(z) \mathcal G^\text{Tr}(z) \\
 &\hspace{2cm} + \frac{1}{6} \left( \gamma_R(\Delta) R \right) \left[ \int_0^\frac{1}{4} \text{d}u \, \measure{0, \frac{3}{2}, 3}{u} \, \Delta \, \mathcal R^\text{Tr}(u\Delta) \, \mathcal G^\text{Tr}(u\Delta) + \int_0^\infty \text{d}z \, \mathcal R^\text{Tr}(z) \mathcal G^\text{Tr}(z) \right] \, R \Bigg] \, .
\end{aligned}
\end{equation}

\subsubsection{\texorpdfstring{$\mathcal F_2$}{F2}}

\paragraph{\texorpdfstring{$\mathcal F_2^1$}{F21}}

This contribution can be computed analogously to $\mathcal F_1^{\text{Tr},1}$. With an obvious substitution, we directly get
\begin{equation}
\begin{aligned}
 \mathcal F_2^1 &= \frac{3}{16\pi^2} \int \text{d}^4x \, \sqrt{g} \Bigg[ \int_0^\infty \text{d}z \, z^2 \, \gamma_R(z) \, \mathcal R^\text{Tr}(z) \mathcal G^\text{Tr}(z) + \frac{1}{6} R \int_0^\infty \text{d}z \, z \, \gamma_R(z) \, \mathcal R^\text{Tr}(z) \mathcal G^\text{Tr}(z) \\
 &\hspace{3cm} + \frac{1}{32} R \int_0^\frac{1}{4} \text{d}u \, \measure{1, 4, 4}{u} \, u \, \Delta \, \gamma_R(u\Delta) \, \mathcal R^\text{Tr}(u\Delta) \mathcal G^\text{Tr}(u\Delta) \, R \\
 &\hspace{3cm} + S^{\mu\nu} \int_0^\frac{1}{4} \text{d}u \, \measure{0,0,1}{u} \, u \, \Delta \, \gamma_R(u\Delta) \, \mathcal R^\text{Tr}(u\Delta) \mathcal G^\text{Tr}(u\Delta) \, S_{\mu\nu} \Bigg] \, .
\end{aligned}
\end{equation}

\paragraph{\texorpdfstring{$\mathcal F_2^2$}{F22}}

Similarly, this contribution is directly related to $\mathcal F_1^{\text{Tr},2}$, and we get
\begin{equation}
\begin{aligned}
 \mathcal F_2^2 &= -\frac{1}{16\pi^2} \int \text{d}^4x \, \sqrt{g} \Bigg[ R \int_0^\infty \text{d}z \, z \, \gamma_R(z) \, \mathcal R^\text{Tr}(z) \mathcal G^\text{Tr}(z) \\
 &\hspace{1.5cm} + \frac{1}{6} R \left[ \int_0^\frac{1}{4} \text{d}u \, \measure{0, \frac{3}{2}, 3}{u} \, \Delta \, \gamma_R(u\Delta) \, \mathcal R^\text{Tr}(u\Delta) \, \mathcal G^\text{Tr}(u\Delta) + \int_0^\infty \text{d}z \, \gamma_R(z) \, \mathcal R^\text{Tr}(z) \mathcal G^\text{Tr}(z) \right] \, R \Bigg] \, .
\end{aligned}
\end{equation}

\subsubsection{\texorpdfstring{$\mathcal F_3$}{F3}}

This contribution is again related to $\mathcal F_1^{\text{Tr},2}$, but the transformations are slightly more complicated due to the double inverse Laplace transform. Note first that there is no contribution to $R$, since there is always at least one Laplacian acting on $R$ already. We thus have
\begin{equation}
\begin{aligned}
 \mathcal F_3 = \frac{3}{2} \int_0^\infty \text{d}t \, \mathcal L^{-1}\left[ \gamma_R(\Delta) \right](t) \, t \, \int_0^1 \text{d}\alpha &\int_0^\infty \text{d}s \, \mathcal L^{-1}\left[ e^{-t\alpha\Delta}\mathcal R^\text{Tr}(\Delta) \mathcal G^\text{Tr}(\Delta) \right](s) \times \\
 \left( \frac{1}{4\pi s} \right)^2 &\int \text{d}^4x \, \sqrt{g} \, \left\{ \Delta e^{-t(1-\alpha)\Delta} R \right\} \, s \, \HKbracket{\frac{1}{4},\frac{1}{2}}{s\Delta} R \, .
\end{aligned}
\end{equation}
To bring this into a useful form, we first perform the integral over $s$ and get
\begin{equation}
\begin{aligned}
 \mathcal F_3 &= \frac{3}{32\pi^2} \int \text{d}^4x \, \sqrt{g} \, \int_0^\infty \text{d}t \, \mathcal L^{-1}\left[ \gamma_R(\Delta) \right](t) \, t \, \int_0^1 \text{d}\alpha \left\{ \Delta e^{-t(1-\alpha)\Delta} R \right\} \times \\
 &\hspace{3cm} \frac{1}{6} \left[ \int_0^\frac{1}{4} \text{d}u \, \measure{0, \frac{3}{2}, 3}{u} \, e^{-t\alpha u \Delta} \, \Delta \, \mathcal R^\text{Tr}(u\Delta) \, \mathcal G^\text{Tr}(u\Delta) + \int_0^\infty \text{d}z \, e^{-t\alpha z} \, \mathcal R^\text{Tr}(z) \, \mathcal G^\text{Tr}(z) \right] \, R \, .
\end{aligned}
\end{equation}
We can now do a partial integration on $\Delta e^{-t(1-\alpha)\Delta}$ and perform the integral over $\alpha$. This results in
\begin{equation}
\begin{aligned}
 \mathcal F_3 &= -\frac{1}{64\pi^2} \int \text{d}^4x \, \sqrt{g} \, \int_0^\infty \text{d}t \, \mathcal L^{-1}\left[ \gamma_R(\Delta) \right](t) \, R \, \times \\
 &\hspace{1cm} \left[ \int_0^\frac{1}{4} \text{d}u \, \measure{0, \frac{3}{2}, 3}{u} \, \frac{e^{-t \Delta} - e^{-t u \Delta}}{1-u} \, \mathcal R^\text{Tr}(u\Delta) \, \mathcal G^\text{Tr}(u\Delta) + \int_0^\infty \text{d}z \, \frac{e^{-t\Delta} - e^{-tz}}{\Delta - z} \, \mathcal R^\text{Tr}(z) \, \mathcal G^\text{Tr}(z) \right] \, \Delta \, R \, .
\end{aligned}
\end{equation}
Note that the extra factor of $t$ has cancelled, and we have pulled out a factor of $-1/6$. We can finally do the integral over $t$, which is trivial, to get the result
\begin{equation}
\begin{aligned}
 \mathcal F_3 &= -\frac{1}{64\pi^2} \int \text{d}^4x \, \sqrt{g} \, R \, \Bigg[ \int_0^\frac{1}{4} \text{d}u \, \measure{0, \frac{3}{2}, 3}{u} \, \frac{\gamma_R(\Delta) - \gamma_R(u \Delta)}{1-u} \, \mathcal R^\text{Tr}(u\Delta) \, \mathcal G^\text{Tr}(u\Delta) \\
 &\hspace{7cm} + \int_0^\infty \text{d}z \, \frac{\gamma_R(\Delta) - \gamma_R(z)}{\Delta - z} \, \mathcal R^\text{Tr}(z) \, \mathcal G^\text{Tr}(z) \Bigg] \, \Delta \, R \, .
\end{aligned}
\end{equation}

\subsubsection{\texorpdfstring{$\mathcal F_4$}{F4}}

\paragraph{\texorpdfstring{$\mathcal F_4^1$}{F41}}

This contribution is similar to $\mathcal F_1^{\text{TL},1}$, and we can directly infer
\begin{equation}
\begin{aligned}
 \mathcal F_4^1 &= \frac{1}{32\pi^2} \int \text{d}^4x \, \sqrt{g} \Bigg[ 9 \int_0^\infty \text{d}z \, z^2 \, \gamma_S(z) \, \mathcal R^\text{TL}(z) \mathcal G^\text{TL}(z) - \frac{9}{2} R \int_0^\infty \text{d}z \, z \, \gamma_S(z) \, \mathcal R^\text{TL}(z) \mathcal G^\text{TL}(z) \\
 &\hspace{3cm} + \frac{1}{32} R \int_0^\frac{1}{4} \text{d}u \, \measure{17, -60, 36}{u} \, u \, \Delta \, \gamma_S(u\Delta) \, \mathcal R^\text{TL}(u\Delta) \mathcal G^\text{TL}(u\Delta) \, R \\
 &\hspace{3cm} + 3 \, S^{\mu\nu} \int_0^\frac{1}{4} \text{d}u \, \measure{1,4,3}{u} \, u \, \Delta \, \gamma_S(u\Delta) \, \mathcal R^\text{TL}(u\Delta) \mathcal G^\text{TL}(u\Delta) \, S_{\mu\nu} \Bigg] \, .
\end{aligned}
\end{equation}

\paragraph{\texorpdfstring{$\mathcal F_4^2$}{F42}}

The computation of this contribution is analogous to that of $\mathcal F_1^{\text{TL},2}$, and we find
\begin{equation}
\begin{aligned}
 \mathcal F_4^2 &= -\frac{5}{192\pi^2} \int \text{d}^4x \, \sqrt{g} \Bigg[ 9 \, R \int_0^\infty \text{d}z \, z \, \gamma_S(z) \, \mathcal R^\text{TL}(z) \mathcal G^\text{TL}(z) \\
 &\hspace{1.5cm} + \frac{9}{2} R \left[ \int_0^\frac{1}{4} \text{d}u \, \measure{0, -\frac{5}{6}, 1}{u} \, \Delta \, \gamma_S(u\Delta) \, \mathcal R^\text{TL}(u\Delta) \, \mathcal G^\text{TL}(u\Delta) - \int_0^\infty \text{d}z \, \gamma_S(z) \mathcal R^\text{TL}(z) \mathcal G^\text{TL}(z) \right] \, R \Bigg] \, .
\end{aligned}
\end{equation}

\paragraph{\texorpdfstring{$\mathcal F_4^3$}{F43}}

This contribution is different from previous contributions. The only contribution to the order considered comes from the endomorphism:
\begin{equation}
\begin{aligned}
 \mathcal F_4^3 &= \frac{1}{2} \int_0^\infty \text{d}s \, \mathcal L^{-1}\left[ \gamma_S(\Delta_2) \mathcal R^\text{TL}(\Delta_2) \mathcal G^\text{TL}(\Delta_2) \right](s) \, \left( \frac{1}{4\pi s} \right)^2 \int \text{d}^4x \, \sqrt{g} \, \mathbbm C \, 2 \, s \, f(s \Delta) \, \mathbbm C \\
 &= \frac{1}{16\pi^2} \int \text{d}^4x \, \sqrt{g} \int_0^\infty \text{d}s \, \mathcal L^{-1}\left[ \gamma_S(\Delta_2) \mathcal R^\text{TL}(\Delta_2) \mathcal G^\text{TL}(\Delta_2) \right](s) \, \frac{3}{4s} C^{\mu\nu\rho\sigma} f(s\Delta) C_{\mu\nu\rho\sigma} \\
 &= \frac{3}{64\pi^2} \int \text{d}^4x \, \sqrt{g} \int_0^\infty \text{d}s \, \mathcal L^{-1}\left[ \gamma_S(\Delta_2) \mathcal R^\text{TL}(\Delta_2) \mathcal G^\text{TL}(\Delta_2) \right](s) \, \left[ \frac{1}{s} \Euler{} - \frac{1}{6s} R \, f(s\Delta) \, R + \frac{2}{s} S^{\mu\nu} \, f(s\Delta) \, S_{\mu\nu} \right] \, .
\end{aligned}
\end{equation}
Using the standard rules, this can be written as
\begin{equation}
\begin{aligned}
 \mathcal F_4^3 &= \frac{3}{64\pi^2} \int \text{d}^4x \, \sqrt{g} \Bigg[ \Euler{} \, \int_0^\infty \text{d}z \, \gamma_S(z) \, \mathcal R^\text{TL}(z) \, \mathcal G^\text{TL}(z) \\
 &\hspace{1.5cm} - \frac{1}{6} R \, \left( \int_0^\frac{1}{4} \text{d}u \, \measure{0,1}{u} \, \Delta \, \gamma_S(u\Delta) \, \mathcal R^\text{TL}(u\Delta) \, \mathcal G^\text{TL}(u\Delta) + \int_0^\infty \text{d}z \, \gamma_S(z) \, \mathcal R^\text{TL}(z) \, \mathcal G^\text{TL}(z) \right) \, R \\
 &\hspace{1.5cm} + 2 S^{\mu\nu} \, \left( \int_0^\frac{1}{4} \text{d}u \, \measure{0,1}{u} \, \Delta \, \gamma_S(u\Delta) \, \mathcal R^\text{TL}(u\Delta) \, \mathcal G^\text{TL}(u\Delta) + \int_0^\infty \text{d}z \, \gamma_S(z) \, \mathcal R^\text{TL}(z) \, \mathcal G^\text{TL}(z) \right) \, S_{\mu\nu} \Bigg] \, .
\end{aligned}
\end{equation}

\paragraph{\texorpdfstring{$\mathcal F_4^4$}{F44}}

It turns out that the traces for this contribution vanish identically,
\begin{equation}
 \mathcal F_4^4 = 0 \, .
\end{equation}

\paragraph{\texorpdfstring{$\mathcal F_4^5$}{F45}}

This contribution can be split into two parts. The first comes from the derivatives acting on the exponential of the world function, and gives
\begin{equation}
\begin{aligned}
 \mathcal F_4^{5,1} &= \int_0^\infty \text{d}s \, \mathcal L^{-1}\left[ \gamma_S(\Delta_2) \mathcal R^\text{TL}(\Delta_2) \mathcal G^\text{TL}(\Delta_2) \right](s) \, \text{Tr}_\text{TL} \left[ \mathbbm X^{\alpha\beta} \left( -\frac{1}{2s} g_{\alpha\beta} \right) e^{-s\Delta_2} \right] \\ 
 &= -\frac{1}{2} \int_0^\infty \text{d}s \, \mathcal L^{-1}\left[ \gamma_S(\Delta_2) \mathcal R^\text{TL}(\Delta_2) \mathcal G^\text{TL}(\Delta_2) \right](s) \, \frac{1}{s} \, \text{Tr}_\text{TL} \left[ e^{-s\Delta_2} \right] \, .
\end{aligned}
\end{equation}
This follows from
\begin{equation}
 \mathbbm X^{\alpha\beta} g_{\alpha\beta} = \Pi^\text{TL} \, .
\end{equation}
This contribution is very similar to $\mathcal F_1^{\text{TL},1}$. We have
\begin{equation}
\begin{aligned}
 \mathcal F_4^{5,1} = -\frac{1}{2} &\int_0^\infty \text{d}s \, \mathcal L^{-1}\left[ \gamma_S(\Delta_2) \mathcal R^\text{TL}(\Delta_2) \mathcal G^\text{TL}(\Delta_2) \right](s) \times \\
 & \frac{1}{s} \, \left( \frac{1}{4\pi s} \right)^2 \int \text{d}^4x \, \sqrt{g} \Bigg[ 9 - \frac{9}{2} s \, R + \frac{21}{20} s^2 \, \Euler{} + s^2 \, R \, \HKbracket{\frac{17}{32},-\frac{15}{8},\frac{9}{8}}{s\Delta} R + s^2 \, S^{\mu\nu} \, \HKbracket{3,12,9}{s\Delta} S_{\mu\nu} \Bigg] \, .
\end{aligned}
\end{equation}
This converts into
\begin{equation}
\begin{aligned}
 \mathcal F_4^{5,1} &= -\frac{1}{32\pi^2} \int \text{d}^4x \, \sqrt{g} \Bigg[ \frac{9}{2} \int_0^\infty \text{d}z \, z^2 \, \gamma_S(z) \, \mathcal R^\text{TL}(z) \, \mathcal G^\text{TL}(z) \\
 & - \frac{9}{2} R \int_0^\infty \text{d}z \, z \, \gamma_S(z) \, \mathcal R^\text{TL}(z) \, \mathcal G^\text{TL}(z) + \frac{21}{20} \Euler{} \int_0^\infty \text{d}z \, \gamma_S(z) \, \mathcal R^\text{TL}(z) \, \mathcal G^\text{TL}(z) \\
 & + \frac{69}{80} R \, \Bigg\{ \int_0^\frac{1}{4} \text{d}u \, \measure{0, \frac{85}{138}, -\frac{50}{23}, \frac{30}{23}}{u} \, \Delta \, \gamma_S(u\Delta) \, \mathcal R^\text{TL}(u\Delta) \, \mathcal G^\text{TL}(u\Delta) + \int_0^\infty \text{d}z \, \gamma_S(z) \, \mathcal R^\text{TL}(z) \, \mathcal G^\text{TL}(z) \Bigg\} \, R \\
 & + \frac{23}{20} S^{\mu\nu} \, \Bigg\{ \int_0^\frac{1}{4} \text{d}u \, \measure{0, \frac{60}{23}, \frac{240}{23}, \frac{180}{23}}{u} \, \Delta \, \gamma_S(u\Delta) \, \mathcal R^\text{TL}(u\Delta) \, \mathcal G^\text{TL}(u\Delta) + \int_0^\infty \text{d}z \, \gamma_S(z) \, \mathcal R^\text{TL}(z) \, \mathcal G^\text{TL}(z) \Bigg\} \, S_{\mu\nu} \Bigg] \, .
\end{aligned}
\end{equation}
For the second part, inserting the trace and simplifying gives
\begin{equation}
\begin{aligned}
 \mathcal F_4^{5,2} &= \int_0^\infty \text{d}s \, \mathcal L^{-1}\left[ \gamma_S(\Delta_2) \mathcal R^\text{TL}(\Delta_2) \mathcal G^\text{TL}(\Delta_2) \right](s) \, \left( \frac{1}{4\pi s} \right)^2 \int \text{d}^4x \, \sqrt{g} \\
 &\hspace{0.25cm} \Bigg[ \frac{3}{8} R - \frac{133}{480} s \, R^2 - \frac{151}{120} s \, S^{\mu\nu} S_{\mu\nu} - \frac{9}{40} s \, \Euler{} \\
 &\hspace{0.5cm} + s \, R \, s \, \Delta \, \HKbracket{-\frac{17}{512},-\frac{47}{256},\frac{121}{128},-\frac{9}{64}}{s\Delta} R + s \, S^{\mu\nu} \, s \, \Delta \, \HKbracket{-\frac{3}{16},-\frac{13}{8},-\frac{1}{16},-\frac{9}{8}}{s\Delta} S_{\mu\nu} \Bigg] \\
 &= \frac{1}{16\pi^2} \int_0^\infty \text{d}s \, \mathcal L^{-1}\left[ \gamma_S(\Delta_2) \mathcal R^\text{TL}(\Delta_2) \mathcal G^\text{TL}(\Delta_2) \right](s) \, \int \text{d}^4x \, \sqrt{g} \\
 &\hspace{0.25cm} \Bigg[ \frac{3}{8s^2} R - \frac{133}{480s} R^2 - \frac{151}{120s} S^{\mu\nu} S_{\mu\nu} - \frac{9}{40s} \Euler{} \\
 &\hspace{0.5cm} + R \, \Delta \, \HKbracket{-\frac{17}{512},-\frac{47}{256},\frac{121}{128},-\frac{9}{64}}{s\Delta} R + S^{\mu\nu} \, \Delta \, \HKbracket{-\frac{3}{16},-\frac{13}{8},-\frac{1}{16},-\frac{9}{8}}{s\Delta} S_{\mu\nu} \Bigg] \, .
\end{aligned}
\end{equation}
With the standard formulas, this gives the contribution
\begin{equation}
\begin{aligned}
 \mathcal F_4^{5,2} = \frac{1}{16\pi^2} &\int \text{d}^4x \, \sqrt{g} \, \Bigg[ \frac{3}{8} R \, \int_0^\infty \text{d}z \, z \, \gamma_S(z) \, \mathcal R^\text{TL}(z) \, \mathcal G^\text{TL}(z) - \frac{133}{480} R^2 \, \int_0^\infty \text{d}z \, \gamma_S(z) \, \mathcal R^\text{TL}(z) \, \mathcal G^\text{TL}(z) \\
 & - \frac{151}{120} S^{\mu\nu} S_{\mu\nu} \, \int_0^\infty \text{d}z \, \gamma_S(z) \, \mathcal R^\text{TL}(z) \, \mathcal G^\text{TL}(z) - \frac{9}{40} \Euler{} \, \int_0^\infty \text{d}z \, \gamma_S(z) \, \mathcal R^\text{TL}(z) \, \mathcal G^\text{TL}(z) \\
 & + R \, \int_0^\frac{1}{4} \text{d}u \, \measure{-\frac{17}{512}, -\frac{47}{256}, \frac{121}{128}, -\frac{9}{64}}{u} \, \Delta \, \gamma_S(u\Delta) \, \mathcal R^\text{TL}(u\Delta) \, \mathcal G^\text{TL}(u\Delta) \, R \\
 & + S^{\mu\nu} \, \int_0^\frac{1}{4} \text{d}u \, \measure{-\frac{3}{16}, -\frac{13}{8}, -\frac{1}{16}, -\frac{9}{8}}{u} \, \Delta \, \gamma_S(u\Delta) \, \mathcal R^\text{TL}(u\Delta) \, \mathcal G^\text{TL}(u\Delta) \, S_{\mu\nu} \Bigg] \, .
\end{aligned}
\end{equation}

\subsubsection{\texorpdfstring{$\mathcal F_5$}{F5}}

\paragraph{\texorpdfstring{$\mathcal F_5^1$}{F51}}

As previously noted, this contribution vanishes,
\begin{equation}
 \mathcal F_5^1 = 0 \, .
\end{equation}

\paragraph{\texorpdfstring{$\mathcal F_5^2$}{F52}}

For this contribution, we find
\begin{equation}
\begin{aligned}
 \mathcal F_5^2 &= -\int_0^\infty \text{d}t \, \mathcal L^{-1}\left[ \gamma_S(\Delta_2) \right](t) \, t \int_0^1 \text{d}\alpha \int_0^\infty \text{d}s \, \mathcal L^{-1}\left[ e^{-t\alpha\Delta_2} \mathcal R^\text{TL}(\Delta_2) \mathcal G^\text{TL}(\Delta_2) \right](s) \times \\
  & \left( \frac{1}{4\pi s} \right)^2 \int \text{d}^4x \, \sqrt{g} \, \left\{ D_{(\gamma} D_{\delta)} e^{-t(1-\alpha)\Delta_2} S_{\alpha\beta} \right\} \, \left[ \text{tr} \left\{ A_2^{\alpha\beta\gamma\delta} \Pi^\text{TL} \right\} \, s \, \HKbracket{-\frac{5}{12},\frac{1}{2}}{s\Delta} \, R + s \, \HKbracket{2}{s\Delta} \, \text{tr} \left\{ A_2^{\alpha\beta\gamma\delta} \, \mathbbm C \right\} \right] \, .
\end{aligned}
\end{equation}
The two needed contractions are
\begin{align}
 \left\{ D_{(\gamma} D_{\delta)} e^{-t(1-\alpha)\Delta_2} S_{\alpha\beta} \right\} \text{tr} \left\{ A_2^{\alpha\beta\gamma\delta} \Pi^\text{TL} \right\} &\simeq -\frac{3}{2} \Delta e^{-t(1-\alpha)\Delta} R \, , \\
 \left\{ D_{(\gamma} D_{\delta)} e^{-t(1-\alpha)\Delta_2} S_{\alpha\beta} \right\} \, f(s\Delta) \, \text{tr} \left\{ A_2^{\alpha\beta\gamma\delta} \, \mathbbm C \right\} &\simeq -\frac{3}{2} S^{\mu\nu} \, \Delta \, f(s\Delta) \, e^{-t(1-\alpha)\Delta} \, S_{\mu\nu} + \frac{1}{8} R \, f(s\Delta) \, \Delta \, e^{-t(1-\alpha)\Delta} \, R \, ,
\end{align}
where we commuted derivatives and integrated them by parts freely. With this, we find
\begin{equation}
\begin{aligned}
 \mathcal F_5^2 &= -\int_0^\infty \text{d}t \, \mathcal L^{-1}\left[ \gamma_S(\Delta_2) \right](t) \, t \int_0^1 \text{d}\alpha \int_0^\infty \text{d}s \, \mathcal L^{-1}\left[ e^{-t\alpha\Delta_2} \mathcal R^\text{TL}(\Delta_2) \mathcal G^\text{TL}(\Delta_2) \right](s) \times \\
  &\hspace{0.5cm} \left( \frac{1}{4\pi s} \right)^2 \int \text{d}^4x \, \sqrt{g} \, \Bigg[ S^{\mu\nu} \, s \, \Delta \, \HKbracket{-3}{s\Delta} \, e^{-t(1-\alpha)\Delta} \, S_{\mu\nu} + R \, s \, \Delta \, \HKbracket{\frac{7}{8},-\frac{3}{4}}{s\Delta} e^{-t(1-\alpha)\Delta} \, R \Bigg] \, .
\end{aligned}
\end{equation}
Next, we perform the integral over s and get
\begin{equation}
\begin{aligned}
 \mathcal F_5^2 &= -\frac{1}{16\pi^2} \int \text{d}^4x \, \sqrt{g} \int_0^\infty \text{d}t \, \mathcal L^{-1}\left[ \gamma_S(\Delta_2) \right](t) \, t \int_0^1 \text{d}\alpha \\
  &\hspace{1cm} \Bigg[ -3 S^{\mu\nu} \, \Delta \, \Bigg\{ \int_0^\frac{1}{4} \text{d}u \, \measure{0, 1}{u} \, \Delta \, e^{-t\alpha u \Delta} \, \mathcal R^\text{TL}(u\Delta) \, \mathcal G^\text{TL}(u\Delta) \\
  &\hspace{8cm} + \int_0^\infty \text{d}z \, e^{-t\alpha z} \, \mathcal R^\text{TL}(z) \, \mathcal G^\text{TL}(z) \Bigg\} \, e^{-t(1-\alpha)\Delta} \,S_{\mu\nu} \\
  &\hspace{1cm} + R \, \Delta \, \Bigg\{ \int_0^\frac{1}{4} \text{d}u \, \measure{0, \frac{7}{8}, -\frac{3}{4}}{u} \, \Delta \, e^{-t\alpha u \Delta} \, \mathcal R^\text{TL}(u\Delta) \, \mathcal G^\text{TL}(u\Delta) \\
  &\hspace{8cm} + \int_0^\infty \text{d}z \, e^{-t\alpha z} \, \mathcal R^\text{TL}(z) \, \mathcal G^\text{TL}(z) \Bigg\} \, e^{-t(1-\alpha)\Delta} \, R \Bigg] \, .
\end{aligned}
\end{equation}
Next, we perform the integral over $\alpha$:
\begin{equation}
\begin{aligned}
 \mathcal F_5^2 &= \frac{1}{16\pi^2} \int \text{d}^4x \, \sqrt{g} \int_0^\infty \text{d}t \, \mathcal L^{-1}\left[ \gamma_S(\Delta_2) \right](t) \\
  &\hspace{1cm} \Bigg[ -3 S^{\mu\nu} \, \Delta \, \Bigg\{ \int_0^\frac{1}{4} \text{d}u \, \measure{0, 1}{u} \, \frac{e^{-t\Delta} - e^{-t u \Delta}}{1-u} \, \mathcal R^\text{TL}(u\Delta) \, \mathcal G^\text{TL}(u\Delta) \\
  &\hspace{8cm} + \int_0^\infty \text{d}z \, \frac{e^{-t \Delta} - e^{-t z}}{\Delta - z} \, \mathcal R^\text{TL}(z) \, \mathcal G^\text{TL}(z) \Bigg\} \,S_{\mu\nu} \\
  &\hspace{1cm} + R \, \Delta \, \Bigg\{ \int_0^\frac{1}{4} \text{d}u \, \measure{0, \frac{7}{8}, -\frac{3}{4}}{u} \, \frac{e^{-t\Delta} - e^{-t u \Delta}}{1-u} \, \mathcal R^\text{TL}(u\Delta) \, \mathcal G^\text{TL}(u\Delta) \\
  &\hspace{8cm} + \int_0^\infty \text{d}z \, \frac{e^{-t \Delta} - e^{-t z}}{\Delta - z} \, \mathcal R^\text{TL}(z) \, \mathcal G^\text{TL}(z) \Bigg\} \, R \Bigg] \, .
\end{aligned}
\end{equation}
Finally, performing the integral over $t$, we find
\begin{equation}
\begin{aligned}
 \mathcal F_5^2 &= \frac{1}{16\pi^2} \int \text{d}^4x \, \sqrt{g} \, \Bigg[ - 3 S^{\mu\nu} \, \Delta \, \Bigg\{ \int_0^\frac{1}{4} \text{d}u \, \measure{0, 1}{u} \, \frac{\gamma_S(\Delta) - \gamma_S(u \Delta)}{1-u} \, \mathcal R^\text{TL}(u\Delta) \, \mathcal G^\text{TL}(u\Delta) \\
 &\hspace{8cm} + \int_0^\infty \text{d}z \, \frac{\gamma_S(\Delta) - \gamma_S(z)}{\Delta - z} \, \mathcal R^\text{TL}(z) \, \mathcal G^\text{TL}(z) \Bigg\} \,S_{\mu\nu} \\
  &\hspace{3cm} + R \, \Delta \, \Bigg\{ \int_0^\frac{1}{4} \text{d}u \, \measure{0, \frac{7}{8}, -\frac{3}{4}}{u} \, \frac{\gamma_S(\Delta) - \gamma_S(u \Delta)}{1-u} \, \mathcal R^\text{TL}(u\Delta) \, \mathcal G^\text{TL}(u\Delta) \\
  &\hspace{8cm} + \int_0^\infty \text{d}z \, \frac{\gamma_S(\Delta) - \gamma_S(z)}{\Delta - z} \, \mathcal R^\text{TL}(z) \, \mathcal G^\text{TL}(z) \Bigg\} \, R \Bigg] \, .
\end{aligned}
\end{equation}

\paragraph{\texorpdfstring{$\mathcal F_5^3$}{F53}}

The heat kernel splits this contribution into two. The first is related to $\mathcal F_5^2$, and reads
\begin{equation}
\begin{aligned}
 \mathcal F_5^{3,1} &= \int_0^\infty \text{d}t \, \mathcal L^{-1}\left[ \gamma_S(\Delta_2) \right](t) \, t \int_0^1 \text{d}\alpha \int_0^\infty \text{d}s \, \mathcal L^{-1}\left[ e^{-t\alpha\Delta_2} \mathcal R^\text{TL}(\Delta_2) \mathcal G^\text{TL}(\Delta_2) \right](s) \times \\
  & \left( \frac{1}{4\pi s} \right)^2 \int \text{d}^4x \, \sqrt{g} \, \left\{ D_\delta D_\gamma e^{-t(1-\alpha)\Delta_2} S_{\alpha\beta} \right\} \, \left[ \text{tr} \left\{ A_3^{\alpha\beta\gamma\delta} \Pi^\text{TL} \right\} \, s \, \HKbracket{-\frac{5}{24},\frac{1}{4}}{s\Delta} R + s \, \HKbracket{1}{s\Delta} \, \text{tr} \left\{ A_3^{\alpha\beta\gamma\delta} \, \mathbbm C \right\} \right] \, .
\end{aligned}
\end{equation}
The second part reads
\begin{equation}
\begin{aligned}
 \mathcal F_5^{3,2} &= \int_0^\infty \text{d}t \, \mathcal L^{-1}\left[ \gamma_S(\Delta_2) \right](t) \, t \int_0^1 \text{d}\alpha \int_0^\infty \text{d}s \, \mathcal L^{-1}\left[ e^{-t\alpha\Delta_2} \mathcal R^\text{TL}(\Delta_2) \mathcal G^\text{TL}(\Delta_2) \right](s) \times \\
  &\hspace{1cm} \left( \frac{1}{4\pi s} \right)^2 \int \text{d}^4x \, \sqrt{g} \, \left\{ D^\kappa D_\gamma e^{-t(1-\alpha)\Delta_2} S_{\alpha\beta} \right\} s \, \HKbracket{0,1}{s\Delta} \, \text{tr} \left\{ A_3^{\alpha\beta\gamma\delta} \mathcal F_{\delta\kappa} \right\} \, .
\end{aligned}
\end{equation}
We need the following contractions:
\begin{align}
 \left\{ D_\delta D_\gamma e^{-t(1-\alpha)\Delta_2} S_{\alpha\beta} \right\} \text{tr} \left\{ A_3^{\alpha\beta\gamma\delta} \Pi^\text{TL} \right\} &\simeq -\frac{7}{4} \Delta e^{-t(1-\alpha)\Delta} R \, , \\
 \left\{ D_\delta D_\gamma e^{-t(1-\alpha)\Delta_2} S_{\alpha\beta} \right\} \, f(s\Delta) \, \text{tr} \left\{ A_3^{\alpha\beta\gamma\delta} \, \mathbbm C \right\} &\simeq -2 S^{\mu\nu} \, \Delta \, f(s\Delta) \, e^{-t(1-\alpha)\Delta} \, S_{\mu\nu} \nonumber \\
 &\hspace{2cm} + \frac{1}{6} R \, \Delta \, f(s\Delta) \, e^{-t(1-\alpha)\Delta} \, R \, , \\
 \left\{ D^\kappa D_\gamma e^{-t(1-\alpha)\Delta_2} S_{\alpha\beta} \right\} \, \frac{f(s\Delta)-1}{s\Delta} \, \text{tr} \left\{ A_3^{\alpha\beta\gamma\delta} \, \Pi^\text{TL} \, \mathcal F_{\delta\kappa} \, \Pi^\text{TL} \right\} &\simeq 4 S^{\mu\nu} \, \Delta \, \frac{f(s\Delta)-1}{s\Delta} \, e^{-t(1-\alpha)\Delta} \, S_{\mu\nu} \nonumber \\
 &\hspace{2cm} - \frac{1}{2} R \, \Delta \, \frac{f(s\Delta)-1}{s\Delta} \, e^{-t(1-\alpha)\Delta} \, R \, .
\end{align}
where once again we have freely integrated by parts and commuted derivatives. Inserting these traces gives
\begin{equation}
\begin{aligned}
 \mathcal F_5^3 &= -\int_0^\infty \text{d}t \, \mathcal L^{-1}\left[ \gamma_S(\Delta_2) \right](t) \, t \int_0^1 \text{d}\alpha \int_0^\infty \text{d}s \, \mathcal L^{-1}\left[ e^{-t\alpha\Delta_2} \mathcal R^\text{TL}(\Delta_2) \mathcal G^\text{TL}(\Delta_2) \right](s) \times \\
  &\hspace{0.5cm} \left( \frac{1}{4\pi s} \right)^2 \int \text{d}^4x \, \sqrt{g} \, \Bigg[ S^{\mu\nu} \, s \, \Delta \, \HKbracket{2,-4}{s\Delta} e^{-t(1-\alpha)\Delta} \, S_{\mu\nu} + R \, s \, \Delta \, \HKbracket{-\frac{17}{32},\frac{15}{16}}{s\Delta} e^{-t(1-\alpha)\Delta} \, R \Bigg] \, .
\end{aligned}
\end{equation}
Once again we perform the different integrals step by step. The integral over $s$ gives
\begin{equation}
\begin{aligned}
 \mathcal F_5^3 &= -\frac{1}{16\pi^2} \int \text{d}^4x \, \sqrt{g} \int_0^\infty \text{d}t \, \mathcal L^{-1}\left[ \gamma_S(\Delta_2) \right](t) \, t \int_0^1 \text{d}\alpha \\
  &\hspace{1cm} \Bigg[ \frac{8}{3} S^{\mu\nu} \, \Delta \, \Bigg\{ \int_0^\frac{1}{4} \text{d}u \, \measure{0, \frac{3}{4}, -\frac{3}{2}}{u} \, \Delta \, e^{-t\alpha u \Delta} \, \mathcal R^\text{TL}(u\Delta) \, \mathcal G^\text{TL}(u\Delta) \\
  &\hspace{8cm} + \int_0^\infty \text{d}z \, e^{-t\alpha z} \, \mathcal R^\text{TL}(z) \, \mathcal G^\text{TL}(z) \Bigg\} \, e^{-t(1-\alpha)\Delta} \,S_{\mu\nu} \\
  &\hspace{1cm} + \frac{11}{16} R \, \Delta \, \Bigg\{ \int_0^\frac{1}{4} \text{d}u \, \measure{0, -\frac{17}{22}, \frac{15}{11}}{u} \, \Delta \, e^{-t\alpha u \Delta} \, \mathcal R^\text{TL}(u\Delta) \, \mathcal G^\text{TL}(u\Delta) \\
  &\hspace{8cm} - \int_0^\infty \text{d}z \, e^{-t\alpha z} \, \mathcal R^\text{TL}(z) \, \mathcal G^\text{TL}(z) \Bigg\} \, e^{-t(1-\alpha)\Delta} \, R \Bigg] \, .
\end{aligned}
\end{equation}
Performing the integral over $\alpha$ then gives
\begin{equation}
\begin{aligned}
 \mathcal F_5^3 &= \frac{1}{16\pi^2} \int \text{d}^4x \, \sqrt{g} \int_0^\infty \text{d}t \, \mathcal L^{-1}\left[ \gamma_S(\Delta_2) \right](t) \\
  &\hspace{1cm} \Bigg[ \frac{8}{3} S^{\mu\nu} \, \Delta \, \Bigg\{ \int_0^\frac{1}{4} \text{d}u \, \measure{0, \frac{3}{4}, -\frac{3}{2}}{u} \, \frac{e^{-t\Delta} - e^{-t u \Delta}}{1-u} \, \mathcal R^\text{TL}(u\Delta) \, \mathcal G^\text{TL}(u\Delta) \\
  &\hspace{8cm} + \int_0^\infty \text{d}z \, \frac{e^{-t \Delta} - e^{-t z}}{\Delta - z} \, \mathcal R^\text{TL}(z) \, \mathcal G^\text{TL}(z) \Bigg\} \,S_{\mu\nu} \\
  &\hspace{1cm} + \frac{11}{16} R \, \Delta \, \Bigg\{ \int_0^\frac{1}{4} \text{d}u \, \measure{0, -\frac{17}{22}, \frac{15}{11}}{u} \, \frac{e^{-t\Delta} - e^{-t u \Delta}}{1-u} \, \mathcal R^\text{TL}(u\Delta) \, \mathcal G^\text{TL}(u\Delta) \\
  &\hspace{8cm} - \int_0^\infty \text{d}z \, \frac{e^{-t \Delta} - e^{-t z}}{\Delta - z} \, \mathcal R^\text{TL}(z) \, \mathcal G^\text{TL}(z) \Bigg\} \, R \Bigg] \, .
\end{aligned}
\end{equation}
Finally, with the integral over $t$, we arrive at
\begin{equation}
\begin{aligned}
 \mathcal F_5^3 &= \frac{1}{16\pi^2} \int \text{d}^4x \, \sqrt{g} \, \Bigg[ \frac{8}{3} S^{\mu\nu} \, \Delta \, \Bigg\{ \int_0^\frac{1}{4} \text{d}u \, \measure{0, \frac{3}{4}, -\frac{3}{2}}{u} \, \frac{\gamma_S(\Delta) - \gamma_S(u \Delta)}{1-u} \, \mathcal R^\text{TL}(u\Delta) \, \mathcal G^\text{TL}(u\Delta) \\
 &\hspace{8cm} + \int_0^\infty \text{d}z \, \frac{\gamma_S(\Delta) - \gamma_S(z)}{\Delta - z} \, \mathcal R^\text{TL}(z) \, \mathcal G^\text{TL}(z) \Bigg\} \,S_{\mu\nu} \\
  &\hspace{3cm} + \frac{11}{16} R \, \Delta \, \Bigg\{ \int_0^\frac{1}{4} \text{d}u \, \measure{0, -\frac{17}{22}, \frac{15}{11}}{u} \, \frac{\gamma_S(\Delta) - \gamma_S(u \Delta)}{1-u} \, \mathcal R^\text{TL}(u\Delta) \, \mathcal G^\text{TL}(u\Delta) \\
  &\hspace{8cm} - \int_0^\infty \text{d}z \, \frac{\gamma_S(\Delta) - \gamma_S(z)}{\Delta - z} \, \mathcal R^\text{TL}(z) \, \mathcal G^\text{TL}(z) \Bigg\} \, R \Bigg] \, .
\end{aligned}
\end{equation}

\paragraph{\texorpdfstring{$\mathcal F_5^4$}{F54}}

The final trace gets two contributions. The first is from the term where the covariant derivatives of the heat kernel act on the exponential of the world function, which actually vanishes:
\begin{equation}
\begin{aligned}
 \mathcal F_5^{4,1} &= \int_0^\infty \text{d}t \, \mathcal L^{-1}\left[ \gamma_S(\Delta_2) \right](t) \, t \int_0^1 \text{d}\alpha \int_0^\infty \text{d}s \, \mathcal L^{-1}\left[ e^{-t\alpha\Delta_2} \mathcal R^\text{TL}(\Delta_2) \mathcal G^\text{TL}(\Delta_2) \right](s) \times \\
  &\hspace{8cm} \frac{1}{2s} \text{Tr}_\text{TL} \left[ \left\{ e^{-t(1-\alpha)\Delta_2} S_{\alpha\beta} \right\} A_4^{\alpha\beta\gamma\delta} g_{\gamma\delta} e^{-s\Delta_2} \right] \\
  &= 0 \, .
\end{aligned}
\end{equation}
The second contribution reads
\begin{equation}
\begin{aligned}
 \mathcal F_5^{4,2} &= -\int_0^\infty \text{d}t \, \mathcal L^{-1}\left[ \gamma_S(\Delta_2) \right](t) \, t \int_0^1 \text{d}\alpha \int_0^\infty \text{d}s \, \mathcal L^{-1}\left[ e^{-t\alpha\Delta_2} \mathcal R^\text{TL}(\Delta_2) \mathcal G^\text{TL}(\Delta_2) \right](s) \times \\
  &\hspace{1cm} \left( \frac{1}{4\pi s} \right)^2 \int \text{d}^4x \, \sqrt{g} \, \left\{ e^{-t(1-\alpha)\Delta_2} S_{\alpha\beta} \right\} \Bigg[ \HKbracket{0,-1}{s\Delta} \text{tr} \left\{ A_4^{\alpha\beta\gamma\delta} R_{\gamma\delta} \Pi^\text{TL} \right\} \\
  &\hspace{1.5cm} + s \, \HKbracket{-\frac{5}{48},\frac{1}{3},-\frac{3}{4}}{s\Delta} \text{tr} \left\{ A_4^{\alpha\beta\gamma\delta} D_{(\gamma} D_{\delta)} R \, \Pi^\text{TL} \right\} + s \, \HKbracket{\frac{1}{2},-1}{s\Delta} \text{tr} \left\{ A_4^{\alpha\beta\gamma\delta} D_{(\gamma} D_{\delta)} \mathbbm C \right\} \\
  &\hspace{1.5cm} + s \, \HKbracket{0,1}{s\Delta} \text{tr} \left\{ A_4^{\alpha\beta\gamma\delta} D_{(\gamma} D^{\kappa} \mathcal F_{\delta)\kappa} \right\} \Bigg] \, .
\end{aligned}
\end{equation}
For this, we need to compute the traces
\begin{align}
 \left\{ e^{-t(1-\alpha)\Delta_2} S_{\alpha\beta} \right\} \left( -\frac{f(s\Delta)-1}{s\Delta} \right) \text{tr} \left\{ A_4^{\alpha\beta\gamma\delta} R_{\gamma\delta} \Pi^\text{TL} \right\} &\simeq -2 S^{\mu\nu} \, \frac{f(s\Delta)-1}{s\Delta} \, e^{-t(1-\alpha)\Delta} \, S_{\mu\nu} \, , \\
 \left\{ e^{-t(1-\alpha)\Delta_2} S_{\alpha\beta} \right\} B(s\Delta) \, \text{tr} \left\{ A_4^{\alpha\beta\gamma\delta} D_{(\gamma} D_{\delta)} R \, \Pi^\text{TL} \right\} &\simeq -\frac{1}{2} R \, \Delta \, B(s\Delta) \, e^{-t(1-\alpha)\Delta} \, R \, , \\
 \left\{ e^{-t(1-\alpha)\Delta_2} S_{\alpha\beta} \right\} C(s\Delta) \, \text{tr} \left\{ A_4^{\alpha\beta\gamma\delta} D_{(\gamma} D_{\delta)} \mathbbm C \right\} &\simeq \frac{1}{12} R \, \Delta \, C(s\Delta) \, e^{-t(1-\alpha)\Delta} \, R \nonumber \\
 &\hspace{2cm} - S^{\mu\nu} \, \Delta \, C(s\Delta) \, e^{-t(1-\alpha)\Delta} \, S_{\mu\nu} \, , \\
 \left\{ e^{-t(1-\alpha)\Delta_2} S_{\alpha\beta} \right\} \, \frac{f(s\Delta)-1}{s\Delta} \, \text{tr} \left\{ A_4^{\alpha\beta\gamma\delta} D_{(\gamma} D^{\kappa} \mathcal F_{\delta)\kappa} \right\} &\simeq 0 \, .
\end{align}
Due to the different structure in $s$ and $\Delta$, we will discuss the first term individually. It reads
\begin{equation}
\begin{aligned}
 \mathcal F_5^{4,2,1} &= -\int_0^\infty \text{d}t \, \mathcal L^{-1}\left[ \gamma_S(\Delta_2) \right](t) \, t \int_0^1 \text{d}\alpha \int_0^\infty \text{d}s \, \mathcal L^{-1}\left[ e^{-t\alpha\Delta_2} \mathcal R^\text{TL}(\Delta_2) \mathcal G^\text{TL}(\Delta_2) \right](s) \times \\
  &\hspace{5cm} \left( \frac{1}{4\pi s} \right)^2 \int \text{d}^4x \, \sqrt{g} \, \left\{ -2 S^{\mu\nu} \, \HKbracket{0,1}{s\Delta} e^{-t(1-\alpha)\Delta} \, S_{\mu\nu} \right\} \, .
\end{aligned}
\end{equation}
Performing the integral over $s$ gives
\begin{equation}
\begin{aligned}
 \mathcal F_5^{4,2,1} &= \frac{1}{8\pi^2} \int \text{d}^4x \, \sqrt{g} \int_0^\infty \text{d}t \, \mathcal L^{-1}\left[ \gamma_S(\Delta_2) \right](t) \, t \int_0^1 \text{d}\alpha \\
  &\hspace{1cm} S^{\mu\nu} \Bigg\{ \Delta^2 \int_0^\frac{1}{4} \text{d}u \, \measure{0,0,0,1}{u} \, e^{-t\alpha u \Delta} \, \mathcal R^\text{TL}(u\Delta) \, \mathcal G^\text{TL}(u\Delta) \\
  &\hspace{2cm} - \frac{1}{6} \int_0^\infty \text{d}z \, z \, e^{-t\alpha z} \, \mathcal R^\text{TL}(z) \, \mathcal G^\text{TL}(z) + \frac{\Delta}{60} \int_0^\infty \text{d}z \, e^{-t\alpha z} \, \mathcal R^\text{TL}(z) \, \mathcal G^\text{TL}(z) \Bigg\} \, e^{-t(1-\alpha)\Delta} \, S_{\mu\nu} \, .
\end{aligned}
\end{equation}
Next, with the integration over $\alpha$, we get
\begin{equation}
\begin{aligned}
 \mathcal F_5^{4,2,1} &= -\frac{1}{8\pi^2} \int \text{d}^4x \, \sqrt{g} \int_0^\infty \text{d}t \, \mathcal L^{-1}\left[ \gamma_S(\Delta_2) \right](t) \\
  &\hspace{1cm} S^{\mu\nu} \Bigg\{ \Delta \int_0^\frac{1}{4} \text{d}u \, \measure{0,0,0,1}{u} \, \frac{e^{-t\Delta} - e^{-t u \Delta}}{1-u} \, \mathcal R^\text{TL}(u\Delta) \, \mathcal G^\text{TL}(u\Delta) \\
  &\hspace{2cm} - \frac{1}{6} \int_0^\infty \text{d}z \, z \, \frac{e^{-t \Delta} - e^{-t z}}{\Delta - z} \, \mathcal R^\text{TL}(z) \, \mathcal G^\text{TL}(z) + \frac{\Delta}{60} \int_0^\infty \text{d}z \, \frac{e^{-t \Delta} - e^{-t z}}{\Delta - z} \, \mathcal R^\text{TL}(z) \, \mathcal G^\text{TL}(z) \Bigg\} \, S_{\mu\nu} \, .
\end{aligned}
\end{equation}
Finally, the integral over $t$ yields
\begin{equation}
\begin{aligned}
 \mathcal F_5^{4,2,1} &= -\frac{1}{8\pi^2} \int \text{d}^4x \, \sqrt{g} \, S^{\mu\nu} \Bigg\{ \Delta \int_0^\frac{1}{4} \text{d}u \, \measure{0,0,0,1}{u} \, \frac{\gamma_S(\Delta) - \gamma_S(u \Delta)}{1-u} \, \mathcal R^\text{TL}(u\Delta) \, \mathcal G^\text{TL}(u\Delta) \\
  &\hspace{1cm} - \frac{1}{6} \int_0^\infty \text{d}z \, z \, \frac{\gamma_S(\Delta) - \gamma_S(z)}{\Delta - z} \, \mathcal R^\text{TL}(z) \, \mathcal G^\text{TL}(z) + \frac{\Delta}{60} \int_0^\infty \text{d}z \, \frac{\gamma_S(\Delta) - \gamma_S(z)}{\Delta - z} \, \mathcal R^\text{TL}(z) \, \mathcal G^\text{TL}(z) \Bigg\} \, S_{\mu\nu} \, .
\end{aligned}
\end{equation}
The remaining contributions can be easily combined. Inserting the traces,
\begin{equation}
\begin{aligned}
 \mathcal F_5^{4,2,2} &= -\int_0^\infty \text{d}t \, \mathcal L^{-1}\left[ \gamma_S(\Delta_2) \right](t) \, t \int_0^1 \text{d}\alpha \int_0^\infty \text{d}s \, \mathcal L^{-1}\left[ e^{-t\alpha\Delta_2} \mathcal R^\text{TL}(\Delta_2) \mathcal G^\text{TL}(\Delta_2) \right](s) \times \\
  &\hspace{0.5cm} \left( \frac{1}{4\pi s} \right)^2 \int \text{d}^4x \, \sqrt{g} \, \Bigg[ S^{\mu\nu} \, s \, \Delta \, \HKbracket{-\frac{1}{2},1}{s\Delta} e^{-t(1-\alpha)\Delta} \, S_{\mu\nu} + R \, s \, \Delta \, \HKbracket{\frac{3}{32},-\frac{1}{4},\frac{3}{8}}{s\Delta} e^{-t(1-\alpha)\Delta} \, R \Bigg] \, .
\end{aligned}
\end{equation}
The integration over $s$ gives then
\begin{equation}
\begin{aligned}
 \mathcal F_5^{4,2,2} &= -\frac{1}{16\pi^2} \int \text{d}^4x \, \sqrt{g} \int_0^\infty \text{d}t \, \mathcal L^{-1}\left[ \gamma_S(\Delta_2) \right](t) \, t \int_0^1 \text{d}\alpha \\
  &\hspace{1cm} \Bigg[ \frac{2}{3} S^{\mu\nu} \, \Delta \, \Bigg\{ \int_0^\frac{1}{4} \text{d}u \, \measure{0, -\frac{3}{4}, \frac{3}{2}}{u} \, \Delta \, e^{-t\alpha u \Delta} \, \mathcal R^\text{TL}(u\Delta) \, \mathcal G^\text{TL}(u\Delta) \\
  &\hspace{8cm} - \int_0^\infty \text{d}z \, e^{-t\alpha z} \, \mathcal R^\text{TL}(z) \, \mathcal G^\text{TL}(z) \Bigg\} \, e^{-t(1-\alpha)\Delta} \,S_{\mu\nu} \\
  &\hspace{1cm} + \frac{17}{120} R \, \Delta \, \Bigg\{ \int_0^\frac{1}{4} \text{d}u \, \measure{0, \frac{45}{68}, -\frac{30}{17}, \frac{45}{17}}{u} \, \Delta \, e^{-t\alpha u \Delta} \, \mathcal R^\text{TL}(u\Delta) \, \mathcal G^\text{TL}(u\Delta) \\
  &\hspace{8cm} + \int_0^\infty \text{d}z \, e^{-t\alpha z} \, \mathcal R^\text{TL}(z) \, \mathcal G^\text{TL}(z) \Bigg\} \, e^{-t(1-\alpha)\Delta} \, R \Bigg] \, .
\end{aligned}
\end{equation}
Integrating over $\alpha$, we arrive at
\begin{equation}
\begin{aligned}
 \mathcal F_5^{4,2,2} &= \frac{1}{16\pi^2} \int \text{d}^4x \, \sqrt{g} \int_0^\infty \text{d}t \, \mathcal L^{-1}\left[ \gamma_S(\Delta_2) \right](t) \\
  &\hspace{1cm} \Bigg[ \frac{2}{3} S^{\mu\nu} \, \Delta \, \Bigg\{ \int_0^\frac{1}{4} \text{d}u \, \measure{0, -\frac{3}{4}, \frac{3}{2}}{u} \, \frac{e^{-t\Delta} - e^{-t u \Delta}}{1-u} \, \mathcal R^\text{TL}(u\Delta) \, \mathcal G^\text{TL}(u\Delta) \\
  &\hspace{8cm} - \int_0^\infty \text{d}z \, \frac{e^{-t \Delta} - e^{-t z}}{\Delta - z} \, \mathcal R^\text{TL}(z) \, \mathcal G^\text{TL}(z) \Bigg\} \, S_{\mu\nu} \\
  &\hspace{1cm} + \frac{17}{120} R \, \Delta \, \Bigg\{ \int_0^\frac{1}{4} \text{d}u \, \measure{0, \frac{45}{68}, -\frac{30}{17}, \frac{45}{17}}{u} \, \frac{e^{-t\Delta} - e^{-t u \Delta}}{1-u} \, \mathcal R^\text{TL}(u\Delta) \, \mathcal G^\text{TL}(u\Delta) \\
  &\hspace{8cm} + \int_0^\infty \text{d}z \, \frac{e^{-t \Delta} - e^{-t z}}{\Delta - z} \, \mathcal R^\text{TL}(z) \, \mathcal G^\text{TL}(z) \Bigg\} \, R \Bigg] \, .
\end{aligned}
\end{equation}
Finally, the integral over $t$ gives
\begin{equation}
\begin{aligned}
 \mathcal F_5^{4,2,2} &= \frac{1}{16\pi^2} \int \text{d}^4x \, \sqrt{g} \, \Bigg[ \frac{2}{3} S^{\mu\nu} \, \Delta \, \Bigg\{ \int_0^\frac{1}{4} \text{d}u \, \measure{0, -\frac{3}{4}, \frac{3}{2}}{u} \, \frac{\gamma_S(\Delta) - \gamma_S(u \Delta)}{1-u} \, \mathcal R^\text{TL}(u\Delta) \, \mathcal G^\text{TL}(u\Delta) \\
  &\hspace{8.5cm} - \int_0^\infty \text{d}z \, \frac{\gamma_S(\Delta) - \gamma_S(z)}{\Delta - z} \, \mathcal R^\text{TL}(z) \, \mathcal G^\text{TL}(z) \Bigg\} \, S_{\mu\nu} \\
  &\hspace{3cm} + \frac{17}{120} R \, \Delta \, \Bigg\{ \int_0^\frac{1}{4} \text{d}u \, \measure{0, \frac{45}{68}, -\frac{30}{17}, \frac{45}{17}}{u} \, \frac{\gamma_S(\Delta) - \gamma_S(u \Delta)}{1-u} \, \mathcal R^\text{TL}(u\Delta) \, \mathcal G^\text{TL}(u\Delta) \\
  &\hspace{8.5cm} + \int_0^\infty \text{d}z \, \frac{\gamma_S(\Delta) - \gamma_S(z)}{\Delta - z} \, \mathcal R^\text{TL}(z) \, \mathcal G^\text{TL}(z) \Bigg\} \, R \Bigg] \, .
\end{aligned}
\end{equation}

\subsection{Ghost contribution}\label{sec:ghosttrace}

The computation of the ghost trace is straightforward, as we only need the diagonal part of the non-local heat kernel in our setup, \eqref{eq:HKc}. We find
\begin{equation}
\begin{aligned}
 \mathcal F_c = - &\int_0^\infty \text{d}s \, \mathcal L^{-1}\left[ \mathring{\mathcal R}^\text{c}(\Delta_c) \, \mathcal G^\text{c}(\Delta_c) \right](s) \times \\
 & \left( \frac{1}{4\pi s} \right)^2 \int \text{d}^4x \, \sqrt{g} \Bigg[ 4 + \frac{5}{3} s \, R - \frac{11}{180} s^2 \, \Euler{} + s^2 \, R \, \HKbracket{\frac{1}{2},1,\frac{1}{2}}{s\Delta} R + s^2 \, S^{\mu\nu} \, \HKbracket{\frac{1}{2},2,4}{s\Delta} S_{\mu\nu} \Bigg] \, .
\end{aligned}
\end{equation}
This can be transformed in the standard way to yield
\begin{equation}
\begin{aligned}
 \mathcal F_c &= -\frac{1}{16\pi^2} \int \text{d}^4x \, \sqrt{g} \Bigg[ 4 \int_0^\infty \text{d}z \, z \, \mathring{\mathcal R}^\text{c}(z) \, \mathcal G^\text{c}(z) + \frac{5}{3} R \int_0^\infty \text{d}z \, \mathring{\mathcal R}^\text{c}(z) \, \mathcal G^\text{c}(z) - \frac{11}{180} \mathring{\mathcal R}^\text{c}(0) \, \mathcal G^\text{c}(0) \, \Euler{} \\
 &\hspace{1.5cm} + \frac{1}{2} R \int_0^\frac{1}{4} \text{d}u \, \measure{1, 2, 1}{u} \, \mathring{\mathcal R}^\text{c}(u\Delta) \, \mathcal G^\text{c}(u\Delta) \, R + \frac{1}{2} S^{\mu\nu} \int_0^\frac{1}{4} \text{d}u \, \measure{1, 4, 8}{u} \, \mathring{\mathcal R}^\text{c}(u\Delta) \, \mathcal G^\text{c}(u\Delta) \, S_{\mu\nu} \Bigg] \, .
\end{aligned}
\end{equation}

\subsection{Grand total}\label{sec:totaltrace}

We now collect the complete expression for the trace, operator by operator. The volume term of the trace reads
\begin{equation}\label{eq:flow1}
\begin{aligned}
 \mathcal F_{\mathbbm 1} &= \frac{1}{2} \frac{1}{16\pi^2} \int \text{d}^4x \, \sqrt{g} \, \Bigg[ 9 \int_0^\infty \text{d}z \, z \, \left( \mathring{\mathcal R}^\text{TL}(z) + \left( 2 \gamma_g + \frac{z}{2} \gamma_S(z) \right) \mathcal R^\text{TL}(z) \right) \, \mathcal G^\text{TL}(z) \\
 &\hspace{1.5cm} + \int_0^\infty \text{d}z \, z \, \left( \mathring{\mathcal R}^\text{Tr}(z) + 2 \left( \gamma_g + 3z \gamma_R(z) \right) \mathcal R^\text{Tr}(z) \right) \, \mathcal G^\text{Tr}(z) - 8 \int_0^\infty \text{d}z \, z \, \mathring{\mathcal R}^\text{c}(z) \, \mathcal G^\text{c}(z) \Bigg] \, .
\end{aligned}
\end{equation}
For the Einstein-Hilbert term, we find
\begin{equation}\label{eq:flowR}
\begin{aligned}
 \mathcal F_{R} &= \frac{1}{2} \frac{1}{16\pi^2} \int \text{d}^4x \, \sqrt{g} \, \Bigg[ -\frac{9}{2} \int_0^\infty \text{d}z \, \left( \mathring{\mathcal R}^\text{TL}(z) + \left( 2 \gamma_g - 4z \, \gamma_R(0) + \frac{3}{2} z \, \gamma_S(z) \right) \mathcal R^\text{TL}(z) \right) \, \mathcal G^\text{TL}(z) \\
 &\hspace{1.5cm} + \frac{1}{6} \int_0^\infty \text{d}z \, \left( \mathring{\mathcal R}^\text{Tr}(z) + 2 \left( \gamma_g + 6 \, z \, (2\gamma_R(0) - \gamma_R(z)) \right) \mathcal R^\text{Tr}(z) \right) \, \mathcal G^\text{Tr}(z) - \frac{10}{3} \int_0^\infty \text{d}z \, \mathring{\mathcal R}^\text{c}(z) \, \mathcal G^\text{c}(z) \Bigg] \, R \, .
\end{aligned}
\end{equation}
Moving on to the Ricci scalar form factor, we have
\begin{equation}\label{eq:flowFFR}
\begin{aligned}
 \mathcal F_{R^2} &= \frac{1}{2} \frac{1}{16\pi^2} \int \text{d}^4x \, \sqrt{g} \, R \, \Bigg[ \int_0^\frac{1}{4} \text{d}u \, \measure{\frac{17}{32}, -\frac{15}{8}, \frac{9}{8}}{u}  \, \left(\mathring{\mathcal R}^\text{TL}(u\Delta) + 2\gamma_g \mathcal R^\text{TL}(u\Delta)\right) \mathcal G^\text{TL}(u\Delta) \\
 &\hspace{3.75cm} + 9 \Delta \, \gamma_R(\Delta) \, \int_0^\frac{1}{4} \text{d}u \, \measure{0, -\frac{5}{6}, 1}{u}  \, \mathcal R^\text{TL}(u\Delta) \mathcal G^\text{TL}(u\Delta) \\
 &\hspace{3.75cm} - 9 \, \gamma_R(\Delta) \, \int_0^\infty \text{d}z \, \mathcal R^\text{TL}(z) \, \mathcal G^\text{TL}(z) \\
 &\hspace{3.75cm} + \frac{25}{12} \Delta \, \int_0^\frac{1}{4} \text{d}u \, \measure{\frac{51}{1600}, \frac{681}{800}, -\frac{483}{400}, \frac{27}{40}}{u} \, \gamma_S(u\Delta) \, \mathcal R^\text{TL}(u\Delta) \mathcal G^\text{TL}(u\Delta) \\
 &\hspace{3.75cm} + \frac{25}{12} \int_0^\infty \text{d}z \, \gamma_S(z) \, \mathcal R^\text{TL}(z) \, \mathcal G^\text{TL}(z) \\
 &\hspace{3.75cm} + \frac{109}{120} \, \Delta \, \int_0^\frac{1}{4} \text{d}u \, \measure{0, \frac{105}{109}, -\frac{15}{109}, \frac{90}{109}}{u} \, \frac{\gamma_S(\Delta) - \gamma_S(u\Delta)}{1-u} \, \mathcal R^\text{TL}(u\Delta) \, \mathcal G^\text{TL}(u\Delta) \\
 &\hspace{3.75cm} + \frac{109}{120} \, \Delta \, \int_0^\infty \text{d}z \, \frac{\gamma_S(\Delta) - \gamma_S(z)}{\Delta - z} \, \mathcal R^\text{TL}(z) \, \mathcal G^\text{TL}(z) \\
 &\hspace{3.75cm} + \int_0^\frac{1}{4} \text{d}u \, \measure{\frac{1}{32}, \frac{1}{8}, \frac{1}{8}}{u}  \, \left(\mathring{\mathcal R}^\text{Tr}(u\Delta) + 2\gamma_g \mathcal R^\text{Tr}(u\Delta)\right) \mathcal G^\text{Tr}(u\Delta) \\
 &\hspace{3.75cm} + \frac{1}{3} \Delta \, \gamma_R(\Delta) \, \int_0^\frac{1}{4} \text{d}u \, \measure{0, \frac{3}{2}, 3}{u}  \, \mathcal R^\text{Tr}(u\Delta) \mathcal G^\text{Tr}(u\Delta) \\
 &\hspace{3.75cm} + \frac{1}{3} \, \gamma_R(\Delta) \, \int_0^\infty \text{d}z \, \mathcal R^\text{Tr}(z) \, \mathcal G^\text{Tr}(z) \\
 &\hspace{3.75cm} + \frac{1}{3} \Delta \, \int_0^\frac{1}{4} \text{d}u \, \measure{\frac{9}{64}, -\frac{21}{32}, \frac{15}{16}, \frac{45}{8}}{u} \, \gamma_R(u\Delta) \, \mathcal R^\text{Tr}(u\Delta) \mathcal G^\text{Tr}(u\Delta) \\
 &\hspace{3.75cm} - \frac{1}{3} \int_0^\infty \text{d}z \, \gamma_R(z) \, \mathcal R^\text{Tr}(z) \, \mathcal G^\text{Tr}(z) \\
 &\hspace{3.75cm} + \frac{1}{2} \, \Delta \, \int_0^\frac{1}{4} \text{d}u \, \measure{0, -\frac{3}{2}, -3}{u} \, \frac{\gamma_R(\Delta) - \gamma_R(u\Delta)}{1-u} \, \mathcal R^\text{Tr}(u\Delta) \, \mathcal G^\text{Tr}(u\Delta) \\
 &\hspace{3.75cm} - \frac{1}{2} \, \Delta \, \int_0^\infty \text{d}z \, \frac{\gamma_R(\Delta) - \gamma_R(z)}{\Delta - z} \, \mathcal R^\text{Tr}(z) \, \mathcal G^\text{Tr}(z) \\
 &\hspace{3.75cm} + \int_0^\frac{1}{4} \text{d}u \, \measure{-1,-2,-1}{u} \, \mathring{\mathcal R}^\text{c}(u\Delta) \, \mathcal G^\text{c}(u\Delta) \Bigg] \, R \, .
\end{aligned}
\end{equation}
Here we used that we can absorb extra factors of $u$ in the measure, $\mu(\{a_i\},u) u = \mu(\{b_i\},u)$ for suitably chosen $b_i$, to bring all integrals into a uniform form.
Similarly, for the tracefree Ricci tensor, we have
\begin{equation}
\begin{aligned}
 \mathcal F_{S^2} &= \frac{1}{2} \frac{1}{16\pi^2} \int \text{d}^4x \, \sqrt{g} \, S^{\mu\nu} \, \Bigg[ \int_0^\frac{1}{4} \text{d}u \, \measure{3,12,9}{u}  \, \left(\mathring{\mathcal R}^\text{TL}(u\Delta) + 2\gamma_g \mathcal R^\text{TL}(u\Delta)\right) \mathcal G^\text{TL}(u\Delta) \\
 &\hspace{3.75cm} + \frac{2}{3} \Delta \, \int_0^\frac{1}{4} \text{d}u \, \measure{\frac{9}{16}, \frac{15}{8}, \frac{195}{16}, \frac{135}{8}}{u} \, \gamma_S(u\Delta) \, \mathcal R^\text{TL}(u\Delta) \mathcal G^\text{TL}(u\Delta) \\
 &\hspace{3.75cm} - \frac{2}{3} \int_0^\infty \text{d}z \, \gamma_S(z) \, \mathcal R^\text{TL}(z) \, \mathcal G^\text{TL}(z) \\
 &\hspace{3.75cm} - \frac{31}{15} \, \Delta \, \int_0^\frac{1}{4} \text{d}u \, \measure{0, \frac{45}{31}, \frac{90}{31}, \frac{60}{31}}{u} \, \frac{\gamma_S(\Delta) - \gamma_S(u\Delta)}{1-u} \, \mathcal R^\text{TL}(u\Delta) \, \mathcal G^\text{TL}(u\Delta) \\
 &\hspace{3.75cm} - \frac{31}{15} \, \Delta \, \int_0^\infty \text{d}z \, \frac{\gamma_S(\Delta) - \gamma_S(z)}{\Delta - z} \, \mathcal R^\text{TL}(z) \, \mathcal G^\text{TL}(z) \\
 &\hspace{3.75cm} + \frac{2}{3} \, \int_0^\infty \text{d}z \, z \, \frac{\gamma_S(\Delta) - \gamma_S(z)}{\Delta - z} \, \mathcal R^\text{TL}(z) \, \mathcal G^\text{TL}(z) \\
 &\hspace{3.75cm} + \int_0^\frac{1}{4} \text{d}u \, \measure{0,0,1}{u}  \, \left(\mathring{\mathcal R}^\text{Tr}(u\Delta) + 2\gamma_g \mathcal R^\text{Tr}(u\Delta)\right) \mathcal G^\text{Tr}(u\Delta) \\
 &\hspace{3.75cm} + \Delta \, \int_0^\frac{1}{4} \text{d}u \, \measure{0,0,\frac{3}{2},15}{u} \, \gamma_R(u\Delta) \, \mathcal R^\text{Tr}(u\Delta) \mathcal G^\text{Tr}(u\Delta) \\
 &\hspace{3.75cm} + \int_0^\frac{1}{4} \text{d}u \, \measure{-1,-4,-8}{u} \, \mathring{\mathcal R}^\text{c}(u\Delta) \, \mathcal G^\text{c}(u\Delta) \Bigg] \, S_{\mu\nu} \, .
\end{aligned}
\end{equation}
This can be rewritten in the following form:
\begin{equation}\label{eq:flowFFS}
\begin{aligned}
 \mathcal F_{S^2} &= \frac{1}{2} \frac{1}{16\pi^2} \int \text{d}^4x \, \sqrt{g} \, S^{\mu\nu} \, \Bigg[ \int_0^\frac{1}{4} \text{d}u \, \measure{3,12,9}{u}  \, \left(\mathring{\mathcal R}^\text{TL}(u\Delta) + 2\gamma_g \mathcal R^\text{TL}(u\Delta)\right) \mathcal G^\text{TL}(u\Delta) \\
 &\hspace{3.75cm} + \frac{2}{3} \Delta \, \int_0^\frac{1}{4} \text{d}u \, \measure{\frac{9}{16}, \frac{15}{8}, \frac{195}{16}, \frac{135}{8}}{u} \, \gamma_S(u\Delta) \, \mathcal R^\text{TL}(u\Delta) \mathcal G^\text{TL}(u\Delta) \\
 &\hspace{3.75cm} - \frac{31}{15} \, \Delta \, \int_0^\frac{1}{4} \text{d}u \, \measure{0, \frac{45}{31}, \frac{90}{31}, \frac{60}{31}}{u} \, \frac{\gamma_S(\Delta) - \gamma_S(u\Delta)}{1-u} \, \mathcal R^\text{TL}(u\Delta) \, \mathcal G^\text{TL}(u\Delta) \\
 &\hspace{3.75cm} - \frac{7}{5} \, \Delta \, \int_0^\infty \text{d}z \, \frac{\gamma_S(\Delta) - \gamma_S(z)}{\Delta - z} \, \mathcal R^\text{TL}(z) \, \mathcal G^\text{TL}(z) \\
 &\hspace{3.75cm} - \frac{2}{3} \, \gamma_S(\Delta) \, \int_0^\infty \text{d}z \, \mathcal R^\text{TL}(z) \, \mathcal G^\text{TL}(z) \\
 &\hspace{3.75cm} + \int_0^\frac{1}{4} \text{d}u \, \measure{0,0,1}{u}  \, \left(\mathring{\mathcal R}^\text{Tr}(u\Delta) + 2\gamma_g \mathcal R^\text{Tr}(u\Delta)\right) \mathcal G^\text{Tr}(u\Delta) \\
 &\hspace{3.75cm} + \Delta \, \int_0^\frac{1}{4} \text{d}u \, \measure{0,0,\frac{3}{2},15}{u} \, \gamma_R(u\Delta) \, \mathcal R^\text{Tr}(u\Delta) \mathcal G^\text{Tr}(u\Delta) \\
 &\hspace{3.75cm} + \int_0^\frac{1}{4} \text{d}u \, \measure{-1,-4,-8}{u} \, \mathring{\mathcal R}^\text{c}(u\Delta) \, \mathcal G^\text{c}(u\Delta) \Bigg] \, S_{\mu\nu} \, .
\end{aligned}
\end{equation}
Finally, for the Euler density, we find
\begin{equation}\label{eq:flowE}
 \mathcal F_{\Euler{}} = \frac{1}{32\pi^2} \int \text{d}^4x \, \sqrt{g} \, \Bigg[ \frac{21}{20} \left(\mathring{\mathcal R}^\text{TL}(0) + 2\gamma_g \mathcal R^\text{TL}(0)\right) \mathcal G^\text{TL}(0) + \frac{1}{180} \left(\mathring{\mathcal R}^\text{Tr}(0) + 2\gamma_g \mathcal R^\text{Tr}(0)\right) \mathcal G^\text{Tr}(0) + \frac{11}{90} \, \mathring{\mathcal R}^\text{c}(0) \, \mathcal G^\text{c}(0) \Bigg] \, \Euler{} \, .
\end{equation}
Remarkably, all integral contributions to this vanish, as do all contributions from $\gamma_R$ and $\gamma_S$.

\end{widetext}

\bibliography{general_bib}

\end{document}